# An MML-based tool for evaluating the complexity of (stochastic) logic theories

Héctor Castillo Andreu

Departament de Sistemes Informàtics i Computadors
Universitat Politècnica de València

Advisors: José Hernández-Orallo,
Maria José Ramírez-Quintana and Cèsar Ferri
September 2012

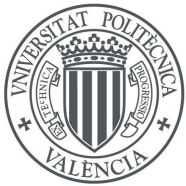

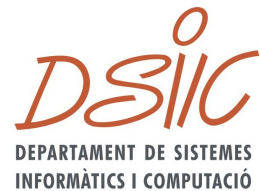

# Abstract

Theory evaluation is a key problem in many areas: machine learning, scientific discovery, inverse engineering, decision making, software engineering, design, human sciences, etc. If we have a set of theories that are able to explain the same set of phenomena, we need a criterion to choose which one is best. There are, of course, many possible criteria. Model simplicity is one of the most common criteria in theory evaluation. The Minimum Message Length (MML) is a solid approach to evaluate theories relative to a given evidence or data.

Theories can be expressed in specific or general (Turing-complete) languages. First-order logic, and logic programming in particular, is a Turing-complete language. Evaluating the simplicity of a theory or program described in a Turing-complete language is much more difficult than just counting the number of lines or bits. It is, in fact, the problem of calculating its Kolmogorov complexity, which is uncomputable. Few works in the literature have been able to present accurate and effective approximations for a Turing-complete language.

In this work, we present the first general MML coding scheme for logic programs. With this scheme, we can quantify the bits of information required to code (or send) a theory, a set of data or the same data given the theory.

Moreover, we extend the expressiveness of the language to stochastic logic programs, which are not only able to model the truth value of any set of phenomena, but also their probability. As a result, we extend the coding scheme to stochastic logic programs.

This opens up the applicability of model selection to many different problems which have a stochastic or probabilistic character, such as games, social phenomena, language processing, Markov processes, etc.

As a realization of the above-mentioned schemes, we present a software tool which is able to code and evaluate a set of alternative (stochastic) theories (programs) against a set of examples. We illustrate the application of the tool to a variety of non-probabilistic and probabilistic scenarios.

# List of Contents

# List of Figures and Tables

# List of Source Code

# Chapter 1

# Introduction

Induction involves the generalization of facts into patterns. Patterns can be expressed in the form of rules, equations or other kind of concepts. In the process of generalizing the data (or evidence), patterns try to eliminate redundancy, a key issue in inference processes such as inductive learning and abstraction. Different reasoning systems, individuals, contexts or procedures lead to different models or theories (see Figure 1.1), with varied syntactic and semantic properties. Hence, the evaluation of theories with respect to data is a very common problem.

Imagine now that we have a group of models for a given phenomena. The question is to find which one of these models is best, that is, has the lowest redundancy, the best condensation of the problem. The purpose of this work is a system that numerically evaluates a model, by using a criterion known as the Minimum Message Length (MML) [56], which we will describe later on.

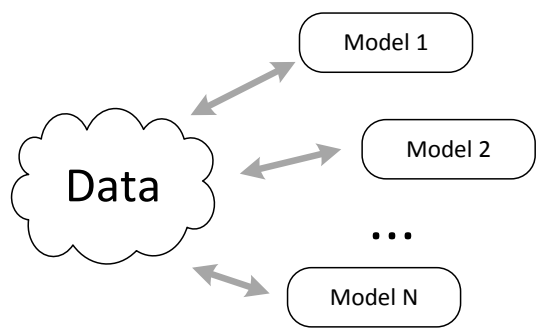

Figure 1.1: Same data could apply to different models

This number must represent something: one pervasive preference criterion is Ockham's principle of parsimony, which states that simple explanations are preferable, provided anything else is similar. This principle was incarnated in terms of information theory as the cost in bits of transferring the evidence using a model from a sender to a receiver. This cost (or information) is the MML value in bits, and the best theory will be the one which leads to the shortest message.

We will use first-order logic as the language for expressing theories and models. In particular, we will use logic programs, which are Turing-complete. This will allow us to evaluate any possible theory: provided we express it as a logic program. We will present an MML coding such that, given a set of examples (the evidence) and a logic program (a theory), it will output a number representing how much information (in bits) is required in order to transmit the theory and the evidence using the theory.

## 1.1   A simple example

Imagine we have a normal deck of French cards with four suits and usual ranks, and that we use the predicate `card/2` for representing the whole deck:

```
suit(clubs).
suit(hearts).
suit(diamonds).
suit(spades).

rank(ace).
rank(2).
...
rank(10).
rank(jack).
rank(queen).
rank(king).

card(R,S):- suit(S), rank(R).
```

If the deck is not rigged, a model that represents the suit and the rank of the deck using labeled clauses with probabilities is:

```
0.25 :: card(clubs,_).
0.25 :: card(hearts,_).
0.25 :: card(spades,_).
0.25 :: card(diamonds,_).

0.0769 :: card(_,ace). % 1/13
...
```

The previous program leads to a probability of 1/52 for every card in the pack. But let us see what happens if we introduce some tricks in the deck. For example, imagine that we remove three cards (8 of hearts, 7 of diamonds and 6 of spades) and we repeat the experiment. With this alteration, a possible alternative model (but certainly not accurate):

```
0.2653 :: card(clubs,_).    % 13/49
0.2448 :: card(hearts,_).   % 12/49
0.2448 :: card(spades,_).   % 12/49
0.2448 :: card(diamonds,_).% 12/49

0.0816 :: card(_,10).       %  4/49
...
0.0612 :: card(_, 8).       %  3/49
0.0612 :: card(_, 7).       %  3/49
0.0612 :: card(_, 5).       %  3/49
```

The only difference between the two models lies in the probabilities, because we do not have completely removed any rank or suit, so the rules in the program are the same. This means that a MML evaluation of these two models will give lower or higher scores depending on the frequency of appearance of the cards. The optimal situation is when the probabilities given by the program are higher for the events that we have observed, because in this case fewer bits are invested for examples that do not appear.

Consider now a more altered deck without clubs:

```
0.3333 :: card(hearts,_).   % 13/39
0.3333 :: card(spades,_).   % 13/39
0.3333 :: card(diamonds,_). % 13/39

0.1026 :: card(_,10).       %  4/39
...
```

In this case, not only do the rule probabilities change, but also the rules themselves: we do not have the rule `card(clubs,_)`. In general, we would need to measure the cost of the program and the cost of coding the examples given that program. The best compromise leads to the shortest coding and hence the best MML choice.

## 1.2  Objectives

Many possible models using predicate `card/2` could be derived by a human expert or an inductive inference system. As a result, if we have more than a model that applies for a set of observations, the problem is then to determine the best one. It is not a question of computational (execution) efficiency, but rather a question of compactness and coverage of the evidence. The main objective of this work is **to develop a MML coding scheme for logic and stochastic logic programs**, and **to implement a software tool** **able to calculate the cost using this coding for several programs and sets of data**.

From this general objective, some more specific objectives follow:

- To develop the most appropriate MML coding schema for logic programs. For this, it will be compulsory to first review the state of the art about the MML principle and related notions, and also to review some other previous work on coding logic programs.

- To develop a tool that implements that coding scheme. To achieve this objective, the subgoals will be:

  1. To select a logic language (Prolog) to make the implementation.
  2. To analyze the different ways to implement probabilities.
  3. To insert them into the selected language.
  4. To develop the tool using it.

- To analyze and test the developed tool in different scenarios, some specific examples will be studied and tested.

- To determine the advantages of the developed tool.

The ultimate purpose of achieving these goals is to have a tool to apply the MML principle to a universal (Turing-complete) language. Also there is another important advantage: by having a tool we can look for new areas where the principle can be applied.

## 1.3   Organization

The thesis is organized in six chapters and an appendix with five sections. The document also includes a list of references at the end.

The following chapter reviews concepts like logic programming, induction, stochasticity, logic models and the MML principle. This chapter also includes a comparison of previous approximations to the problem of theory evaluation in general, and specific approaches known as Model Complexity ($\mathcal{MC}$) and Proof Complexity ($\mathcal{PC}$) in particular.

In the third chapter, we deal with the coding scheme. This coding scheme has been developed in an incremental way. So, first we will show how to do this coding for non-stochastic programs (also analyzing the different parts: heads, bodies, variables, signature...) and later on we will deal with the stochastic programs and the examples.

The fourth chapter is focused on the description of the implemented tool: the transformations done to the programs of the model that we want to analyze, how we have introduced the probabilities in Prolog, how the evidence and the probabilities together lead to the development of a meta-interpreter to determine the probability of each example, and finally, a section for the work conducted to manage free variables in the clauses.

The fifth chapter will give three examples of application, covering different areas where logic programming can be used as a representation/modeling language, with and without probabilities. With these examples we see how the tool works and we anticipate the range of possible applications.

The last chapter has the conclusions and the proposal for future work.

Appendix A is dedicated to the way we code numbers: real, rational or integer.

Appendix B explains the basic rules of the Prolog language.

Appendix D appendix explains the file organization of the tool, and how to install it.

Following it, Appendix E includes the source code of the most complete example we describe in this thesis.

# Chapter 2

# Background

In this chapter, we will first explain theory evaluation, the MML principle and its use for model selection. We will also introduce some fundamentals of logic programming, inductive logic programming and stochastic logic programming. At the end of the chapter, we will review some previous approaches for coding logical theories, such as the Model Complexity ($\mathcal{MC}$) and the Proof Complexity ($\mathcal{PC}$).

## 2.1   Theory evaluation

In the 1970s, Len Bickman, Carol Weiss and Joseph Wholey started to use the concept of "model" in the quest of a simplification of the huge amount of data in the international aid and development area. Since that initial proposal, many refinements and variations have been added to the basic concept, as we can see in [59].

A logic model from a given domain (a collection of entities) gives us an appropriate semantic entity, using a function which, for each symbol in the domain, specifies an appropriate semantic value. It is an abstract representation of a domain of information (a program, but also a project, a policy, a machine...) that is intended to produce particular results. The key is to extract the relationships (the outputs) among the resources (the inputs) that are used, and the benefits or changes that result, as a sequence of events. We can consider a specific stage of an activity, or the whole. We could analyze the process from left to right, or from top to bottom. Any of these analyses could generate a logic model that summarizes the behavior of a system in the form of patterns (not only rules) that govern it.

Many models take some inputs and generate some output data. A model must describe the semantics of those inputs and return also the outputs in the same type of representation. Since those representations are created by human beings, we must consider that the same domain of information might have many different models, and all of them could be valid.

As a result, one of the big issues in modeling, design and any other abstraction processes is to determine how well the model represents the domain. This problem is known as "theory evaluation", in very general terms.

Theory evaluation (or selection) has been addressed from many different points of view. In philosophy of science, for instance, we may find criteria for theory evaluation, such as accuracy, performance, generalizability [19] [14] [11] [54], explanatory power [52], coherence [53], consilience [24] [1], reinforcement [25], information gain [22], aesthetics [33] and, of course, simplicity [26]. Similar criteria can be found when models are used in engineering and social sciences. This usually contrasts to the view of a theory or model in software engineering and database design, especially in software requirement analysis [21], where the criteria are, apparently, different. Nonetheless, the notion of simplicity has also been advocated in software engineering as a way of reducing maintenance costs [23].

Even though the notions of accuracy and simplicity are the most common criteria for model selection and evaluation, both concepts are understood in many different ways. Most especially, the relation between accuracy and simplicity is always at stake, because a very simple model can be highly inaccurate, and a very detailed, accurate model can be extremely complex. As a result, the idea of a compromise between accuracy and simplicity has been a recurrent issue in philosophy, science and engineering.

The basic idea can be traced back to Ockham's Razor, which says that the simplest model is the best one, all the other things being equal. The original assert is attributed to William of Ockham (14th-century): "Entia non sunt multiplicanda praeter necessitatem" (Entities must not be multiplied beyond necessity). It is a good principle to explain the phenomena by the simplest possible hypothesis.

Around 1960, Ray Solomonoff founded the theory of *universal inductive inference*, which set the basics for a formal notion of inductive inference, based upon Ockham's Razor and Kolmogorov complexity (an area of research that despite its name, was endeavored by Solomonoff himself).

Kolmogorov or algorithmic complexity [32] is a measure of the amount of computational resources needed to describe the object, given a descriptive language for the strings or sequences. In other words: given a Turing machine $T$, the algorithmic complexity of a string $S$, denoted by $K_T(S)$, is the length of the shortest input given to $T$ which causes $T$ to output $S$. It is also known as algorithmic entropy. It is named after the Russian mathematician Andrey Kolmogorov.

Still, simplicity can be understood in many other ways, and not only as the length of description. For example, in computer science, and programming in particular, we are usually interested in the economy of the computational resources needed by a program, such as time and space complexity. This has led to an enormous amount of work and results in algorithmic and theoretical computer science.

---

[1] The integration of principles from different disciplines especially when forming a comprehensive theory.

Even if we restrict to "program size", the calculation of the "size", "length" or descriptional complexity of a model or theory $T$, denoted by $L(T)$, has been attempted in the past for many different kinds of representation languages: graphs can be evaluated by a function of the number of nodes and arrows, trees by the number of leaves, *propositional formulae* by the number of terms in their canonical CNF [2] expression, computer programs by the number of lines, variables or loops, etc.

As we have already said, in this work we focus on logical models (i.e., logic programs) as representation language. Logic programming is a declarative Turing-complete language, which has a simple syntax and a clear semantics. All this makes logic programs very suitable for knowledge representation.

The applications of properly evaluating the simplicity of logic programs are countless, since logic programs can be used to represent models of any parcel of the world, specifications in software engineering, including database models, cryptographic models, biological models, social models, etc.

Even though logic programs are Turing-complete, there is an important limitation when modeling stochastic processes and events, which are very frequent phenomena. Many real-world problems appear with stochastic or probabilistic evidence, where the data from which we can construct a model is probabilistic in nature. Domains such as meteorology, social processes, transportation, games, networks, just to name a few, are stochastic in nature and are better described by stochastic models.

Stochastic, probabilistic and Bayesian logic programs are extensions of logic programming to consider these domains. These approaches tighten better for phenomena where the examples in the evidence are just a sample from a probability distribution. Thus, the theory must represent not only the things that are possible and impossible, but also how likely the phenomena are. For instance, a logic program can state whether the fact `rains('10/10/2011',valencia)` is true, but a stochastic logic program can state the probability of `rains(X,valencia) :- October(X)`. In fact, the view of a logic program as defining a distribution on examples, makes it very easy to define the cost of coding an evidence $E$ given a theory $T$ as $L(E|T) = -log_{e \in E} p(e|T)$. In this way, the cost will be higher the lower the probabilities that the theory assigns to the examples and lower otherwise. This makes the coding of theory and data given the theory: $L(T) + L(E|T)$ fully in accordance with Shannon's information theory, as we will see below.

---

[2]A formula is in conjunctive normal form (CNF) if it is a conjunction of clauses, where a clause is a disjunction of literals.

## 2.2   The MML principle

The Minimum Message Length (MML) principle can be seen as a proper restatement of Ockham's Razor in the area of information theory: since models are not usually equal in goodness of fit accuracy to the observed data, the one generating the shortest overall message is more likely to be correct (including the cost of coding the exceptions or details not well covered by the theory). The message is composed of the set of rules of the model plus the evidence concisely encoded using that model. The MML principle was invented by Chris Wallace in 1968 [57].

The MML principle is related to Kolmogorov complexity [58] [32]. In fact, we know that the calculation of the shortest representation for a set of data (i.e., its Kolmogorov complexity) is undecidable in general. Most MML coding schemes can be seen, then, as robust approximations of the Kolmogorov complexity of the theory, $K(T)$ plus the Kolmogorov complexity of the data given the theory, $K(D|T)$.

Let us see now how the MML principle derives from first principles. This will help us to understand it better. If we have one hypothesis $H$ that covers the problem data $D$, and we have a way to determine (or estimate) the probabilities of each of the facts of the theory, we have enough information to apply Bayes' Theorem:

$$P(H|D) = \frac{P(H) \cdot P(D|H)}{P(D)}$$

$$\text{where} \quad \begin{aligned} &P(H) \text{ is the prior probability of hypothesis } H \\ &P(D) \text{ is the prior probability of the evidence } D \\ &P(H|D) \text{ is the posterior probability of} \\ &\qquad \text{hypothesis } H \text{ given the data} \\ &P(D|H) \text{ is the likelihood of } H \end{aligned} \quad (2.1)$$

Shannon[3] says that, in an optimal code, the message length of an event $E$, $MsgLen(E)$, where $E$ has probability $P(E)$, is given by:

$$MsgLen(E) = -log_2(P(E)) \qquad (2.2)$$

So, we have:

$$MsgLen(H|D) = MsgLen(H) + MsgLen(D|H) - MsgLen(D) \quad (2.3)$$

---

[3]From the well-known theorem in "Mathematical Theory of Communication" of Claude E. Shannon [51]

$MsgLen(H)$ can usually be estimated. The same happens with the calculation of $MsgLen(D|H)$. However, deriving $MsgLen(D)$ is more problematic.

However, we can see that for two rival hypotheses $H$ and $H'$:

$$MsgLen(H|D) - MsgLen(H'|D) = \begin{aligned} & MsgLen(H) + MsgLen(D|H) \\ & -MsgLen(H') - MsgLen(D|H') \end{aligned} \tag{2.4}$$

Clearly, the $MsgLen(D)$ is a constant that vanishes when comparing two or more theories. So if we have a situation where we need to transmit the hypothesis and the data using a Shannon's communication channel, and both sender and receiver have previously agreed on a coding for data and the hypothesis, the best hypothesis $H$ will be the one that transmits both the hypothesis and the data given that theory ($MsgLen(H) + MsgLen(D|H)$) using fewer bits.

Then the Minimum Message Length encoding gives a trade-off between hypothesis complexity, $MsgLen(H)$, and the goodness of fit to the data, $MsgLen(D|H)$. The MML principle is one way to justify and realize Ockham's razor, as we have seen before.

So, even when hypotheses are not equal in goodness of fit to the observed data, the one generating the shortest overall message is more likely to be correct (where the message consists of a statement of the model, followed by a statement of data encoded concisely using that model). This is the trade-off found by the MML principle, which is better, the better the coding is.

Using the same Ockham's Razor formalization, Jorma Rissanen in 1978 defined the concept of the Minimum Description Length (MDL) [49], which says that the best hypothesis for a given set of data is the one that minimizes the length of the theory plus the length of the data coded by the theory. Let us remark first that this concept appeared ten years after Wallace's MML principle, but MDL principle has become more popular than the MML principle.

The difference between MDL and MML is very subtle and it is a source of ongoing confusion, especially because the MDL principle has been changing over time. For a good account of the differences, see [56, sec. 10.2] [12, sec. 6.7] [11, sec. 4] . Part of the confusion comes from the way the source paper of the MML principle [57] is cited by the source paper of MDL [49]. In the rest of this work, we will just use the term MML, as being the oldest original term, except for papers which explicitly use the term MDL, but with the same philosophy.

## 2.3   Logic programs

A logic program, based on Horn-clause logic, is a finite set of rules or clauses
of the form:

$$A : -B_1, \ldots, B_n \tag{2.5}$$

Where $n \geq 0$, $A$ is an atom called *head*, and the symbols $B_i$ are literals
that together constitute the *body* of the rule. If $n = 0$, the rule is a *fact* in
which the symbol $: -$ is dropped in the representation and assumed present.

An *atom* is a predicate term together with its arguments, each argument
being also a term. An atom is called *ground* if it does not contain any
variables. A *literal* is defined as an atom or the negation of an atom (using
the predicate `not/1` or the symbol $\neg$).

We can have different kinds of terms:

- A *predicate term*: an expression $p(t_1, \ldots, t_m)$, with $m \geq 0$. The arity of
  the predicate symbol is $m$, so in the expression $p(t_1, \ldots, t_m)$ we say that
  $f$ is an $m$-ary *predicate symbol*. Each $t_i$ is again a term (for instance,
  another variable or function symbol, but not a predicate term). The
  expression $p(t_1, \ldots, t_m)$ must have an interpretation or definition in
  the logic program.

- A *variable*: a term which can be assigned to any value in the logic
  program. By convention they are represented by uppercase letters
  ($A, B, C...$, or more commonly, $X, Y, Z...$). The arity of a variable is 0.

- A *function term*: an expression of the form $f(t_1, \ldots, t_m)$, with $m \geq 0$.
  It is like a predicate term that cannot appear at the top level of an
  atom. The symbol $f$ is known as the function symbol or functor.

- A *constant*: with a fixed value (numbers, for instance). We could
  consider them as function symbols with arity 0.

For example, the small program `P1` below (Listings 2.1) defines the pred-
icates `uncle/2`, `father/2` and `brother/2`, has the variables `X,Y,Z` and con-
stants `adam`, `peter`, `tom`. This is known as the signature of the program.
Note that the number of terms that can be constructed with this signature
is finite.

Listing 2.1: Program P1

```
uncle(X,Y):-    father(Z,Y),brother(X,Y).
brother(X,Y):-  brother(Y,X).

brother(adam,peter).
father(tom,adam).
```

But if we look at program `P2` (Listings 2.2), the introduction of the function term `suc/1` makes the number of possible terms infinite.

Listing 2.2: Program P2

```
odd(suc(0)).
odd(suc(suc(X))):-
    odd(X).
```

The *Herbrand Base* $B(P)$ for a logic program $P$ is the set of all possible ground atoms formed by using its signature. On the other hand, the set of all ground terms constructed only from functors and constants is called the *Herbrand Universe* $U(P)$.

Listing 2.3: Program P3

```
p(0).
q(suc(X)):-p(X).
```

Let us explain this with an example, using programs `P2` and `P3`. The Herbrand Universe and the Herbrand Base of `P2` is:

$$U(P2)=\{0,\text{suc}(0),\text{suc}(\text{suc}(0)),\text{suc}(\text{suc}(\text{suc}(0))),\dots\}\ B(P2)$$
$$=\{\text{odd}(0),\text{odd}(\text{suc}(0)),\text{odd}(\text{suc}(\text{suc}(0))),\dots\}$$

Since here we have only the predicate `odd/1`, there are no big differences between the Herbrand Base and the Herbrand Universe in this case. But, let us see program `P3`:

$$U(P3)=\{0,\text{suc}(0),\text{suc}(\text{suc}(0)),\dots\}$$
$$B(P3)=\{\ \text{p}(0),\text{p}(\text{suc}(0)),\text{p}(\text{suc}(\text{suc}(0))),\dots,$$
$$\text{q}(0),\text{q}(\text{suc}(0)),\text{q}(\text{suc}(\text{suc}(0))),\dots\ \}$$

A *Herbrand interpretation* `I` is a specific subset of the Herbrand Base. One possible interpretation for the program `P2` is this:

$$I_{P2}=\{\text{odd}(\text{suc}(0)),\text{odd}(\text{suc}(\text{suc}(\text{suc}(0))))\}$$

A *Herbrand model* is a Herbrand interpretation which is a model of every formula in its own set. For example, `{odd(suc(0)),odd(suc(suc(0)))}` is a model of `{odd(suc(0))}` for program `P2`. Notice that the Herbrand base of a program is always a Herbrand model of the program.

A *Herbrand model* is *minimal* if no proper subset of it is also a model. The usual terminology for any partial ordering is that least means unique minimal. So sometimes we speak of the *least Herbrand model*. More formally, we will define the *least Herbrand model* (if it exists) as the intersection of all Herbrand models for a program $P$, denoted by $M(P)$, and it is unique.

In our example, for programs `P2` and `P3`, we have:

```
M(P2)={0,odd(suc(0)),odd(suc(suc(0))),odd(suc(suc(suc(0)))),...}
                M(P3)={p(0),q(suc(0))}
```

Every definite clause program `P` has an operator $T_P$ (the immediate consequence operator) associated with it, which maps subsets of atoms from `P` to subsets of atoms from `P`. Or, what is the same, it maps Herbrand interpretations of the program to Herbrand interpretations of the program.

Our operator $T_P$ has a *fixpoint* interpretation $I$ where $T_P(I) = I$, and also we can define the *least fixpoint* as $T_P \uparrow^\omega$.

With this basis, the process that allows the inference of facts from definite programs [4] can use one of these two different strategies:

- *Top-down*: it produces a resolvent between a head of clause and a goal. Using again the program `P2` seen in the previous example, to solve `odd(suc(suc(X))):-odd(X)`, we unify the rule with `odd(suc(0))`, using the substitution `{X= suc(0)}`. Then we recursively solve the new goal obtained from the right-hand side of the rule (if there is). The *top-down* approach corresponds to SLD-resolution, a concept that we will introduce later on in section 2.5.

- *Bottom-up*: The resolvent is obtained between a fact and the body of a clause. This is also known as the hyper-resolution. In our example, we start from `odd(suc(0))` and we apply that in the right of the other clause. The *bottom-up* approach corresponds to hyper-resolution (sometimes called unit-resolution).

## 2.4  Inductive Logic Programming (ILP)

Deductive inference derives consequences $E$ from a prior theory $T$. Similarly, inductive inference derives a general belief $T$ from $E$, but since it is a hypothesis we are going to name it as $H$. In both deduction and induction $T$ ($H$) and $E$ must be consistent and $H \models E$. It is usual to separate the above elements into examples ($E$), background knowledge ($B$), and hypothesis ($H$), and state that $B \wedge H \models E$.

$E$ consists of ground clauses, and can be separated into positive examples ($E^+$, composed by definite clauses) and negative examples ($E^-$, composed by ground unit headless Horn clauses). The problem is defined as finding the theory that is always true with each example in $E^+$ and also false with each in $E^-$. On some occasions, we only have positive examples, and we only consider $E^+$ .

---

[4] A definite program is a program composed only by definite clauses. A definite clause is a clause where we have exactly one positive literal. These representations are equivalent: $\neg p_1 \vee \neg p_2 \vee \ldots \vee \neg p_n \vee u$ $u \leftarrow p_1 \wedge p_2 \wedge \ldots \wedge p_n$ but we are going to use the second one.

So, the inductive inference will generate a hypothesis $H$ subject to the following requirements:

$$
\begin{array}{lll}
\text{necessity} & B \nvdash E^+ \\
\text{consistency} & B \nvdash e \text{ for every } e \text{ in } E^- & (2.6) \\
\text{coverage} & B \cup H \vdash E^+
\end{array}
$$

The necessity condition ensures that the background knowledge does not already entail the examples (because then there would be no necessity of the new hypothesis). The consistency condition asserts that the negation of the hypothesis is not entailed by the background knowledge and examples (because then the hypothesis would be against what we already know) and the coverage condition ensures that the background knowledge and the hypothesis entail the observations.

Application areas of ILP are many and include: learning drug structure-activity rules, learning rules for predicting protein secondary structure, learning rules from databases [15], learning rules for finite element mesh design, learning qualitative models of transport system and many others [41].

The idea of carrying out induction by inverting deduction was first mathematically investigated by Stanley Jevons [37], solving by tabulation the "inductive Problem" involving two propositional symbols. But the initial research in ILP was conducted by G.D. Plotkin [44] with the *lgg* and *rlgg* operators. This is an example of a *bottom-up* approach, where hypothesis are obtained by generalizing examples.

One of the first operative approaches, the Duce system, is also *bottom-up*: it was presented in 1987 by Stephen Muggleton [35]. It suggests high-level domain features to the user (or oracle on the basis of a set of example object description. Six transformation operators are used to successively compress the given examples by generalization and feature construction:

- *Inter-construction*: It takes a group of rules, like:

$$
\begin{array}{ll}
X \leftarrow B \wedge C \wedge D \wedge E \\
Y \leftarrow A \wedge B \wedge D \wedge F & (2.7)
\end{array}
$$

  and replaces them with the new rules (where the new one is the most specific generalization of the other two rules):

$$
\begin{array}{ll}
X \leftarrow C \wedge E \wedge Z \\
Y \leftarrow A \wedge F \wedge Z & (2.8) \\
Z \leftarrow B \wedge D
\end{array}
$$

- *Intra-construction*: The application of the distribution law of Boolean equations:

$$
\begin{array}{ll}
X \leftarrow B \wedge C \wedge D \wedge E \\
Y \leftarrow A \wedge B \wedge D \wedge F & (2.9)
\end{array}
$$

replaced with

$$X \leftarrow B \wedge D \wedge Z$$
$$Z \leftarrow C \wedge E \qquad (2.10)$$
$$Z \leftarrow A \wedge F$$

- *Absorption*: Profitable to generalize rule sets:

$$X \leftarrow A \wedge B \wedge C \wedge D \wedge E$$
$$Y \leftarrow A \wedge B \wedge D \qquad (2.11)$$

  replaced by

$$X \leftarrow Y \wedge D \wedge E$$
$$Y \leftarrow A \wedge B \wedge C \qquad (2.12)$$

- *Identification*: A set of rules which all have the same head, the body of at least one of them contains exactly one symbol not found within the other rules, such as

$$X \leftarrow A \wedge B \wedge C \wedge D \wedge E$$
$$X \leftarrow A \wedge B \wedge Y \qquad (2.13)$$

  can be replaced by

$$X \leftarrow A \wedge D \wedge Y$$
$$Y \leftarrow C \wedge D \wedge E \qquad (2.14)$$

- *Dichotomization*: This operator works on sets of mixed positive and negative examples. Thus a set of rules which contain positive and negative heads, and which all have some common symbols within the bodies, such as

$$X \leftarrow A \wedge B \wedge C \wedge D$$
$$\neg X \leftarrow A \wedge C \wedge J \wedge K \qquad (2.15)$$
$$\neg X \leftarrow A \wedge C \wedge C \wedge L$$

  can be replaced by

$$X \leftarrow A \wedge C \wedge \neg Z$$
$$\neg X \leftarrow A \wedge C \wedge \neg Z \qquad (2.16)$$

- *Truncation*: The truncation operator, like dichotomization, is intended for the use with rule sets containing positive and negative examples of the same concept. However, truncation generalizes by dropping conditions. A set of rules which all contain the same head, such as

$$X \leftarrow A \wedge B \wedge C \wedge D$$
$$X \leftarrow A \wedge C \wedge J \wedge K \qquad (2.17)$$

  can be replaced by

$$X \leftarrow A \wedge C \qquad (2.18)$$

In 1990, Stephen Muggleton and Cao Feng presented GOLEM. The principle of the system is again based on Plotkin's *relative least general generalizations* (*rlggs*). Consequently, it is a *bottom-up* approach.

In the general case, the *rlgg* may contain infinitely many literals. Therefore, GOLEM imposes some restrictions on the background knowledge and hypothesis language which ensure that the length of *rlggs* grows at worst polynomially with the number of positive examples. The background knowledge of GOLEM is required to consist of ground facts. For the hypothesis language, the determinacy restriction applies, that is, for given values of the head variables of a clause, the values of the arguments of the body literals are determined uniquely. The complexity of GOLEM's hypothesis language is further controlled by two parameters which limit the number and depth of body variables in a hypothesis clause.

GOLEM learns Horn clauses with functors. It may be run as a batch learner or in interactive mode where the induction can be controlled manually. GOLEM is able to learn from positive examples only. Negative examples are used for clause reduction in the post-processing step, as well as input/output mode declarations for the predicates the user may optionally supply. For dealing with noisy data, GOLEM provides a system parameter enabling the user to define a maximum number of negative examples a hypothesis clause is allowed to cover.

In 1993 J.R. Quinlan and R.M. Cameron-Jones presented FOIL [47], a system for learning intensional concept definitions from relational tuples. This is one of the first *top-down* systems. It avoids searching a large hypothesis space for consistent hypotheses (as happened with GOLEM). Stephen Muggleton introduced a new system in 1996 called Progol [37], which implements the theoretical framework of inverting entailment (IE). Since our purpose is just to introduce the basics of ILP, for a better understanding of ILP or a complete list of ILP systems, please refer to [16, sec. 16.3.2], or to the Website of ILPnet2 `http://www-ai.ijs.si/~ilpnet2/systems`.

## 2.5 Stochastic Logic Programs

In general, the integration of probability theory with first order logic is not straightforward. There are numerous proposals for probabilistic logics. Also, there are some proposals for the inclusion of probabilities in logic programs.

One of these proposals introduced the so-called Stochastic Logic Programs (SLPs) [38]. They have been shown to be a generalization of Hidden Markov Models (HMMs), stochastic context-free grammars and directed Bayes' nets.

A stochastic logic program is composed of clauses of the form $(p : C)$ where $p$ represents the probability associated to $C$ ($0 \leq p \leq 1$) and $C$ is a *range-restricted definite clause* [5].

Given a stochastic program $S$, the sum $\sum_p$ of the labels $p_i$ of all clauses $C_i$ of $S$ whose heads share the predicate symbol $p$ must be one. These programs are said to be normalized. In what follows, *normalized* applies to *complete* and *pure* programs [6].

A complete program is a program which is self-contained and it does not depend on external definitions. In the other hand, a program is pure if these both statements about it hold:

- The program always evaluates the same output given the same inputs. It cannot depend on any hidden information or state that may change as program execution proceeds or between different executions of the program, nor depend on any external consideration (so those programs with `random` are impure).

- Evaluation of the result does not cause any semantically observable side effect or output, such as mutation of mutable objects or output to I/O devices.

The SLD-resolution (Selective Linear Definite clause resolution) is the basic inference rule used in logic programming. It is a refinement of resolution, which is both sound and refutation complete for Horn clauses. The terms SLD-derivation and SLD-tree are applied to the non-stochastic programs, and we are going to introduce them shortly.

An SLD-derivation of a program $P$ and a goal $G$ [7] is a sequence of goals $(G_0, G_1, ...)$ where $G = G_0$, a sequence $(C_1, C_2, ...)$ of variants of the program clauses $P$ and a sequence $(\theta_1, \theta_2, ...)$ of MGUs[8].

An SLD-refutation is a finite SLD-derivation which has the empty goal as its last goal. An SLD-tree is a tree containing all possible derivations from a program $P$ and a goal $G$, via some computation rule. The SLD-tree can be infinite or finite.

The terms SLD-refutation, SLD-derivation and SLD-tree will apply in the same way to the Stochastic Logic Programs, but we are going to prefix them with an extra "S". So, from now on, we are going to use the term SSLD-tree for stochastic SLD-tree and SSLD-resolution for SLD-resolution.

The operational semantics of stochastic programs is an extension of the operational semantics of non-stochastic where the labels are used in the SLD-

---

[5]A *range-restricted definite clause* is a clause where all the variables that appear in the head also appear in its body at least once.

[6]A pure program is a program which has not negations as failure or negated clauses.

[7]A goal is a clause $p_1 \wedge p_2 \wedge ... \wedge p_n$. If $n = 0$, it is the empty goal and denoted by $\square$).

[8] A substitution $\sigma$ is the Most General Unifier (MGU) of $p$ and $q$ if: $\sigma$ (a) is a unifier (it is a substitution that makes $p$ and $q$ syntactically identical, and remember that we can only substitute variables) and (b) for every other unifier $\Theta$ of $p$ and $q$ there must exist a third substitution $\lambda$ such that $\Theta = \sigma\lambda$.

derivation to obtain a resolvent with programs and an attached probability label, calculated as:

- For the initial goal $G_0$ the SSLD-derivation is $(1 : G_0)$.

- Given an intermediate labeled goal $(q : G_i)$ in the SSLD-derivation such that the selected atom of $G_i$ unifies with the head of the clauses $(p_1 : C_1, \ldots, p_n : C_n)$ and $(p_i : C_i)$ is the clause chosen, then the next SSLD-tree step between $(q : G_i)$ and the labeled clause $(p : C_i)$ produces the goal $(m : R)$ as resolvent, where $m$, the probability of this SSLD-derivation step, is the product of $q$ and $\frac{p_i}{p_1 + \ldots + p_n}$, and $R$ is the resolvent obtained from $G_i$ and $C_i$ by SLD-resolution.

The SSLD-refutation is the SLD-derivation that ends in the goal $(\sum p_i :\leftarrow)$ where $\sum p_i$, the answer probability, is the summation of the probabilities of the given steps in the refutation.

To build the signature of stochastic programs, we use the method used for non-stochastic ones. The operational mechanism of SLPs described in [38] can be used to define a probability distribution over the Herbrand Base constructed from the signature as follows:

- The labels with the probabilities of the clauses of a SLP are not taken into consideration for the signature.

- Given an atom $a \in B_P$, let $Answ(a, P) = \{p | P \cup \{\leftarrow a\} \longrightarrow_{SLD}^* (p :\leftarrow )\}$ the set of answer probabilities of $a$ wrt. $P$.

- From here, the probability of $a$ is defined as $p(a|P) = \sum_{p \in Answ(a,P)} p$.

For example, with the program `P4` (the notation is in Prolog syntax, with a special functor `::/2` to introduce the probability on the left):

Listing 2.4: Program P4
```
0.3   ::  p(X,Y):-q(X),q(Y).        [1]
0.3   ::  p(X,Y):-q(X).             [2]
0.4   ::  p(0,1).                   [3]
0.25  ::  q(0).                     [4]
0.25  ::  q(1).                     [5]
0.50  ::  q(2).                     [6]
```

The process to obtain the probability of an example like `p(0,1)` is shown at the left part of the Figure 2.1. The SSLD-refutation for `p(0,1)` is composed by three successfully branches, each one labeled with the probabilities of the predicate immediately under the value.

The evidence `p(0,1)` has a probability of 0.49375, obtained as the sum of the three SSLD-refutations:

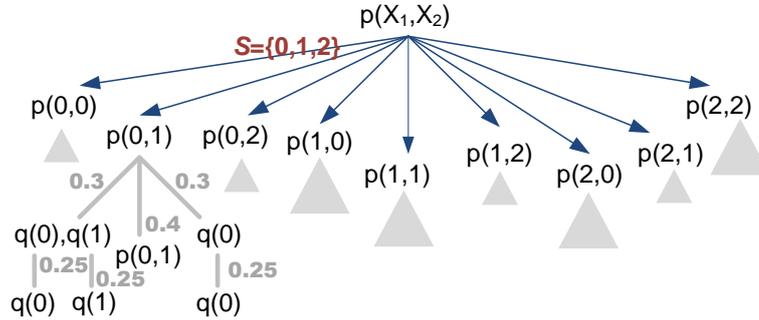

Figure 2.1: Estimating the probability of `p(0,1)`. The derivation of other examples is not illustrated here.

- Using the rule 1 and then applying rules 4 and 5, we obtain a probability of $0.3 \cdot (0.25 \cdot 0.25) = 0.01875$.

- Using the rule 2 and then the rule 4, we obtain a probability of $0.3 \cdot 0.25 = 0.075$.

- Using the rule 3 we obtain a probability of 0.4.

Despite the fact that we are going to focus our attention to SSLDs, there are other important approaches that have tried to integrate probabilities into logical programming. For example, Bayesian Networks [42] express the probability distribution over a finite set of random variables with two components: one *qualitative* component that encodes the local influences among the random variables using an acyclic graph, and a *quantitative* component that encodes the probability densities over these local influences. The most important limitation of Bayesian Networks is that they use a propositional representation that limits us to express interactions. An evolution of this approach are the MML Bayesian nets [3] [4, Chapter 11, pages 265–294], or the Relational Bayesian Networks of Manfred Jäger [28]. Other approaches are Poole's Independent Choice Logic [45] and Pfeffer's Probabilistic Relational Models [20].

### 2.5.1  SSLD-derivation approaches

Constructing a mechanism to derive accurate probabilities for a general family of stochastic logic programs is not an easy task. This is the reason-why the construction of a stochastic logic programming interpreter calculating the probabilities is a challenging problem. For instance, for the program `P4`, if we calculate the probabilities of the facts `p(0,0)`, `p(0,1)`, `p(1,0)`, etc. we will see that the sum of their probabilities is much greater than one. Consequently, this way of calculating the probabilities does not really generate (in general) a probability distribution and hence it is useless for our MML coding.

**Muggleton's approach**

The first approach using SSLD-derivation is due to Muggleton. The underlying idea is the same we have introduced above, and it is to obtain the probability of a fact in the evidence as the sum of the probabilities of all successful refutations in the SSLD-tree. The examples are facts for a single predicate and the goal is to learn the corresponding predicate definition annotated with probabilities incrementally, from each individual example. The probability of the predicate definition is estimated as we go along, in each step.

In his approach, Muggleton [39] starts with the initial objective of a goal composed of the example with all the arguments replaced by variables. So, in our program `P4` (Listings 2.4), we will solve $p(X_1, X_2, X_3)$ to obtain the probability of `p/2`: that means eight different combinations `p(0,1)` using the signature `{0,1,2}`. Before we have calculated the probability of `p(1,0)`, but we will need also `p(2,0)`, `p(0,2)`, etc. The cost of each of this branches (and also of the seven not unfolded) is added up to obtain the cost for the predicate `p/2`.

Unfortunately, this mechanism is so simple that is only able to manage ground facts as evidence, and there are many stochastic logic programs not suitable to apply this technique. Muggleton also supposes that there is no overlapping between the rules of the stochastic logic program (and it is possible, as we are going to see), so with this technique, some predicates in the evidence could have a value greater than 1.

**De Raedt's approach**

De Raedt [31] introduced the normalization with hindsight of the predicates: the underlying idea to apply a posteriori normalization of the values is to correct the final probability of a predicate to be exactly 1. The problem with this approach is involved with the huge amount of calculations we will need to determine the whole SSLD-tree for each predicate and also the normalization process.

### 2.5.2 Our method for SSLD-derivation

Basically we are going to follow Muggleton's approach together with the normalization of De Raedt. We will also introduce some extensions to have an approach able to cover stochastic logic programs of any nature.

The first modification we are going to apply to Muggleton's approach is involved with the substitutions. We will introduce it directly with an exam-ple. Let us use the example $sum(s(0), s(s(0)), s(s(s(0))))$ with program `P5` (Listing 2.5). We depart from $sum(X_1, X_2, X_3)$ as the root of our demonstration, and we apply the two rules once: the second one and next the first one, so $\frac{1}{2} \times \frac{1}{2} = \frac{1}{4}$. We will assign to the branch a probability of $\frac{1}{4}$.

Listing 2.5: Program P5

```
0.5  ::  sum(0,W,W).
0.5  ::  sum(s(X),Y,s(Z)):-sum(X,Y,Z).
```

But we have not finished, because we have to solve the substitution $W/s(s(0))$. To estimate this last substitution, we will need the signature (which is $\{0/0, s/1\}$). As we have only two function symbols, each time we will select one of them we will have a probability of $\frac{1}{2}$. In $W/s(s(0))$ we have used this selection three times, so the probability of this last substitution is $\frac{1}{2^3} = \frac{1}{8}$.

Thus, with our approximation the probability for this example is $\frac{1}{4} \times \frac{1}{8} = \frac{1}{32}$.

In this case, since there are no overlapping rules and no other exceptional issue, we get the right probability for this example on program `P5`.

**Overlapping heads**

If we look again at the program `P4` (Listings 2.4), we can determine the probabilities of predicate `p/2` calculating the probabilities of all the combinations, to discover that the sum is greater than 1:

|          | Obtained probability | Normalized probability |
|----------|---------------------:|-----------------------:|
| p(0,0)   | 0.09375              | 0,06117                |
| p(0,1)   | 0.49375              | 0,32216                |
| p(0,2)   | 0.07875              | 0,05138                |
| p(1,0)   | 0.09375              | 0,06117                |
| p(1,1)   | 0.09375              | 0,06117                |
| p(1,2)   | 0.07875              | 0,05138                |
| p(2,0)   | 0.1875               | 0,12234                |
| p(2,1)   | 0.1875               | 0,12234                |
| p(2,2)   | 0.225                | 0,14681                |
| Total    | 1.53262              | 1                      |

Table 2.1: Probabilities before normalization of `p/2` in program `P4`

We have *overlapping heads* in program `P4`. This is a specific problem of the stochastic programs, where the probabilities of the assigned rules can hide different paths for the same fact. The solution here will be to apply an *a posteriori* normalization process for the predicates. For instance, with `p/2`, we have determined a sum of 1.53262 (Table 2.1), so we will divide by this value to have the real probability of each partial probability (for instance, `p(2,2)` it will be $0.225/1.53262 = 0,14681$).

**The problem with free variables**

There could be SSLD-derivations that have variables without instantiation. So, the estimation of the probability of a predicate with this kind of derivations will generate an infinite loop in our proposed approach if we do not introduce some corrections. For example in this simple program, the variable $Z_1$ is not bound with the head of the rule. The same happens with the variable $Z_2$ with the body of the second rule. We give $Z_1$ and $Z_2$ the name of "free variables":

```
p(X,Y):-r(X),r(Z1).
q(X,Z2):-r(X).
r(_).
```

If we try to generate the SLD-tree for p($X_1$, $X_2$) and for q($X_1$, $X_2$), we can see that our solver is unable to find a valid assignment for the free variables:

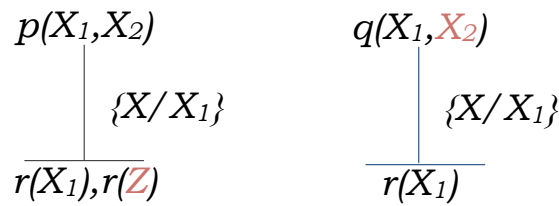

Figure 2.2: An example of SLD-tree with free-variables

In fact, it is impossible to determine a priori which the value of probability is for all the derivations of those predicates using the solver described in the previous chapter. Since it is a non-assigned variable (a free variable), its possible assignments depend on the different symbols that we have in the signature (if the free variable is in the head of the rule) or on the probability of the predicate that remains with a free variable, if the predicate is in the body of the rule.

In our example (Figure 2.2) the free variables that remain without assignation after the unification/matching process of Prolog are two: `Y` in the head of `p/2` and `Z` in the body of `r/1`.

In this situation, the success of the unification is not being discussed. The only problem is that the probability of that assignment is affected by this non-assigned variable, which could take any value from the signature of the program as we have said. So, we need to apply some corrections to the solver.

**The solution for free variables**

The changes necessary in the process of solving a predicate differ if the free variable is in the head, in the body or also if the predicate of the head is also present in the body of that rule (due to the recursion). Let us see each of them separately:

- Free variables in the head of the rule:

    - If there is a free variable in the head of a rule the different combinations that we could have are all the different we can build with the signature (because there is no limitation in the rule to the apparition of any symbol here, because there is no bound between the free variable and any rule).

    - So, the number of different possibilities to instantiate a value to the free variable is given by the division by the number of elements of the signaturem if the arity of the function symbols is equal to zero. That is, if we have `q(X,Y):-s(X)` and a signature $S$=`{a/0, b/0, c/0}`, it is clear that we will have only `q(_,0)`, `q(_,1)` and `q(_,c)` as possibilities.

    - But maybe the signature could contain a function symbol with a greater arity, for example $S$=`{0/0, s/1}`. Here the possibilities for the given are infinite combinations of that symbols for that variable: `0`, `s(0)`, `s(s(0))`, etc...

    - So generalizing, the solution will be to trace from the beginning where the free variables are in the head of a rule. This is because if we postpone this analysis after solving it against the Prolog interpreter, the arguments of the head will have variables that maybe are bound with the body of the rule, maybe not. And we need to separate those two groups of variables to compute the cost.

        So, the probability of the evidence will be the result of dividing the probability of the rule by the number of elements of the signature, raised to the power of the number of different symbols that are in the assignment of that free variable in the example:

$$ProbExample = Prob(Pred) \prod_{k=1...a} \frac{1}{length(S)}$$

where

$$
\begin{aligned}
Prob(Pred) &= \text{Probability of the predicate} \\
length(S) &= \text{Length of the signature} \\
&\quad \text{(only function symbols)} \\
a &= \text{Number of different elements} \\
&\quad \text{in the free position}
\end{aligned}
\qquad (2.19)
$$

For example, we have determined that the probability of the example `r(0,t(0,s(0)))` using the goal `r(X1,X2)` as root of the tree, over the program P6 (Listing 2.6) is 0.8, using the assignment `X1/0`.

Listing 2.6: Program P6

```
r(X,_):-p(X).
0.8  ::  p(0).
0.2  ::  p(s(_)).

0.3   ::  q(X):-X<1,r(X,Y).
0.7   ::  q(1).
0.25  ::  r(0,1).
0.25  ::  r(0,2).
0.5   ::  r(1,2).
```

Variable `X2` remains as a free variable. If we have the signature $S=\{0/0, s/1,t/2\}$, with the example `r(0,t(0,s(0)))` (notice that the second argument is using exactly *four* symbols from the signature) the estimated probability for this example would be:

$$ProbExample = 0.8\tfrac{1}{3}\tfrac{1}{3}\tfrac{1}{3}\tfrac{1}{3} = 0.009876$$

- Free variables in the body of the rule:

  - The other problem is when there are free variables in the body: in this case, the probability is not going to depend on the number of symbols of the signature. Although we have a free variable, all the combinations done with the signature are not going to be valid, because this predicate is defined and it has rules that make to success or not some combinations we can made with the signature.

    For instance, in Program P6 (Listings 2.6) we have a free variable `Y` in `q(X):-r(X,Y)`. So, we must analyze the valid subtrees of predicate $r(X_1,X_2)$ that can be applied here. On the one hand, there

are invalid combinations of the signature (`r(2,0)`,`r(2,1)`,...). On the other hand, the first bound variable of the predicate `r/2` makes that only `r(0,1)` and `r(0,2)` are valid subtrees.

– So, we need to check the rule after each solving step, checking if there are new variables after the substitutions. If there are, we will need to obtain the probability of the whole predicate that has free variables, in the same way we are calculating a new example. For instance, if we have:

```
0.7 :: r(A,B):-B is A,p(C).
0.3 :: r(A,0).
...   % we do not include the rules for p/1.
```

To calculate $r(X_1, X_2)$, we will need first $p(X)$ because in the rule `r/2` this predicate appears with a free variable. If we suppose that `p/1` has an estimated probability of 0.25, then we will multiply by this factor the result for the rule `r/2` to obtain the real probability of that evidence.

Although, a limitation in the depth of SSLD-derivations must be considered, since they could be infinite. If we look again at program P5 (Listings 2.5), the function symbol `s/1` makes infinite the growth of the SSLD-tree for predicate `sum/3`, as we show in Figure 2.3.

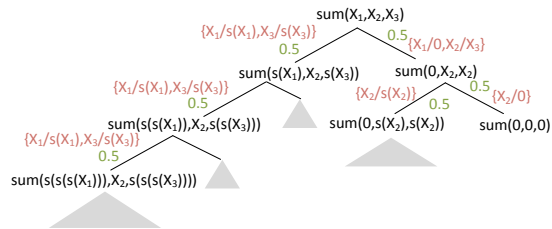

Figure 2.3: An example of SLD-tree with infinite derivations

In this SSLD-tree, if we use a low value for the maximum depth inside the tree, the sum of the probabilities for the predicate is going to be nearby 1, but not all the stochastic logic programs are going to have a so simple structure in which the prune of the SSLD-tree is not going to trim important facts that need more depth in the derivation.

## 2.6   Previous approaches for coding logical theories

Measuring the length or "cost" of a theory expressed as a logic program is not a new area of research. As we have pointed out at the beginning of this chapter, there are other previous works for coding logic programs. However, most approaches have been based on the semantics of the program, rather than its syntax. Two exceptions are the so-called Model Complexity ($\mathcal{MC}$) [8] and the Proof Complexity ($\mathcal{PC}$) [36], which present a way of coding $L(D|T)$ and $L(T)$. Here, however, the complexity of the program, $L(T)$, is measured in a very simplistic way, just counting the number of rules, literals, etc., in a program, so disregarding repetitions, structure and other redundancies in how the program is expressed. One of the crucial issues in MML is that the coding must be efficient. Otherwise, the evaluation of the "length" of the program might be misleading.

Also, there are several problems about how the $\mathcal{MC}$ and $\mathcal{PC}$ approaches calculate the cost of the examples given the theory, $L(D|T)$, especially for cases where the program covers an infinite number of examples, which is a very common situation.

Let us start with the coding of the program, $L(T)$, which is common for both approaches. Two conditions are assumed for these approaches:

- Examples must be ground.

- Function-free logic programs are used, in order to avoid infinite least Herbrand models.

With these conditions, assuming that well-formed logic programs have an unambiguous probabilistic context-free grammar, the length in bits of the coding $T$ is calculated as $-log_2 Pr(T)$, where $Pr(T)$ is the product of the production probabilities used in the derivation of $T$. For each atom in the theory $T$ with arity $n$, the code length of the atom is $log_2|P|+n{\cdot}log_2(|X|+|F|)$, where $P$ represents the predicate terms of the theory, $X$ the variables and $F$ the function terms. In addition to this, to identify the number of variables $v$ needed to express $T$, we must send a code of length $log_2(X)+1$ bits before.

To sum up, the number of bits required to code a logic program is the sum of:

- $log_2(v+1)$ bits for the variables, where $v = |X|$.

- 1 bit per program.

- 2 bits per rule in the program.

- 2 bits per literal in the body of each rule.

- The bits for coding all atoms in the program, using the coding scheme seen above.

The previous scheme works for function-free programs. In the next chapter we will extend this to programs including functions, using a MML-coding scheme.

In the following two subsections we will analyze how examples are coded given the program, i.e. $L(D|T)$, in both the model complexity and proof complexity approaches.

### 2.6.1   The Model Complexity Approach

The Model Complexity ($\mathcal{MC}$) [8] was presented by D. Conklin and I. H. Witten in 1995. In it, the best theory for a concept is defined to be the one that minimizes the number of bits required to communicate the examples, that involves the encoding and transmission of the theory $T$ and the same examples using that theory $T$, while they refer to this as a MDL approach, it certainly follows the MML philosophy.  First, the coding requires the definition of the content $Q(T)$, defined as:

$$Q(T) = O \cap M_T \, , \ \ E \subseteq O \tag{2.20}$$

where $O$ are the possible observations and $M_T$ is the least Herbrand model of the logic program $T$ ($M_T$ is finite).  $Q(T)$ is just the set of ground facts derived from $T$. With this, there are three possible situations:

1. $E \not\subset Q(T)$.  There are examples not covered by the theory.  In this case we must increase the theory to cover those examples, so leading to cases 2 and 3.

2. $E \subseteq Q(T)$.  The theory covers all the examples, and perhaps other observable atoms: this is the situation that the $\mathcal{MC}$ considers, sending a code of length (in bits):

$$L_{\mathcal{MC}}(E|T) = \log_2 \left( \begin{array}{c} |Q(T)| \\ |E| \end{array} \right) \tag{2.21}$$

   which is large enough for identifying a subset of $Q(T)$ of either size $|E|$ or size $|Q(T)|$.  One extra bit allows the communication of which of these is the case.

3. $E = Q(T)$.  The theory covers all and only all the examples: the ideal but not usual situation. This is a particular case of 2.

There are, however, some problems with this approach:

1. It cannot be applied when $Q(T)$ is infinite.  This happens frequently if we have function symbols.

2. As expected, adding a new positive example to the evidence would need fewer bits than adding an example as a rule to the theory, but this asymmetry depends highly on the size of the evidence.

The disadvantages are partially solved with the Evidence Complexity ($\mathcal{EC}$) approach, which we are going to introduce in chapter 3.3.

### 2.6.2  The Proof Complexity Approach

The other approach for $L(E|T)$ was proposed by Muggleton, Srinivasan and Bain in 1992 [36] called the "proof complexity" measure ($\mathcal{PC}$), defined as the bits required to code the proof of each example given the theory. The proof complexity of an example $e$ is calculated as the sum of the logarithm of the choice-points involved in the SLD-refutation of $e$. In each step of a refutation, if $G$ is the actual goal, the choice-points are the number of program clauses whose heads unify with the atom of $G$ selected by the selection function. Therefore,

$$L_{\mathcal{PC}}(E|T) = \sum_{A \in E} L_{\mathcal{PC}}(A|T) \qquad (2.22)$$

where $L_{\mathcal{PC}}(A|T)$ is the code length of an atom $A$ wrt. a theory $T$.

In this case, only the given evidence is coded, never the absent examples (those elements in $Q(T)$ which are not in $E$). This is counter-intuitive, since $L_{\mathcal{PC}}(E|T) > 0$ when $Q(T) = E$. Even with a perfect-covering program, we have that $L_{\mathcal{PC}}(E|T)$ is not zero.

The following chapter is then devoted to a coding scheme which considers programs with function symbols (for both $L(T)$ and $L(E|T)$) and tries to solve the problems of both the $\mathcal{MC}$ and $\mathcal{PC}$ approaches.

Also, one of the important things of a coding scheme, especially if it is used for theory selection, is that it must be as efficient as possible. The $\mathcal{MC}$ and $\mathcal{PC}$ approaches, as we will see, contain several sources of redundancy in their codes.

# Chapter 3

# Coding scheme

> In this chapter we will detail the process of coding theory and data in the form of (stochastic) logic programs, using the MML principle. The organization of the chapter is as follows: we start with the coding of non-stochastic programs. Next we analyze how to extend the coding to stochastic programs. Finally, we focus on the coding of examples. In the non-stochastic section, we split the coding of the signature from the coding of the parts of the rules (heads, body, variables). There is also a subsection dedicated to how background knowledge should be handled.

## 3.1  Coding of non-stochastic programs

A preliminary version of this scheme was presented in [13] in 2007. The goal of the first scheme was to overcome some of the problems found in the $\mathcal{MC}$ and $\mathcal{PC}$ approaches seen in the previous chapter. However, the original version contained some glitches and did not properly address the necessary normalization for stochastic programs. Consequently, we present a new scheme based on the previous one, but including an important number of changes and extensions.

To determine the MML cost of a program we will need to analyze the terms of each rule: the head, the body and also the variables that are used.

For a non-empty program $P$ with $n_r$ rules $r_1, r_2, \ldots, r_m$, the cost of encoding $P$ is done by the formula:

$$cost(P) = CodeLength(n_r) + cost(\Sigma_P) + \sum_{1 \leq i \leq n_r} cost(r_i) \qquad (3.1)$$

Where $\Sigma_P$ represents the signature, and $CodeLength$ the cost of transferring the number of rules, which is calculated as:

$$CodeLength(n_r) = \log^* n_r + C \qquad (3.2)$$

Here $C$ is a normalization constant chosen to satisfy $\sum_{n=1}^{\infty} 2^{-\log^* n} = 1$ and whose value is $\log_2 2.8665064$. And besides, $\log^*$ represents the *iterated logarithm* (usually read "log star"). The iterated logarithm is explained in Appendix A.1. Basically, it is used in *CodeLength* for coding unbounded natural numbers and it is the number of times the logarithm function must be iteratively applied before the result is less than or equal to 1.

Continuing with the cost of a program $P$, for each rule $r_i$ of the form $H_i : -B_i$, its cost is calculated as:

$$cost(r_i) = cost(H_i) + cost(B_i) + cost(V_i) \qquad (3.3)$$

Where the last term represents the cost of coding the variables appearing in the rule (if there are, or 1 bit otherwise). The rules without body (heads) have $cost(B_i) = 0$.

### 3.1.1   Cost of the signature: $cost(\Sigma_P)$

Consider a program $P$ which has $n_p$ predicates and $n_f$ function symbols: We need to transmit that list of elements (which might be zero), but the only relevant information to transmit about them is the arity of each element $i$ (because both function symbols and predicate symbols have arity). The order in which that signature is transmitted is not relevant.

For example: If we have the predicate symbols $\{p/3, q/0, r/2, s/1\}$, it does not matter if we transmit $\{3, 1, 2, 0\}$, $\{2, 0, 3, 1\}$, $\{1, 0, 3, 2\}$ or any of the possible twelve different permutations. We only need to transmit the arities in an optimal way.

Let us then start finding the maximum arity of all the predicate symbols $p_i$ and all the function symbols $f_i$:

$$\begin{aligned} max_p &= max_{i=1\ldots n_p} arity(p_i) \\ max_f &= max_{j=1\ldots n_f} arity(f_j) \end{aligned} \qquad (3.4)$$

And now, for each possible value in the intervals $1..max_p$ and $1..max_f$ we will count how many symbols we have to transmit with each arity. So, the cost of transferring the whole signature will be:

$$\begin{aligned} cost(\Sigma_P) = \ & CodeLength(1 + max_p) + \\ & CodeLength(1 + max_f) + \\ & \sum_{0 \le k < max_p} CodeLength\left(1 + card\{p_i : arity(p_i) = k\}_{i=1\ldots n_p}\right) + \\ & \sum_{0 \le l < max_f} CodeLength\left(1 + card\{f_j : arity(f_j) = l\}_{j=1\ldots n_f}\right) \end{aligned} \qquad (3.5)$$

Note that $k$ and $l$ do not need to range until $max_p$ and $max_f$, since the number of predicate and function symbols for the maximum arities can be inferred from the already coded information.

It is easier to explain this with an example: if we have the signature $S = \{p/3, q/1, r/3, f/0\}$, where only $f/0$ is a function symbol, the cost of this signature would be:

$$
\begin{aligned}
max_p = \ & 3 \\
max_f = \ & 0 \\
\\
Cost = \ & \quad CodeLength(3+1) + CodeLength(0+1) \\
& + CodeLength(1 + \{0 \ p_i \text{ with arity } 0\}) \\
& + CodeLength(1 + \{1 \ p_i \text{ with arity } 1\}) \\
& + CodeLength(1 + \{0 \ p_i \text{ with arity } 2\}) \\
& + CodeLength(1 + \{2 \ p_i \text{ with arity } 3\}) \\
& + CodeLength(1 + \{1 \ f_j \text{ with arity } 0\}) \\
= \ & \quad CodeLength(4) + CodeLength(1) + CodeLength(1) \\
& + CodeLength(2) + CodeLength(1) + CodeLength(3) \\
& + CodeLength(2) \\
= \ & \quad 4.51929 + 1.51929 + 1.51929 \\
& + 2.51929 + 1.51929 + 3.76870 + 2.51929 \\
= \ & \ 17.88444
\end{aligned}
\tag{3.6}
$$

Note that this coding is inefficient when there are many "unused" arities. In logic programs, this is not usually the case, since we have many predicates and functions with low arities.

### 3.1.2 Cost of coding rule heads: $cost(H)$

The first thing to know about the head of a rule is which predicate symbol appears in it. So for the total $n_r$ rules of a program this will cost $n_r \times \log_2(n_p)$ to code, where $n_r$ are the total number of rules and $n_p$ is the total number of predicates. Note that the order of rules matters, so we cannot subtract any redundancy to this term.

For each argument of the head's atom (if the term has arguments, because this could also be zero), it is necessary to know whether it is a variable or a function symbol, so we will add extra $\log_2(2)$ bits (assuming a probability of $\frac{1}{2}$). If it is a function symbol, we need to add also an extra $\log_2(n_f)$ for coding which function it is (considering a uniform distribution for probabilities), and this is applied in a recursive calculation if the arity of that argument is greater than zero.

So, summarizing:

$$cost(H) = n_r \log_2 n_p + \sum_{i=1}^{n_r} \sum_{j=1}^{n_{arg,i}} cost_t(r_{i,j})$$

$$
\begin{aligned}
\text{where } cost_t(t) =& \quad \text{cost of the term } t \\
n_r =& \quad \text{number of rules} \\
n_p =& \quad \text{number of predicates} \\
n_{arg,i} =& \quad \text{number or arguments of predicate } i \\
r_{i,j} =& \quad \text{argument } j \text{ of the predicate term in the head of rule } i
\end{aligned}
\tag{3.7}
$$

Remember that with each $r_{i,j}$, if it is a function symbol, we must proceed recursively. The cost of each term is:

$$cost_t(t) = \log_2 2 + \begin{cases} 0 & \text{if } t \text{ is a variable} \\[2mm] \log_2 n_f + \sum_{j=1}^{n_{arg}} cost_t(r_j) & \text{if } t \text{ is a function term} \end{cases} \tag{3.8}$$

$$
\begin{aligned}
where \quad n_f =& \quad \text{number of functions} \\
n_{arg} =& \quad \text{number of arguments of the function term } t \\
r_j =& \quad \text{argument } j \text{ of the function term } t
\end{aligned}
$$

Let us use again an example to facilitate the understanding. Suppose that we have this small program:

Listing 3.1: Program P7

```prolog
pred1(s(X),Y):-pred2(s(X)),pred2(Y).
pred2(s(s(X))):-pred2(X).
pred2(0).
```

Applying the explained scheme for this example, the first term of the formula ($n_r \log_2(n_p)$) is $3 \log_2 2 = 3$ bits, because we have 3 rules of a total of 2 predicates. In the program there are a two function symbols, `0/0` and `s/1`), so when a function symbol appears, we must add $\log_2 2$.

Let us focus on the first rule, whose head predicate has two arguments. The first argument of this rule contains a function symbol and a variable inside (so $\log_2 2 + \log_2 2 + \log_2 2$), and the second argument has only one variable ($\log_2 2$). So, if we add up everything, we have a total of $4 \log_2 2 = 4$ bits for this first rule.

Applying the same procedure to the second rule we will obtain $2 \log_2 2 + 2 \log_2 2 + \log_2 2 = 5$ bits.

Finally, for the third rule we have $2 \log_2 2 = 2$ bits.

So, for the whole program, the cost of the heads is $cost(H) = 3 + 4 + 5 + 2 = 14$ bits.

### 3.1.3 Cost of the bodies: $cost(B)$

The next step is to encode the bodies of the rules. Here the order of the rules and also the order of the literals inside each rule matters. Again, we will use $n_p$ as the number of different predicates, and $n_r$ as the number of rules. We will introduce now $lit_i$ as the number of literals inside the rule $i$.

For each literal, we will add a bit $(\log_2(2))$ if it is variable or $\log_2(n_f)$ if it is a function symbol, to inform which function it is. We will proceed recursively in the same way with its arguments (if there are), as $cost_t$.

So, summarizing:

$$Cost(B) = \sum_{i=1}^{n_r} \left[ CodeLength(1+m_i) + \sum_{j=1}^{m_i} \left( \log_2 n_p + \sum_{k=1}^{nargs_{i,j}} cost_t(t_{i,j,k}) \right) \right]$$

where $n_r =$ number of rules
$\quad m_i =$ number or literals of rule $r_i$
$\quad t_{i,j,k} =$ term $k$ of the literal $j$ of rule $i$
$\quad nargs_{i,j} =$ number of arguments in the literal $i, j$

(3.9)

Let us use again the first rule of the example P7 (Listing 3.1):

```
pred1(s(X),Y):-pred2(s(X)),pred2(Y)
```

The cost of coding the body of this rule (`pred2(s(X)),pred2(Y)`) is 9.76870 bits, due to:

- It has two literals, so $m_i = 2$, so we need $CodeLength(1+2)$ bits.

- We need to code the predicate symbol of each first literal $(\log_2 2)$ and the first literal contains also a function symbol `s/1` and a variable inside, so we need $3\log_2 2$.

- The second literal contains only a variable, so we need $\log_2 2$ for coding the predicate and $\log_2 2$ for saying it is a variable.

- All together we have:

$$\begin{aligned} Cost(B_1) =\ & CodeLength(1+2) + [\log_2 2 + 3\log_2 2] \\ & + [\log_2 2 + \log_2 2] \\ =\ & CodeLength(3) + [3+1] + [1+1] = \\ =\ & 3.76870 + 6 = 9.76870 \text{ bits} \end{aligned}$$

(3.10)

### 3.1.4 Cost of the variables: $cost(V)$

After the previous information is sent, the receiver can know the number of different variables existing in each rule. So far, we have not coded yet the position of them and how many times each of them appears in the rule (remember that the name is not relevant). Fortunately, in a logical program, all the variables are local to each rule.

So, we need to consider whether they are the same or different variable, and the number of appearances of each one.

Computing the position is a combinatorial problem but with some considerations. Let us suppose that $n_v$ is the number of different variables in a rule, and $d$ is the number of positions inside the rule.

Since the order of the variables matters we will have different $(n_v)^d$ combinations at the most, instead of $\binom{n_v}{d}$, because `p(X,Y)` is not equal to `p(Y,X)`.

Moreover, we know that at least each variable appears at least once, so there are a few combinations in $(n_v)^d$ we could remove: those that do not contain the $n_v$ variables. We will subtract the sum of the ways we can fill the $d$ positions with $1 \ldots n_v - 1$ variables only.

For example, if we have 3 different variables and 10 positions in the rule, we will subtract two amounts: The first one, the number of ways that two variables of the three can fill all the positions: $\binom{3}{2}(2^{10} - 2)$. The second amount is more intuitive: the 3 different combinations that one variable of the three can fill all the positions. So, the different possibilities we have with 3 variables and 10 positions are finally $3^{10} - \binom{3}{2}(2^{10} - 2) - 3 = 55{,}980$ combinations.

Generalizing:

$$
\begin{aligned}
cost(V) =\ & \textstyle\sum_{i=1}^{n_r} \log_2 F(d_i, n_{v,i}) \\[4pt]
F(d, n_v) =\ & (n_v)^d - \textstyle\sum_{1 \le i \le (n_v-1)} \binom{n_v}{i} * F(d, i) \\
F(d, 0) =\ & 1
\end{aligned}
\tag{3.11}
$$

$$
\begin{aligned}
\text{where } n_{v,i} =\ & \text{number of different variables for the rule } i \\
n_r =\ & \text{number of rules}
\end{aligned}
$$

We do not need to code the number of positions with variables in the rules because from section 3.1.2 and section 3.1.3 we can infer that information. Let us apply the formula using the same rule of program P7 (Listing 3.1):

```
pred1(s(X),Y):-pred2(s(X)),pred2(Y).
```

The cost of the variables of the rule is 5.80735 bits:

- It has four different positions, so $d = 4$.

- It has two different variables, so $n_{v,i} = 2$.

- So we have for this rule $cost(V_1) = \log_2(F(4,2)) = \log_2 14 = 3.80735$ bits.

### 3.1.5   Cost of the background knowledge

The *background knowledge*, *previous knowledge* or *knowledge base* (KB in its short form) is the collection of rules (not included in the program) that we assume that are known by both sender and receiver. We can think of many examples: the predicates that allow us to show results on the screen, the predicates that deal with the storage of results in disk (I/O predicates), libraries shared and commonly used, or knowledge that we assume as known. All of this is not really part of the model we want to analyze, since this previous knowledge is shared by sender and receiver, we do not need to transfer it, and then, to code it.

If the *knowledge base* was really relevant and we wanted to consider in the calculations, one way it would be to include it as part of the program and analyze as a whole, summing the values for its heads, bodies and variables as we have described till now.

But if we prefer not to join part of the KB to the program, we can analyze it as if it was part of the program but as a separate entity. The reasons to do this can be many: maybe the previous knowledge is shared between different models and it will be faster not to copy the *knowledge base* for each model, or maybe we want to evaluate more complex models and programs in which other models and programs take part in, but we want a separate assessment. In that situation, the *knowledge base* will be managed as if it was part of the same program, increasing the cost because its heads, bodies and variables.

Considering the KB or not is important on how the programs are coded, and this influence is not necessarily constant for all programs. The reason is that if we sum up the *knowledge base* it will also affect the signature, and will impact also in the operations done with the models or programs analyzed. The signature obtained from the whole KB will be merged and analyzed with the one obtained from the code introduced as program. Since the number or predicates $n_p$ and functions $n_f$ is increased, it will affect also to the cost in bits generated in the programs or models.

So, as an example, if we add this two predicates as a *knowledge base* to program P7 (Listing 3.1):

```
pred3(1,2).
pred4(t(1)).
```

As program P7 has not used any of this predicate and function symbols (they are new), it will affect the global cost because we have increased the signature with new predicate and function symbols `{t/1,1/0,2/0}` ($n_f$ is increased in 3), and also two new rules ($n_p$ is increased in 2). That extra cost will be calculated as we have explained before for heads and signature, and the same if there were variables or bodies.

## 3.2   Coding of stochastic programs

Stochastic programs can be coded in the same way as non-stochastic programs, but we need to send the bits for probabilities of each rule. These probabilities are real numbers but they will be approximated and transformed to rational numbers.

A real number $x$ can be approximated by a rational number. Of course, since rational numbers are dense on the real line, we can make the difference between $x$ and its rational approximation $\frac{p}{q}$ as small as we wish. The problem is that, as we try to make $\frac{p}{q}$ closer and closer to $x$, we may have to use larger and larger $p$ and $q$ and, clearly, more bits. So, the reasonable question to ask here is how well we can approximate $x$ by rational denominators which are not too large. The idea has been to use a precision threshold.

The way real numbers are transformed to rational numbers and the way they are later coded it is fully explained in Appendix A. Basically, we find a close rational approximation and we code the denominator and ignore the numerator (among all the values which are not reducible). As an example, if we have this program:

```
0.25 :: p(0).
0.25 :: p(1).
0.50 :: p(X):-X > 20.
```

The cost of coding this stochastic program is the same as this other equivalent program without probabilities but increasing it with the cost of coding the three real numbers:

```
p(0).
p(1).
p(X):-X > 0.
```

Using the procedures we have seen till now, the MML cost of the program without probabilities is 31.43476 bits. Now, we will add the extra probabilistic cost of 11.03707 due to the coding of the probabilities of the three rules, because:

- The first probability has a cost of $Cost(\frac{1}{4}) = CodeLength(4) + \log_2 2 = 4.51929 + 1 = 5.51929$ bits.

- The second clause the same.

- The third probability $\frac{1}{2}$ has a cost of $Cost(\frac{1}{2}) = CodeLength(2) + \log_2 1 = 2.51929$ bits. However, we actually do not need to communicate this last probability of $\frac{1}{2}$ because it is redundant since the probabilities are supposed to be normalized, so they must add up to 1.

- So, we obtain summing up the probabilities of first and second clauses the value of 11.03858 bits.

## 3.3 Coding the examples ($cost(E|T)$)

The evidence consists of facts. The coding cost for the evidence will depend on the probability of each example observed, deriving the cost of coding it as the $-\log_2$ of the probability. This leads to this definition of the Evidence Complexity approach ($\mathcal{EC}$):

$$Cost(E|T) = -\log_2 p(E|T) \qquad (3.12)$$

Where $p(E|T)$ represents the probability of the evidence, given the program $T$. This does not solve the problem by itself, because there are many different ways to estimate $p(E|T)$. For stochastic programs, this probability is derived by the specific method used for deriving the probabilities. However, for non-stochastic programs we need to derive a way to get the probabilities. Basically, as we detail regard below, we consider a non-stochastic program as a stochastic one where the probabilities are set uniformly.

Then, the idea is to use the program as a stochastic example generator and compare $E$ (the examples) with $M(T)$ (the Minimal Herbrand Model of $T$). This is highly related to the $\mathcal{PC}$ approach, but we have to derive the probabilities with some conditions.

The first condition is that the probability for each rule with the same predicate on the head must have a uniform distribution. We will suppose that $\forall e \in Q(T) : p(e|T) > 0$ (if not, that example can be removed). $Q(T)$, as seen in equation 2.20, is the set of ground facts derived from $T$.

The second condition is:

$$\sum_{e \in Q(T)} p(e|T) = 1 \qquad (3.13)$$

In case this second condition does not hold we can normalize the probabilities.

For instance, for the small program P8:

Listing 3.2: Program P8
```
even(0).
even(s(s(X))) :- even(X).
```

Each rule will have a probability of 0.5. The probability of each possible observable fact is computed by SSLD-resolution: $p(\texttt{even(0)}) = 0.5$, $p(\texttt{even(s(s(0)))}) = 0.25$, $p(\texttt{even(s(s(s(s(0)))))}) = 0.125$, ...

Another consideration is that the evidence could have or not repeated examples. Also we may cover other facts that are not at the evidence. Examples not covered by the program are not allowed: if there are examples not covered by the program, we will add these examples to the program. Therefore, we have four different combinations:

1. No repeated examples and $E = Q(T)$: This is the simplest case, since $Q(T)$ covers all examples in $E$ and nothing else, from the set of logic consequences of $Q(T)$ we will obtain $E$, and therefore we do not need to transmit $L(E|T)$.

2. No repeated examples and $E \subset Q(T)$: In this case we have two options: To code $E$ wrt. $Q(T)$ using the probabilities, or to code the exceptions, i.e. $Q(T) - E$. We select the cheapest option and then we add an additional bit to inform about our selection.

3. Repeated examples and $E = Q(T)$: From $Q(T)$ we can deduce $E$, but in this situation some of the elements of $E$ appear more than once, therefore we need to code the number of times that every element appears ($N_e$). This costs $\sum_{e \in E} CodeLength(N_e)$.

4. Repeated examples and $E \subset Q(T)$: This case is very similar to the previous one, but here we have that some logical consequences of $T$ do not appear in $E$, then we need to code $E$ wrt. $Q(T)$ using the probabilities. Another option could be to code the number of times that every element appears ($N_e$) where $N_e$ can be 0; this costs $\sum_{e \in E} CodeLength(N_e + 1)$.

Considering these ways of coding the evidence, we select the case that the evidence contains repeated examples (4), because we can use it for both stochastic and non-stochastic programs and evidence. It is also easy to code each example just using their probabilities. To simplify the coding scheme, we have preferred to consider the case with repeated examples, without distinguish whether the number of repetitions is high or not. If we had a huge number of repetitions, a more optimal coding could be found (consider reading [13]).

With this, we first need to code the length of the given evidence as $CodeLength(|E|)$.

Now, we compute the probability of the sample as the product of the single probabilities: $p(E|T) = \prod_{e \in E} p(e|T)$. The cost of coding this is $-\log_2 p(E|T) = \sum_{e \in E} -\log_2 p(e|T)$. We have to correct this expression because the order does not matter. Since the order is not important, there will be some combinatorial term corresponding to the number of possible permissible syntactically different but semantically equivalent orderings. A corresponding term should be subtracted from the message length. In this case, the number of different possible orders is given by the permutations with repetition of the elements of $E$.

To sum up, with the evidence we will proceed as follows:

$$cost(E|T) = \; CodeLength(|E|) + \sum_{e \in E} -\log_2 p(e|T) - \log_2 \left( \frac{|E|!}{\prod_{e_i \in E_{nr}} |e_i|!} \right)$$

where $|e_i| = \;$ number of times that example $e_i$ is in $E$

$\phantom{where } E_{nr} = \;$ the evidence without repeated examples

$$(3.14)$$

Here the term $\log_2 \left( \frac{|E|!}{\prod_{e_i \in E_{nr}} |e_i|!} \right)$ is a reduction because the order does not matter. Let us see the process using an example. If we have the program P8 (Listing 3.2) just seen and the evidence is this:

```
even(0).
even(0).
even(0).
even(s(s(0))).
even(s(s(0))).
```

Then the cost of coding this will be:

$$
\begin{aligned}
Cost(E|P) &= CodeLength(5) + \sum_{e \in E} -\log_2 p(e|P) - \log_2(\tfrac{5!}{3! \cdot 2!}) \\
&= 5.33789 - 3 \cdot log_2 \tfrac{1}{2} - 2 \cdot log_2 \tfrac{1}{2} - log_2 10 \\
&= 5.33789 + 3 + 4 - 3.32193 = 9.01596 \text{ bits}
\end{aligned}
$$

$$(3.15)$$

There is a particular situation that needs a better description. For instance, if we have this program:

```
0.25 :: p(0).
0.75 :: p(X):->X>=0,X<20.
```

Here, the estimated probability of the fact `p(0)` is not just $\frac{1}{4}$, because it matches also with the second rule. In this case, the SSLD-tree obtained for $p(X_1)$ will have leafs with two paths, due to this situations with multiple matching rules. We will refer to this as the *overlapping heads*. The solution is just to add up the probabilities of the paths (here $\frac{1}{4} + \frac{3}{4} \cdot \frac{1}{20}$). If there are more steps to arrive to the refutations (applying more rules), the same process applies.

### 3.3.1 Comparing the different approaches to code the evidence

It is interesting to compare the results of this coding scheme with the other approaches we reviewed in section 2.6, the Model Complexity ($\mathcal{MC}$) and the Proof Complexity ($\mathcal{PC}$). Imagine again the program P5 that we have seen in section 2.5.1, which comes from [13]:

Listing 3.3: Program P5 without probabilities

```
sum(0,X,X).
sum(s(X),Y,s(Z)):-sum(X,Y,Z).
```

And the evidence composed by the following set of examples (with no repetitions):

```
sum(0,0,0).
sum(0,s(s(0)),s(s(0))).
sum(s(0),s(0),s(s(0))).
sum(s(s(0)),s(s(0)),s(s(s(s(0))))).
sum(0,s(0),s(0)).
sum(s(0),s(s(0)),s(s(s(0)))).
sum(s(0),0,s(0)).
```

To code the evidence, we give the same probability for each rule.

So the cost of $L(E|T)$ using our proposal, the ($\mathcal{EC}$), is for this example 22.77 bits. The $\mathcal{MC}$ cannot be applied for this problem, because the program is not finite. For the $\mathcal{PC}$ approach, the cost of coding $L(E|T)$ is 37 bits. It is clear then the advantages of $\mathcal{EC}$ approach against the other two.

Let us then to use an extra example to compare the three different approaches (our $\mathcal{EC}$, the $\mathcal{MC}$, and the $\mathcal{PC}$). This example appears in [13] and also in [8], and it is related with reachability in a network. Suppose we have this knowledge base of which node reaches which other:

```
linked(0,1).        linked(0,3).
linked(1,2).        linked(3,2).
linked(3,4).        linked(4,5).
linked(4,6).        linked(6,8).
linked(7,6).        linked(7,8).
```

One vertex can reach another if there is a path between them in the graph, and we have obtained from this knowledge six different theories that cover the evidence:

| $T_1$ | `reach(X,Y).` | | | |
|---|---|---|---|---|
| $T_2$ | `reach(0,1).` | `reach(0,2).` | `reach(0,3).` | `reach(0,4).` |
| | `reach(0,5).` | `reach(0,6).` | `reach(0,8).` | `reach(1,2).` |
| | `reach(3,2).` | `reach(3,4).` | `reach(3,5).` | `reach(3,6).` |
| | `reach(3,8).` | `reach(4,5).` | `reach(4,6).` | `reach(4,8).` |
| | `reach(6,8).` | `reach(7,6).` | `reach(7,8).` | |
| $T_3$ | `reach(X,Y):- linked(X,Y).` | | | |
| | `reach(0,2).` | `reach(0,4).` | `reach(0,5).` | |
| | `reach(0,6).` | `reach(0,8).` | `reach(3,5).` | |
| | `reach(3,6).` | `reach(3,8).` | `reach(4,8).` | |
| $T_4$ | `reach(X,Y):- linked(X,Y).` | | | |
| | `reach(X,Y):- linked(X,Z).` | | | |
| $T_5$ | `reach(X,Y):- linked(X,Y).` | | | |
| | `reach(X,Y):- linked(X,Z),linked(Z,Y).` | | | |
| | `reach(0,5).` | `reach(0,6).` | | |
| | `reach(0,8).` | `reach(3,8).` | | |
| $T_6$ | `reach(X,Y):- linked(X,Y).` | | | |
| | `reach(X,Y):- linked(X,Z),reach(Z,Y).` | | | |

The Table 3.1 presents the code lengths of these programs using this scheme and using the evidence showed in Listing 3.4. The first five columns show the cost in bits of coding each part of the theories, whose sum is in the sixth column, which represents the cost of coding only $T$. The seventh column includes the cost of expressing $E$ wrt. $T$ for the programs.

Listing 3.4: Evidence for example or reachability

```
reach(0,1).    reach(0,2).    reach(0,3).
reach(0,4).    reach(0,5).    reach(0,6).
reach(0,8).    reach(1,2).    reach(3,2).
reach(3,4).    reach(3,5).    reach(3,6).
reach(3,8).    reach(4,5).    reach(4,6).
reach(4,8).    reach(6,8).    reach(7,6).
reach(7,8).
```

| $T$ | Rules | Lexicon | Heads | Bodies | Variables | $L_{\mathcal{EC}}(T)$ | $L_{\mathcal{EC}}(E|T)$ |
|---|---|---|---|---|---|---|---|
| 1 | 1.51929 | 18.21158 | 3.00000 | 10.68922 | 3.80735 | 37.22744 | 92.70272 |
| 2 | 9.00103 | 18.21158 | 177.45715 | 28.86658 | 0.00000 | 233.53633 | 52.95620 |
| 3 | 7.36570 | 18.21158 | 87.05865 | 19.19294 | 3.80735 | 135.63622 | 68.58149 |
| 4 | 2.51929 | 18.21158 | 6.00000 | 18.45792 | 11.03617 | 56.22497 | 71.27147 |
| 5 | 5.92873 | 18.21158 | 43.35940 | 21.36517 | 12.88417 | 101.74905 | 70.40881 |
| 6 | 2.51929 | 18.21158 | 6.00000 | 15.28800 | 12.88417 | 54.90304 | 66.88270 |

Table 3.1: Code lengths of the programs and their parts using our approach $\mathcal{EC}$

The important thing is to compare the cost of coding $E$ wrt. $T$ with the three different approaches. The following table compares the evaluation of each theory according to the Model Complexity approach, the Proof Complexity approach and the Evidence Complexity approach we are using:

$$
\begin{array}{ccccccc}
\mathcal{MC} & T_6 & T_1 & T_4 & T_5 & T_3 & T_2 \\
\mathcal{PC} & T_1 & T_4 & T_6 & T_5 & T_3 & T_2 \\
\mathcal{EC} & T_6 & T_4 & T_1 & T_5 & T_3 & T_2
\end{array}
$$

The ranking of theories is from cheapest to most expensive, from left to right. The results for $\mathcal{MC}$ and $\mathcal{PC}$ come from [8] and are in Table 3.2 (we have rounded our results for $\mathcal{EC}$ to one decimal digit). These numeric results are not directly comparable, because we have included the numbers as function symbols, and the other approaches do not. But if not, theory $T_1$ had obtained an infinite for $L(E|T)$ (the signature is infinite if we accept any number as values for predicate `reach/2`).

| $T$ | $L_{\mathcal{EC}}$ | $L_{\mathcal{MC}}(T)$ | $L_{\mathcal{MC}}(E|T)$ | $L_{\mathcal{MC}}$ | $L_{\mathcal{PC}}(T)$ | $L_{\mathcal{PC}}(E|T)$ | $L_{\mathcal{PC}}$ |
|---|---|---|---|---|---|---|---|
| 1 | 129.9 | 12.5 | 60.4 | 72.9 | 12.5 | 120.5 | 133.0 |
| 2 | 286.5 | 178.5 | 0.0 | 178.5 | 178.5 | 80.7 | 259.2 |
| 3 | 204.2 | 111.7 | 0.0 | 111.7 | 111.7 | 96.3 | 208.0 |
| 4 | 127.5 | 43.7 | 47.4 | 91.1 | 43.7 | 110.6 | 154.3 |
| 5 | 172.2 | 94.5 | 0.0 | 94.5 | 94.5 | 101.9 | 196.5 |
| 6 | 121.8 | 53.8 | 0.0 | 53.8 | 53.8 | 106.1 | 160.0 |

Table 3.2: Code lengths of the programs using the different approaches. Here $L = L(T) + L(E|T)$.

While $\mathcal{MC}$ and $\mathcal{EC}$ obtain almost identical rankings, $L_{\mathcal{PC}}$ differs in their preferences. Both ($\mathcal{MC}$ and $\mathcal{EC}$) rank the "correct" theory ($T_6$) as the best, and $T_2$ as the worst. $L_{\mathcal{PC}}$ selects the most general theory $T_1$ as the best theory. This fact is mainly because $L_{\mathcal{PC}}$ needs to always encode $E$, even when it can be derived from $T$. For that reason, it gives preference to short theories even though being too general. So, in this example we see that $\mathcal{EC}$ inherits the good properties of $\mathcal{MC}$ when $\mathcal{O}$ (possible observations) is finite.

But the good thing is that $\mathcal{EC}$ is also applicable to problems with function symbols, to problems with repeated examples and also for stochastic programs. We will see more coding and examples in chapter 6.

# Chapter 4

# The tool

In this chapter we present a software tool that implements the MML coding scheme seen in the previous chapter. This chapter is organized as follows: first we will present the interfaces of the tool. Next, we will analyze some problems that have appeared, such as the necessary canonization of the code, some changes applied to special predicates in the compiler and also the control of the execution of the user code. Later on, we will explain how the probabilities have been introduced in the Prolog language, and how a meta-interpreter is used to determine the probability of the examples. There is a final section dedicated to free variables, another difficult problem that we have addressed.

## 4.1  Interfaces of the software tool

The tool is an interpreted Prolog source file that needs to be executed in the Yap Prolog compiler, which load also some extra available libraries available in that compiler.

The results are returned in the standard output (`stdout`).

The tool receives the data to analyze from files referenced in the command line. It will return the cost of each combination in the Cartesian product of two different kinds of files that we could use: the program and the evidence files. The content of each kind of files is:

- The program or programs to analyze (at least one program is compulsory).

- The evidence. It could be composed by different files also, The different examples sources must be compatible as evidence with the different programs.

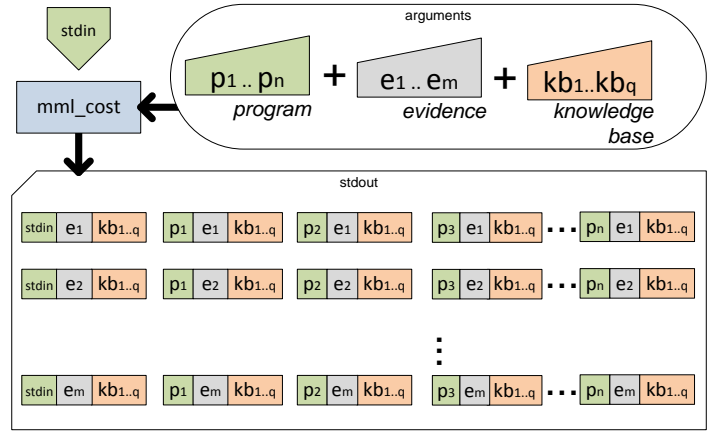

Figure 4.1: Combinations generated by the tool from inputs

- The knowledge base. This kind of files is not going to be included in the Cartesian product to generate the results. Despite the others, if you introduce more than one KB source, they'll be merged together.

In the Figure 4.1, we can see that if we have $p_1 \ldots p_n$ items considered as programs, combined with the $e_1 \ldots e_m$ different evidences, we obtain the matrix of $n \times m$ outputs. If there is also (or only) an input program in the `stdin`, it will also be combined with the $m$ evidence sets.

For instance, if we have these command line arguments:

```
cat p4 | ./mml_cost p1 p2 p3 --examples=e1,e2 --kb=kb1,kb2
```

We will have eight different outputs from the tool (see Table 4.1) using the different program sources (three as arguments `p1`,`p2` and `p3`, and also in the standard input `p4`) with the example files (`e1` and `e2`). The knowledge base will be considered in each one of those eight combinations, and it will contain the direct merge of `kb1` and `kb2`.

Let us introduce some screenshots of the tool in a Windows environment as an example: the Figure 4.2 shows the help that the tool gives to us if we do not use any argument.

Figure 4.3 shows the results of an execution with a program which has evidence and knowledge base. If we had introduced a second program in the command line (compatible with the evidence and knowledge base done) then a second evaluation for that program would have done.

Notice also that the tool analyzes the correctness of the user code ($P$, $B$ and $E$). So, a warning has been issued here due to not assigned variables in the user code.

|      | Program   | Examples | KB      |
|------|-----------|----------|---------|
| #1   | stdin (p4)| e1       | kb1+kb2 |
| #2   | stdin (p4)| e2       | kb1+kb2 |
| #3   | p1        | e1       | kb1+kb2 |
| #4   | p1        | e2       | kb1+kb2 |
| #5   | p2        | e1       | kb1+kb2 |
| #6   | p2        | e2       | kb1+kb2 |
| #7   | p3        | e1       | kb1+kb2 |
| #8   | p3        | e2       | kb1+kb2 |

Table 4.1: Example of number of outputs of the tool using a combination of different files

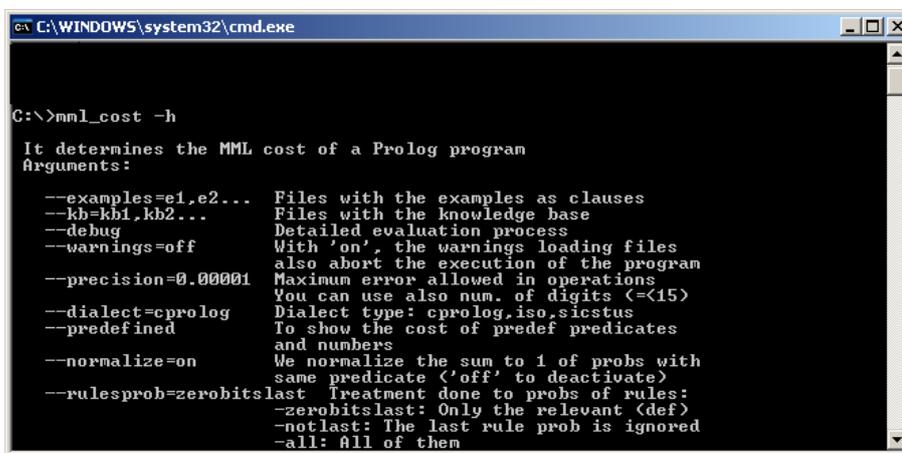

Figure 4.2: Calling the tool without arguments

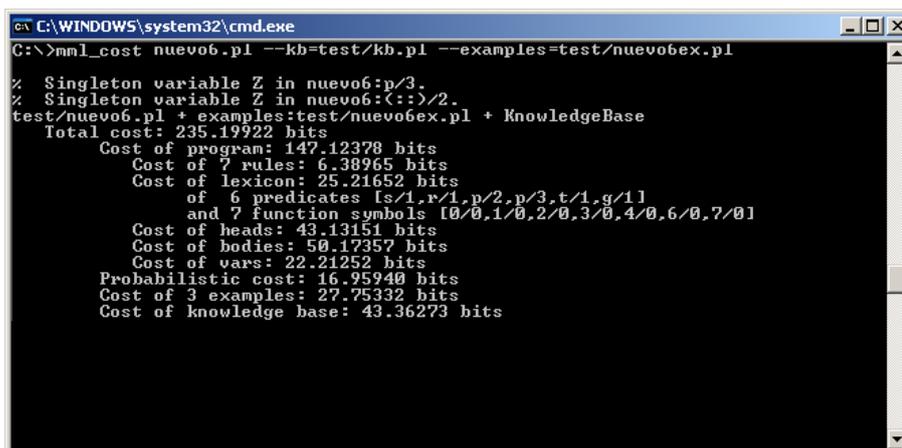

Figure 4.3: Calling the tool with a program

There is also a Web wrapper for the tool for those without a console or an installation. All the options available from the command-line are also available at `http://users.dsic.upv.es/~flip/mml-cost`, whose main
page is shown in Figure 4.4.

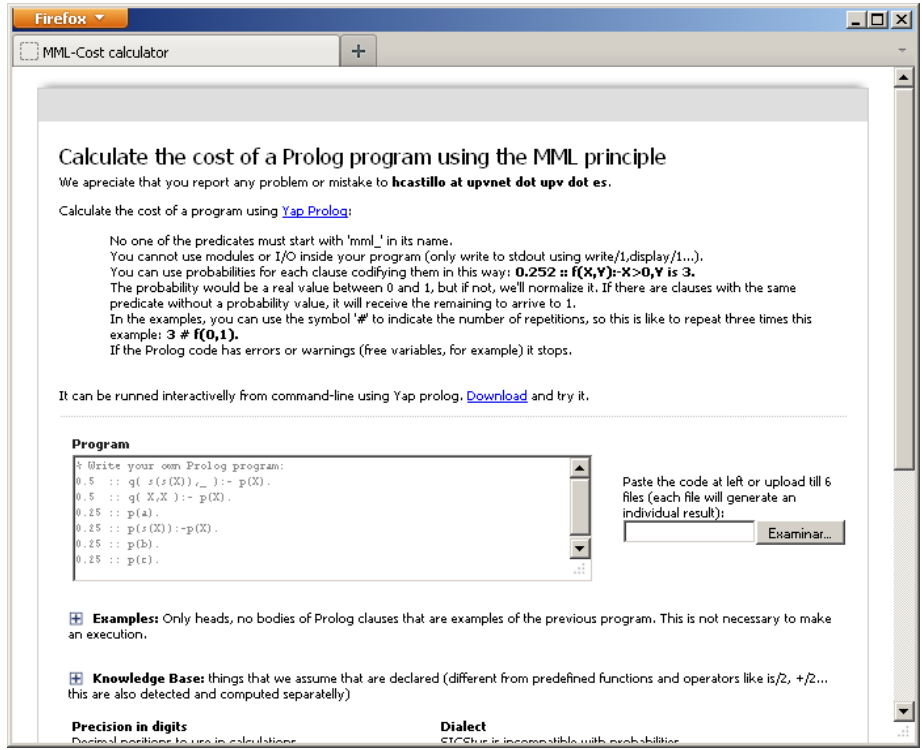

Figure 4.4: Initial main page of the Web based version

The results are shown in Figure refscreenWeb2. They are easier to un-
derstand in this way than for the console application, especially when you
introduce more than a program source or more than an example file, because
the table returned allows us to easily compare the results between them.

There are some parameters that alter the results given by the tool, de-
scribed in section 4.5. They allow us to debug the tool or to format the
results. These are fully explained in Appendix D.3.

## 4.2  Transformations of the user code

In this section we describe some preprocessing and transformations that we
need to apply to the user code.

### 4.2.1  Canonizing the input program

The special predicate `;/2` allows us to synthesize in a single rule many dis-
junctions. This makes the code more readable. For example, if we have:

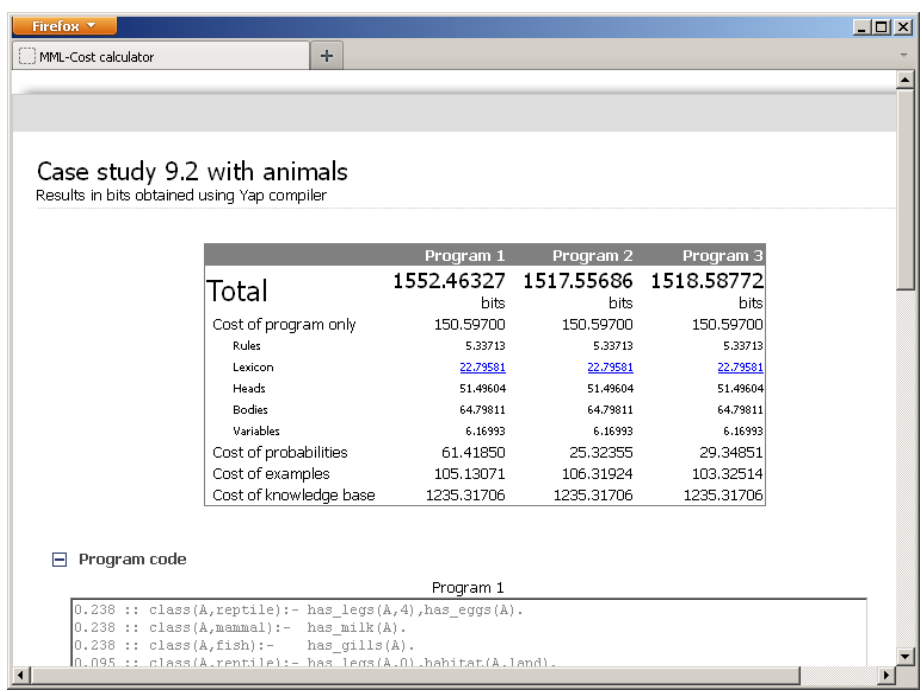

Figure 4.5: Results screen of the Web-based version

```
1/5  ::  p(_):-a,b,c;d,e,(f;g).
```
The coding scheme seen in chapter 3 could not handle this. In order to solve this, we could untie it totally in three equivalent rules. Notice that we have divided the probability uniformly:

```
1/15  ::  p(_):-a,b,c.
1/15  ::  p(_):-d,e,f.
1/15  ::  p(_):-d,e,g.
```

When Yap Prolog stores a rule, it is done in the merged form, and the only solution was to transform it after the user code was loading in a new module where we have replaced the merged rules in new ones with no presence of the `;/2` predicate. But this has not been an easy task: when we access the IDB rules using the predicate `clause/2`, the assigned variables in the head and in the body of a rule could not lose the association, so we must process each rule `Header:-BodiesWithSemicolon` together in any moment to avoid the creation of new variable names and lose the relationship between head and bodies:

```
manage(Head,BodiesWithSemicolon):-(.....)

manage( p(X,Y), (X>0,Y is 5;X<0,Y is 3) ).
%  now Prolog will generate a call in the form:
```

```
%    manage( p(_A,_B), (_C>0,_D is 5;_C<0,_D is
   3)

% So a possible solution could be:
manage(HeadAndBodiesTogether):-(.....)
manage( ( p(X,Y):-X>0,Y is 5;X<0,Y is 3) ).
%  manage( p(_A,_B), (_A>0,_B is 5;_A<0,_B is 3)
```

But, at the same time we were trying to transform a rule dividing it by
its disjunctions, we also need the number of appearances of each different
variable. Storing it was also difficult without loss of the original name (each
time Prolog makes a call or a matching, it assigns new variable names), so
the decision was to transform the rules and their symbols and variables to
strings, analyze them using a grammar and the Prolog predicate `phrase/2`
and store the result in other separated lists with the labeled information we
need: total number of appearances of variables, function symbols, number of
free variables, etc. Finally, when the process is done and the new probabilities
are assigned, we create a new module with the transformed rules generated
from the final transformed list of strings.

Another transformation we are going to do is to convert the rules to polish
notation.  Fortunately we have `format_to_chars/3` with the special
control sequence `'~k'` that uses the `write_canonical/2` predicate:

```
? write_canonical( 3+2+p(3,5) ),nl.
p(+(3,2),[]).
```

We have also decided to modify lists by replacing the predicate `./2` and
the function symbol for the empty list `[]`.  There are two reasons for doing it:
with this transformation, we will also consider the appearances of the func-
tion symbol of the empty list, and it will be easier to divide the disjunctions
and to go through the items of the body of the rules.

### 4.2.2   Solving the problem with predicate `**/2`

Yap Prolog has some particular interpretations about the predicate `**/2` that
could interfere with some user code that comes from other Prolog compilers
or dialects.

Suppose that we have `X is Y**2` (one of the power function symbols to
obtain $Y^2$, because we have also the equal predicate `^/2`).  If we ask Yap
Prolog about `9 is 3**2` the answer will be surprisingly false.

This is because the predicate `**/2` returns always a real number instead of
returning the data type of the operation (however, the predicate `^/2` returns
the correct one).  Float exponentiation is needed for (very) large integers, so
the developers preferred to create a new predicate `^/2` returning the result
in the same kind of values used (real if you are using a real numbers, integer if you are using integers), instead of changing the existing one, since it has been used for exponentiation since the 80's.

But we preferred the other behavior, so during the loading of user code, the occurrences of `**/2` by `^/2` in order to avoid this misinterpretations.

### 4.2.3 Avoiding direct execution of objectives

Finally, another thing we are going to transform during the loading of the program is the execution of objectives (clauses without a head). By default, they are executed immediately after loading the file and they could affect also our program (generating for example an abnormal failure using the predefined predicate `halt/1` that will break our analysis of the user code).

The solution has been to create a new dummy head for those objectives with the dummy predicate `mml-objective/0`, transforming them in a normal rule:

```
user:term_expansion( ( :- Body ) , DefGoal ) :-
    DefGoal = ( mml_objective :- Body ).
```

If we completely remove them, then the analysis of this code is not done and we do not have really the MML cost of the whole program. Refurbishing the code adding a head without variables and predicate is not going to significantly increase the cost and at the same time we are going to have the complete code analyzed.

## 4.3 Introducing probabilities in Prolog

The Prolog language is not thought to work with stochastic logic, and also our purpose with this tool is not to have a full implementation of a stochastic Prolog compiler. The way in which we are going to introduce the probabilities will be a pre-processing of the user code in which we will transform the probabilities in predicates attached as arguments inside the rules.

### 4.3.1 The pre-process of the user code

The Yap Prolog compiler has the possibility of enriching the number of operators with new ones, and also to pre-process all terms read when consulting a file. We are going to take the advantage of these characteristics of Yap Prolog in the same way it has been used by the developers of ProbLog [9]. With a special binary operator `::` with higher priority than the same binary operator `:-`, we will be able to prefix all the clauses with a real number that represents the probability of that rule, in a clear and nice notation:

```
0.33434 :: Head :- Body.
0.1      :: Head.
```

These probabilities are really stored internally in the body of the clause inside the IDB of Yap Prolog, transforming the previous user code in this:

```
Head :- mml_prob(0.33434),Body.
Head :- mml_prob(0.1).
```

That is the reason to forbid the use of the predicate `mml_prob/1` in the user code loaded, and the same with the predicate `::/2`. It is also important to normalize the probabilities later. The user could use any real number higher than zero as a value for the probability of a rule. If the sum of probabilities for all the rules with the same predicate is greater than 1, they are being replaced later by the correct real number that makes the sum of 1 for all the probabilities for the same predicate.

There are two reasons that justify the way we store the probabilities in the IDB. The first one is involved with the access to that number. Since we need to manage many rules, it is easier to always have that value in the body, especially if there are more rules with the same predicate name.

If we have two rules with the same predicate header (and different probabilities), and we do not store the probability in the body, each time we need to use the probability we will need to analyze which is the rule identifying by the body, with the corresponding waste of time.

Another reason is for compatibility with the code for non-probabilistic rules: if a rule does not have a probability, its format was (`Head :- Body`), or in polish notation `:-(Head,Body)`, that is rather different to

$$::(Probability,:-(Head,Body)).$$

So transforming the stochastic rules to the same notation of non-stochastic ones make simpler the coding.

### 4.3.2   Management of the repetitions in the evidence

In the example file (the evidence) we could use a new binary operator `#/2` (again with higher priority than `:-` but incompatible with `::`), created to avoid the necessity of repeating the same example many times. So, if we have this evidence:

```
color(blue).
color(blue).
color(red).
color(red).
color(red).
color(green).
```

We could synthesize this as:

```
2 # color(blue).
3 # color(red).
color(green).
```

And these repetitions are stored internally in the same way we have stored the probabilistic values. We pre-process the examples files when they are loaded in our program replacing the repetitions and introducing them as the first term of each fact of the evidence. So, in the previous example, after loading it in our tool, it will be stored in the IDB as follows:

```
color(blue):-mml_rep(2).
color(red):-mml_rep(3).
color(green).
```

As before, if a fact has no repetitions, it is stored without body (notice that our facts in the evidence could be transformed as rules with body, but this is not going to affect our evaluation of the cost).

If the user has decided not to use the notation with the operator #/2 and has introduced repeated examples as the first example (where `color(blue)` appears twice), then it is not a problem because during the loading of the code this is also detected, remaining only one of each different with the correct term of `mml_rep/1`.

It is important to only allowing the use of positive integer numbers for repetitions: the program will fail with any other value.

### 4.3.3   How we handle real and rational numbers

Our probabilities could be inserted in the program using only real numbers. In order to insert a probability of $\frac{1}{3}$ in a rule, it is necessary to use 0.33333, if our precision is defined with 6 significant decimal positions. The reason for not allowing rational numbers directly is to sustain a compatibility of the code with other previous approaches to stochastic programs. So if the program to analyze is going to have a $\frac{1}{3}$ probability for each clause, it will be necessary to use 0.33333. It is also possible to use proportional amounts that the tool will normalize. For example, in the next program the value of 1 will be transformed to $\frac{1}{3}$ in each rule:

```
1 :: p(a).
1 :: p(b).
1 :: p(c).
```

Even though Yap Prolog has the possibility of managing rational numbers with the predicates `rationalize/1`, `rational/1` and the operator `rdiv/2`, the way it is implemented could produce some strange results, as we see in the following examples:

```
?- A is rational(0.25).
```

```
   A is 1 rdiv 4

?- A is rational(0.1).
   A = 3602879701896397 rdiv 36028797018963968

?- A is rationalize(0.1).
   A = 1 rdiv 10
```

The documentation of the compiler says that the function `rational/1` returns a floating point number, the rational number that exactly represents the float. As floats cannot exactly represent all decimal numbers the results may be surprising. The function `rationalize/1`, unlike the first one, is only accurate within the rounding error of floating point numbers, "generally" producing a much smaller denominator. However, since the denominator is just the most important part of the rational numbers that we are going to use in the estimation of the cost, finally our own implementation for rational numbers was used.

In order to approximate a real number to the nearest rational one, there are many different techniques [18]. The three simplest ones are this:

1. Removing common factors from the fractions. It is always correct but unable to manage 0.33333 in order to get 1/3, because it will be instead approximated by 33333/100000 (that is actually exactly the equivalent to our float number, but not an approximation, as we wanted, and also a bit inefficient).

2. By brute force (sequential testing). If we have the real number $R$:

   (a) We obtain $Q$ as the division of $\frac{1}{R}$.

   (b) Then the float number $R$ is transformed to $\frac{n}{Q \cdot n}$, where $n = 1...k$, till we have $Q \cdot n - round(Q \cdot n) < Precision$.

   This method requires more steps, but it is more precise.

3. And the fastest method, which is the one implemented here: decimal inversion, which is not easy to understand. First we isolate the fractional part of the number, because we need only the decimal part to get the denominator through inversion. After trimming the number, we invert the fractional part and we proceed using this pseudo-code:

```
getDenominator(num) {
    var decimal = num_int(num);
    if (Math.abs(decimal)<Precision)
        return 1;
    else {
```

```
            num = 1/decimal;
            return num*getDenominator(num);
        }
    }
```

With this function we will obtain the denominator, and despite we have only the fractional part of the original number, there is no difference: Trimming the integer part will not affect the denominator q, because the denominator for the number and the decimal are the same. A number 1.4 is represented as $\frac{7}{5}$ before trimming, and after trimming and leave only the decimal part, it is $\frac{2}{5}$, so the denominators for both are the same.

### 4.3.4 Euler's Totient and the $F$ function

In chapter 3, we developed the basics of the cost of coding some aspects of the programs. There were two different functions that we needed to develop in Prolog because they were not part of any library:

- The Euler's Totient (as seen in Appendix A.2) needs the $Phi$ function (also known as Euler's $Phi$ function or $\varphi(n)$) used in the coding of the numerators of rational numbers.

  It is defined for a given integer number $m$ as the amount of numbers from 1 to $m - 1$ without common divisors with $m$. If $m = 1$ then the Euler's Totient is defined as 1 (instead of 0). The formula that represents this is:

  $$\varphi(n) = n \cdot \prod_{p|n} (1 - \frac{1}{p}) \tag{4.1}$$

Where the product is over the distinct prime numbers dividing $n$ ($\prod_{p|n}$ means that $p$ iterates over prime numbers in the range $1 \ldots n$). A direct (but non-efficient) implementation for obtaining this with Prolog:

```
    totientPhi(Number,Result) :-
        totientPhi(Number,Result,1,0).

    totientPhi( _, 1, Acum, Result):-
        Result is Acum + 1.
    totientPhi( M, N, Acum, Result):-
      ( coprime(M,N),
        NewAcum is Acum + 1
        ;
        NewAcum is Acum
      ),
```

```
        NewN is N - 1,
        totientPhi( M, NewN, NewAcum, Result).
```

But a more efficient implementation was the following:

```
    totientPhi(Number,Result) :-
        totientPhi(Number,Result,1,0).

    totientPhi(M,Phi,M,Phi).
    totientPhi(M,Phi,K,C) :-
        K < M, coprime(K,M),!,
        C1 is C + 1, K1 is K + 1,
        totientPhi(M,Phi,K1,C1).
    totientPhi(M,Phi,K,C) :-
        K < M, K1 is K + 1,
        totientPhi(M,Phi,K1,C).
```

- And the other complex thing was to obtain the special function used for coding the variables of the rules (described as "function F" in the code), which is obtained from the number of variables $n_v$:

$$VariableProd(d,n_v) = (n_v)^d - \sum_{i=1} n_v - 1 \binom{n_v}{i} \cdot VariableProd(d,i) \tag{4.2}$$

- Obviously the binomial coefficient of two numbers $n_v$ and $i$ or $\binom{n_v}{i}$ was not implemented in Prolog, so the code used for this was:

```
    binomialCoefficient(N,N,Result):-
        !,factorial(N,Result).
    binomialCoefficient(N,K,Result) :-
        N>=K,
        NMinusK is N-K,
        factorial(NMinusK,FactNMinusK),
        factorial(N,FactN),
        factorial(K,FactK),
        Result is ( FactN / (FactK*FactNMinusK))
            .
```

## 4.4   Analyzing the evidence

To obtain the probability of an example, the simplest method could be to use the Prolog interpreter, and if it success, to use then the labeled probability of the rule to recover the probability. Since there can be different successful paths in the resolution (because different rules can be used), this is not enough for us and will need to solve using a different mechanism: a meta-interpreter, adapting the concepts of the well-known Vanilla meta-interpreter to our needs.

A possible meta-interpreter has only one clause which directly calls Prolog interpreter: `solve(A):-call(Goal).` But it is just the direct execution of code in Prolog what we want to avoid.

The Vanilla meta-interpreter is a meta-interpreter for Logic Programming (and then, for Prolog) specified in rules of Logic-Programming (so, in Prolog). It was first introduced by [30] in 1982, and extended with the years. It will allow us to intercept the solving procedure as we are going to describe. The core of the Vanilla meta-interpreter is this:

Listing 4.1: Vanilla meta-interpreter used to estimate evidence

```prolog
builtin(A is B).
builtin(A = B).
builtin(A >= B).
builtin(read(X)).
builtin(A > B).
builtin(A < B).
builtin(A =:= B).
builtin(A =< B).
builtin(functor(T, F, N)).
builtin(write(X)).

solve(true):- !.
solve((A,B)) :-!, solve(A), solve(B).
solve(A):- builtin(A), !, A.
solve(A) :- clause(A, B), solve(B).
```

The meta-interpreter shown in Listings 4.1 allows us to generate the SSLD-tree of execution of a goal without arrive to execute the Goal in the Prolog interpreter. There is no call to `call/1`, so we are not going never to execute any predicate of the program we are analyzing. Instead of that, we are unfolding the SSLD-tree, obtaining in each step all the different branches, launching again till we success with all the expanded terms, and later summing the returned probabilities of the branches.

To obtain the probabilities then, we only will need to add to the Vanilla meta-interpreter some code able to catch and sum the values we have stored in the clauses of the program with the predicate `mml_prob/1`:

```prolog
solve( ( mml_prob(Number), Other ), ResultProb ):-
    number(Number),!,
    solve( Other, Branch ),
```

```
ResultProb is Number * Branch.
```

Later, in a second visit to the generated tree of probabilities, we will normalize the values (because the probability of each predicate must be 1).

### 4.4.1   Executing the rules when solving an example

There are some predefined predicates in Prolog that could affect the behavior of the calculations. For instance, the `debug/0` predicate will be executed and it interacts with the same MML coding.

So, since those predicates involved with debugging conflict with our purposes (their presence would increase the information and then the cost for the MML coding), we will ignore them during the execution when we are executing the evidence to find its probability.

During that process of executing the rules of the program to analyze an example, we need to analyze whether each term to execute is or is not a predefined predicate (using `predefined_predicate/2` from Yap Prolog). If it is, they will be discarded. Once we have discarded those predicates, the next step is to execute the others using `with_output_to_chars/2`, used to send the output of the execution of the goal to a string instead to `current_output`:

```
executeTransparently(debug):-!.
executeTransparently((spy _)):-!.
[...]
executeTransparently( Term ):-
   copy_term(Term,NewCopy),with_output_to_chars(
      NewCopy,_).
```

With this technique, all these predicates are ignored and not executed during the evaluation of examples:

- `debug/0`, `nodebug/0`, which switches the debugger on/off.

- `spy/1`, `nospyall`, `nospy/1`, which set/unset spy-points on predicates.

- `trace/0`, `notrace/0`, which switch on/off the debugger and also implies the same with the tracing.

- `leash/1`, which sets leashing mode to certain value.

- `source/0`, `no_source/0`: these two predicates are not for debugging, but when we analyze the program we need to ensure that we are obtaining its source. If in the middle of the code a `no_source` appears, the tool will have no more access to the source. To avoid this, we block that predicates also.

- `halt/1` and `halt/0`, to abort the execution of program.

The schema of execution of items of clauses is shown in Figure 4.6.

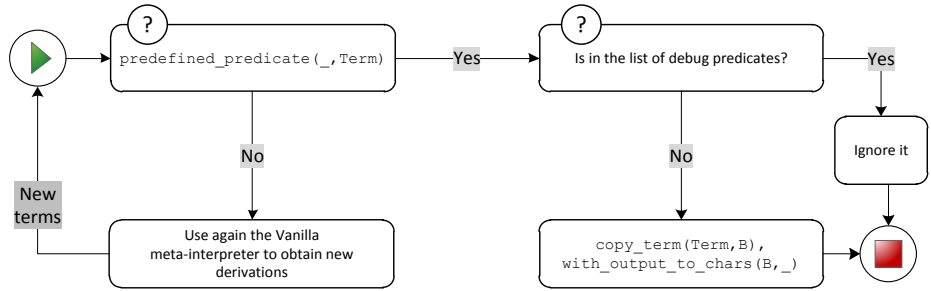

Figure 4.6: Execution schema to check for debug and execution of predefined predicates

### 4.4.2 Obtaining the probabilities of the root predicates

The first step to estimate the probability of a fact is to know the value of all the facts with the same functor name, because we need to normalize. That is:

```
p(3).              --> p(X).
q(3,1).            --> q(X,Y).
```

Till now, we have built a Vanilla meta-interpreter to estimate the probability of the example (for instance `p(3)`). Unfortunately, this meta-interpreter is not going to be useful for estimate `p(X)`, because we need to control some extra problems in the solving process with only variables and no assignments.

As we introduced in subsection , the approach has a maximum depth limit used in the go round SSLD-derivations used to estimate the probability of the root predicates (with the success branches labeled with the probabilities). This is because we can have infinite SSLD-derivations in some programs.

The arbitrary value used in the tool is 20 (it could be changed using the modifier `--maxrecursion`), considered enough to avoid the loss of successful normal branches of the program.

But there is another limitation involved with inequalities and equality predicates. For instance, if we have this two predicates, describing relationships in an intensional way:

```
p(X):-X==a;X\=b.
q(X):-X>0.
```

Our Vanilla meta-interpreter will fail with both predicates, due to the impossibility of assign all the values to the variable if we are deriving $p(X_1)$ or $q(X_1)$. It is possible at least to solve `p/1` with the signature. But `q/1` represents the list of all positive numbers, so the probability of the predicate also it will be problematic.

In our tool, the equalities and inequalities of the bodies of the rules are problematic and they will make not to solve covered examples of a program: they are not going to success. Our Vanilla meta-interpreter does not know how to obtain values to instantiate if there are equalities and inequalities like the one described in the example.

We also need in this second meta-interpreter to control the apparition of new free variables in the rule we are analyzing. In each branch of the SSLD-tree our meta-interpreter must check if there are new free variables and compute them in the probability of the step.

So, the solution will be to use a different Vanilla meta-interpreter enriched to control both free variables and predicates that could fail (all the comparisons).

### 4.4.3   Repeated predicates in the body of the rule without reduction

There is also another problem not really related with free variables, but with a similar treatment. We can have an infinite SSLD-tree if in the body of a rule if it appears the same predicate without any reduction in the body: it is an odd situation, because we cannot refute this program and it will also fail in an interpreter. For example:

```
0.4 :: r(A,B):-r(A,B),B > 0.
0.6 :: r(A,B):-B is A.
```

Here the predicate `r/2` appears again in the first rule at the body, so we have an infinite left-recursion. In this situation, the tool will fail with an error (the same that it will happen if we execute this program with an interpreter), because it has rules without sense (a non-terminating program).

### 4.4.4   The problem with the factorial

In our development of the tool, we looked for a mathematical library able to manage huge numbers. The *GMP* library [1].

But we also will need to use another library to determine a division of floating-point computation that is not possible to attach with Yap Prolog, to determine the cost of the evidence, which we have seen in the previous section 3.1.4. Specifically the problem is the fraction $\log_2 \left( \frac{|E|!}{\prod_{e_i \in E_{nr}} |e_i|!} \right)$.

---

[1] GMP (GNU Multiple Precision Arithmetic Library, `http://gmplib.org`) is a free library for arbitrary precision arithmetic, operating on integers. This library can be included in the Yap interpreter during the compilation, allowing the interpreter to have no limits with integer operations, except for the limitations due to the available memory in the machine GMP runs on.

To solve this fraction we will use the *MPFR* library [2]. If it is not installed in the system, the tool runs without it using an alternative algorithm that obtains the same results when we have less than 900 examples. With more examples in the evidence, the huge values obtained will generate an error in our tool.

We will proceed as follows when the MPFR library is not present in the system. In all the evidence sets, the fraction that causes the problem has a factorial in the numerator that could be simplified:

$$\text{since always } a < b \quad \tfrac{a!}{b!} = a \ldots b$$
$$\text{for instance} \quad \tfrac{15!}{10!} = 15\,14\,13\,12\,11 = 360360$$
(4.3)

The other thing we are going to do is to apply at the same time the multiplications for the factorial in the numerator and in the denominator, inside the recursive function:

$$\tfrac{a!}{b!} = \texttt{factorial(a,b)} = \tfrac{a}{b}\ \texttt{factorial(a-1,b-1)}$$

Both techniques together give us the possibility of having a big numerator before applying the division by a big denominator, making a division each time we apply a multiplication, and also simplifying the operations.

## 4.5   About the cost of stochastic programs

When the tool determines the cost of the probabilities of each rule in a stochastic program, by default it only codes the probabilities of the necessary clauses, since probabilities are assumed to be normalized. This is the coding explained in the previous chapter 3.1, but there are also two more possibilities that are available in the tool using the modifier `--rulesprob=x`.

With an example it will be easier to understand the differences between them. If we have this simple program:

```
0.5  ::  p(X):-X>0,q(X).          [1]
0.5  ::  p(X):-X<0,r(X).          [2]
1    ::  q(3).                    [3]
```

The three options are:

- The default behavior, in which we only sum up the cost of the probabilities of the necessary clauses (without any option, or using the

command-line modifier `--rulesprob=zerobitslast`).  We will code
the probability 0.5 of rule `[1]` only (the probability 0.5 of rule `[2]` is
unnecessary because it is the last rule of `p/1`. Also, the probability of
rule `[3]` is not necessary because it is 1.

- Using the criterion of ignoring only the probabilistic cost of the last
  rule of each predicate (`--rulesprob=notlast`), we will sum up `[1]`
  and `[2]`, even if `q/1` has a probability of 1.

- Using the option to code every probability (`--rulesprob=all`), then
  the three probabilities of the rules will be summed up.

## 4.6   The treatment of the knowledge base

As we have described in section 3.1.5, the knowledge base is not really part
of the model we want to analyze, and since this previous knowledge is shared
between sender and receiver, we do not need to transfer it.

But there are some situations in which maybe the user will find it easier
to have a set of common code for different models. The common code could
then be seen as the knowledge base, and the different models as the programs.
The tool will then estimate the cost of the common code as the knowledge
base, and it will report it in the results separately.

This option is enabled with the command-line option `--kb=files`. The
tool will check that no rule in the knowledge base will appear also defined
with the same predicate name in the program (and the same in the other
direction).

# Chapter 5

# Examples

In order to see this work in practice, we will give some examples where we will apply the MML coding to evaluate and compare several theories using our tool.

We are going to include three small examples in this chapter.

The chapter starts addressing the classical well known problem of animal classification, using different theories, and we will use the tool to compare them and select which one is the most optimal (in terms of the MML principle).

There is a second example where we address a language analysis problem and the possibility of using the MML principle to identify the authorship of a document by determining the model of how a user uses that language.

Finally, the last example also shows the use of probabilities: a probabilistic graph, and a closer look at how the tool works.

## 5.1   Animal classification

This classical logical problem involving classification of animals was first used in 1985 by Winston and Horn [60], and it is a well-known example in ILP. We have eight known attributes of animals, here shown as mode declarations used in Progol:

```
has_milk(+animal).
has_gills(+animal).
has_eggs(+animal).
has_gills(+animal).
habitat(+animal,#habitat).
has_legs(+animal,#nat).
homeothermic(+animal).
has_covering(+animal,#covering).
```

And we also have a set of facts of positive and negative examples (the negative ones are written as objectives, not facts):

Listing 5.1: Evidence (Animal classification)

```
% 16 positive examples:
class(dog,mammal).
class(dolphin,mammal).

class(platypus,mammal).
class(bat,mammal).

class(trout,fish).
class(herring,fish).
class(shark,fish).
class(eel,fish).

class(lizard,reptile).
class(crocodile,reptile).
class(t_rex,reptile).
class(snake,reptile).
class(turtle,reptile).

class(eagle,bird).
class(ostrich,bird).
class(penguin,bird).

% 42 negative examples (used only to generate models):
:-class(trout,mammal).
:-class(herring,mammal).
:-class(shark,mammal).
:-class(lizard,mammal).
:-class(crocodile,mammal).
:-class(t_rex,mammal).
:-class(turtle,mammal).
:-class(eagle,mammal).
:-class(ostrich,mammal).
:-class(penguin,mammal).
:-class(dog,fish).
:-class(dolphin,fish).
:-class(platypus,fish).
:-class(bat,fish).
:-class(lizard,fish).
:-class(crocodile,fish).
:-class(t_rex,fish).
:-class(turtle,fish).
:-class(eagle,fish).
:-class(ostrich,fish).
:-class(penguin,fish).
:-class(dog,reptile).
:-class(dolphin,reptile).
:-class(platypus,reptile).
:-class(bat,reptile).
:-class(trout,reptile).
```

```prolog
:-class(herring,reptile).
:-class(shark,reptile).
:-class(eagle,reptile).
:-class(ostrich,reptile).
:-class(penguin,reptile).
:-class(dog,bird).
:-class(dolphin,bird).
:-class(platypus,bird).
:-class(bat,bird).
:-class(trout,bird).
:-class(herring,bird).
:-class(shark,bird).
:-class(lizard,bird).
:-class(crocodile,bird).
:-class(t_rex,bird).
:-class(turtle,bird).
```

Negative examples are use by ILP systems to infer the model, but we are not going to use them for MML evaluation.

The purpose is to learn the class of those animals using the predicate `class/2`, where its first argument is the animal, and the second argument the result from the set of classes `{mammal,fish,reptile,bird}`.

The knowledge base we are going to use is this:

Listing 5.2: Knowledge base (Animal classification)

```prolog
has_covering(dog,hair).
has_covering(dolphin,none).
has_covering(platypus,hair).
has_covering(bat,hair).
has_covering(trout,scales).
has_covering(herring,scales).
has_covering(shark,none).
has_covering(eel,none).
has_covering(lizard,scales).
has_covering(crocodile,scales).
has_covering(t_rex,scales).
has_covering(snake,scales).
has_covering(turtle,scales).
has_covering(eagle,feathers).
has_covering(ostrich,feathers).
has_covering(penguin,feathers).

has_legs(dog,4).
has_legs(dolphin,0).
has_legs(platypus,2).
has_legs(bat,2).
has_legs(trout,0).
has_legs(herring,0).
has_legs(shark,0).
has_legs(eel,0).
has_legs(lizard,4).
has_legs(crocodile,4).
```

```prolog
has_legs(t_rex,4).
has_legs(snake,0).
has_legs(turtle,4).
has_legs(eagle,2).
has_legs(ostrich,2).
has_legs(penguin,2).

has_milk(dog).
has_milk(dolphin).
has_milk(bat).
has_milk(platypus).

homeothermic(dog).
homeothermic(dolphin).
homeothermic(platypus).
homeothermic(bat).
homeothermic(eagle).
homeothermic(ostrich).
homeothermic(penguin).

habitat(dog,land).
habitat(dolphin,water).
habitat(platypus,water).
habitat(bat,air).
habitat(bat,caves).
habitat(trout,water).
habitat(herring,water).
habitat(shark,water).
habitat(eel,water).
habitat(lizard,land).
habitat(crocodile,water).
habitat(crocodile,land).
habitat(t_rex,land).
habitat(snake,land).
habitat(turtle,water).
habitat(eagle,air).
habitat(eagle,land).
habitat(ostrich,land).
habitat(penguin,water).

has_eggs(platypus).
has_eggs(trout).
has_eggs(herring).
has_eggs(shark).
has_eggs(eel).
has_eggs(lizard).
has_eggs(crocodile).
has_eggs(t_rex).
has_eggs(snake).
has_eggs(turtle).
has_eggs(eagle).
has_eggs(ostrich).
```

```
has_eggs(penguin).

has_gills(trout).
has_gills(herring).
has_gills(shark).
has_gills(eel).
```

This makes a total number of eighty rules in the knowledge base (Listing 5.2). From here, and an ILP system like Aleph or Progol generates the learned hypothesis that fits better with the data:

Listing 5.3: Base model (Animal classification)

```
class(snake,reptile).
class(A,mammal) :- has_milk(A).
class(A,fish) :- has_gills(A).
class(A,bird) :- has_covering(A,feathers).
class(A,reptile) :- has_covering(A,scales), has_legs(A,4).
```

With this theory, all the positive examples are covered, and none of the negative.

Let us obtain more theories. Using TopLog (another ILP solver), it is possible to obtain more models using a boosted classifier (and obviously, losing precision from the evidence). We have selected three only, and added a fourth:

1. The first one, which classifies two negative examples as positive (its accuracy is 96.55%):

Listing 5.4: Model 1 (Animal classification)

```
class(A,mammal) :-
   has_milk(A).
class(A,fish) :-
   has_gills(A).
class(A,reptile) :-
   has_covering(A,scales).
class(A,bird) :-
   has_covering(A,feathers).
```

Since we have not considered the negative examples in our analysis, this will not affect the coding.

2. Another model has an accuracy of 92.45%, and also it does not cover all the positive examples. So, we will complete the theory with the five examples which are not covered:

Listing 5.5: Model 2 (Animal classification)

```
class(A,mammal) :-
   has_milk(A).
class(A,fish) :-
```

```
    has_gills(A).
class(A,bird) :-
    has_covering(A,feathers).
class(lizard,reptile).
class(crocodile,reptile).
class(t_rex,reptile).
class(snake,reptile).
class(turtle,reptile).
```

3. There is another possible model with less accuracy (71.43%). This needs to add a positive example not covered by the model. That is:

Listing 5.6: Model 3 (Animal classification)

```
class(A,mammal) :-
    has_milk(A).
class(A,fish) :-
    has_gills(A).
class(A,bird) :-
    has_covering(A,feathers).
class(A,reptile) :-
    has_covering(A,scales),
    habitat(A,land).
class(turtle,reptile).
```

4. We have added a fourth model with an accuracy of 96.55%, which does not cover any negative example.

Listing 5.7: Model 4 (Animal classification)

```
% class(snake,reptile).
class(A,mammal) :- has_milk(A).
class(A,fish) :- has_gills(A).
class(A,bird) :- has_covering(A,feathers),habitat(A,land)
    .
class(A,reptile) :- has_covering(A,scales), has_legs(A,4)
    .
```

Let us now compare the base model and the three other models we have generated [1]:

In accordance to the results, in terms of total cost, the best one seems to be Model 1. However, we see that models 1 to 3 are inconsistent with some negative examples (so the infinite value with examples in the table, because not all covered evidence is part of the program), so the only possible direct comparative using the MML principle is between models 1 and 4. Like model 4 takes the advantage of having a rule less (despite having extra cost in the body of the last rule but one), it has the best cost (despite the loss of accuracy).

---

[1]In the table, the $\infty$ of cost for examples is not returned by the tool. It only raises a warning indicating that there is evidence not covered by the program and returns the value without considering the negative examples. The returned values are 83.54547 for Model 1, 86.79624 for Model 2, and 101.25119 for Model 3.

| Model | Base Model | Model 1 | Model 2 | Model 3 | Model 4 |
|---|---|---|---|---|---|
| Total cost | 1402.42929 | 1358.30051 | 1424.51142 | 1403.45611 | 1252.03263 |
| Program | 128.95529 | 101.50541 | 164.46556 | 128.95529 | 120.49696 |
| Rules | 5.33789 | 4.51929 | 6.76870 | 5.33789 | 4.51929 |
| Lexicon | 21.18376 | 21.18376 | 21.18376 | 21.18376 | 20.94283 |
| Heads | 54.72518 | 39.81679 | 104.40455 | 54.72518 | 39.22942 |
| Bodies | 47.70847 | 35.98557 | 32.10854 | 47.70847 | 55.80542 |
| Vars | 0.00000 | 0.00000 | 0.00000 | 0.00000 | 0.00000 |
| Examples | 144.47527 | 127.79637 | 131.04713 | 145.50209 | 45.85545 |
| Knowledge Base | 1128.99873 | 1128.99873 | 1128.99873 | 1128.99873 | 1085.68021 |
| No. of $E^+$ covered | 16/16 | 16/16 | 16/16 | 16/16 | 14/16 |
| No. of $E^-$ covered | 0/42 | 2/42 | 0/42 | 0/42 | 0/42 |
| Accuracy | 58/58 | 56/58 | 58/58 | 58/58 | 56/58 |
| | 100.00% | 96.55% | 100.00% | 100.00% | 96.56% |

Table 5.1: Comparison of cost between models covering all the positive evidence

We can also relax the models 2 and 3 removing the rules to cover explicitly positive evidence, to obtain these two new variants:

Listing 5.8: Model 2 modified (Animal classification)

```
class(A,mammal) :-
    has_milk(A).
class(A,fish) :-
    has_gills(A).
class(A,bird) :-
    has_covering(A,feathers).
```

Listing 5.9: Model 3 modified (Animal classification)

```
class(A,mammal) :-
    has_milk(A).
class(A,fish) :-
    has_gills(A).
class(A,bird) :-
    has_covering(A,feathers).
class(A,reptile) :-
    has_covering(A,scales),
    habitat(A,land).
```

So, if we recalculate after the removal of the positive examples, the results are these [2]:

---

[2]Same as we had in Table 5.1: the cost of evidence are not $\infty$ in our tool but 83.54547 for Model 1, 56.78051 for Model 2 and 92.44255 for Model 3. But in accordance with the

| Model | Base Model | Model 1 | Model 2 modified | Model 3 modified | Model 4 |
|---|---|---|---|---|---|
| Total cost | 1402.42929 | 1358.30051 | 1290.35762 | 1373.40119 | 1324.75376 |
| Program | 128.95529 | 101.50541 | 79.32713 | 111.70902 | 120.49696 |
| Rules | 5.33789 | 4.51929 | 3.76870 | 4.51929 | 4.51929 |
| Lexicon | 21.18376 | 21.18376 | 21.18376 | 21.18376 | 20.94283 |
| Heads | 54.72518 | 39.81679 | 29.86259 | 39.81679 | 39.22942 |
| Bodies | 47.70847 | 35.98557 | 24.51208 | 46.18917 | 55.80542 |
| Vars | 0.00000 | 0.00000 | 0.00000 | 0.00000 | 0.00000 |
| Examples | 144.47527 | 127.79637 | 82.03176 | 132.69345 | 118.57658 |
| Knowledge Base | 1128.99873 | 1128.99873 | 1128.99873 | 1128.99873 | 1085.68021 |
| No. of $E^+$ covered | 16/16 | 16/16 | 11/16 | 15/16 | 14/16 |
| No. of $E^-$ covered | 0/42 | 2/42 | 0/42 | 0/42 | 0/42 |
| Accuracy | 58/58 | 56/58 | 53/58 | 57/58 | 56/58 |
|  | 100.00% | 96.55% | 91.38% | 98.28% | 96.55% |

Table 5.2: Comparison of cost between models covering almost the evidence

With this data, the best model now seems to be the modified Model 2, with less cost and only a few less accuracy. And this decision has been possible easily due to this MML tool we are explaining here.

## 5.2   Language analysis: speaker recognition

Another area in which the MML tool could help to decide between models is language analysis. In this area, a model that fits data perfectly is not possible, so in the majority of situations an approximate model is used. Approximate models may have different accuracies and complexities.

Generating a language model from scratch is difficult and time-consuming. However, if we already have a general model, we can specialize it to several contexts.

In general, if we can determine the frequencies of each rule from evidence, we would be able to transform a non-stochastic model (with only rules) into a stochastic one (with probabilities attached to the rules). This will help us (applying the MML tool) to identify which pattern (another set of examples) corresponds to which theory (language or theory).

The example we are going to show is about speeches. If we can generate a model from different speeches, and we have different frequencies of each rule in the model for several speakers, we could apply the tool to determine to which speaker another speech corresponds. The new speech will fit better in terms of cost with the previous calculated model of that speaker. The reason is that each person tends to use some particular type of sentences and vocabulary, so even if the language is the same, analyzing the number of occurrences of each type of sentence or specific words in a text could guide

---

MML principle, the negative examples covered must cost infinite.

us to recognize the speaker. Our purpose is to analyze syntactically the sentences, obtaining the whole syntax of each individual sentence.

Even though it was not the purpose of this thesis to develop a whole parser able to recognize the English language or a different one, a basic analyzer has been developed using Prolog, whose code is available in Appendix E.1. We explain the basics of the parser below.

### 5.2.1 English language parser

Our parser is based on a Definite Clause Grammar (DCG). The DCG notation was developed as the result of research in natural language parsing in 1980 by Fernando Pereira and David Warren [17], and it is widely used in Prolog for language parsing. We have also used a DCG to analyze the Prolog source code. DCGs are closely related to the concept of attribute grammars from which Prolog was originally developed, and usually they are identified only with Prolog. They represent a grammar as a set of definite clauses in first-order logic from which we derive the language:

```
sentence --> noun_phrase , verb_phrase .
noun_phrase --> det , noun .
verb_phrase --> verb , noun_phrase .
det --> [the].
noun --> [cat].
verb --> [eats].
```

In the past the use of Prolog in *natural language processing* (NLP) was common [43], but nowadays the research in this area is focused on other languages and tools, like *NLTK* [2]. Despite this, we have decided to use a *top-down* parser in Prolog because we prefer to obtain the models to use in our MML tool directly in the Prolog language.

The parser is specially constructed to avoid being left-recursive, because Prolog will go into an infinite loop on left-recursive grammars. The state of the art with NLP demonstrates that language analysis is still a big problem, so our capacity of recognition is less than the third of the total amount of sentences with this parser. But it could be considered rather well for this example and our purpose.

The process we have followed to obtain models from speeches using a DCG can be depicted in Figure 5.1 and goes as follows:

- First, we introduce a speech in the English language into `language.pl`, a tool which converts the input stream into a list of sentences suitable for Prolog syntax. Each sentence is also a list of words, transformed into Prolog terms.

- Then, the same `language.pl` program loads the DCG to analyze the

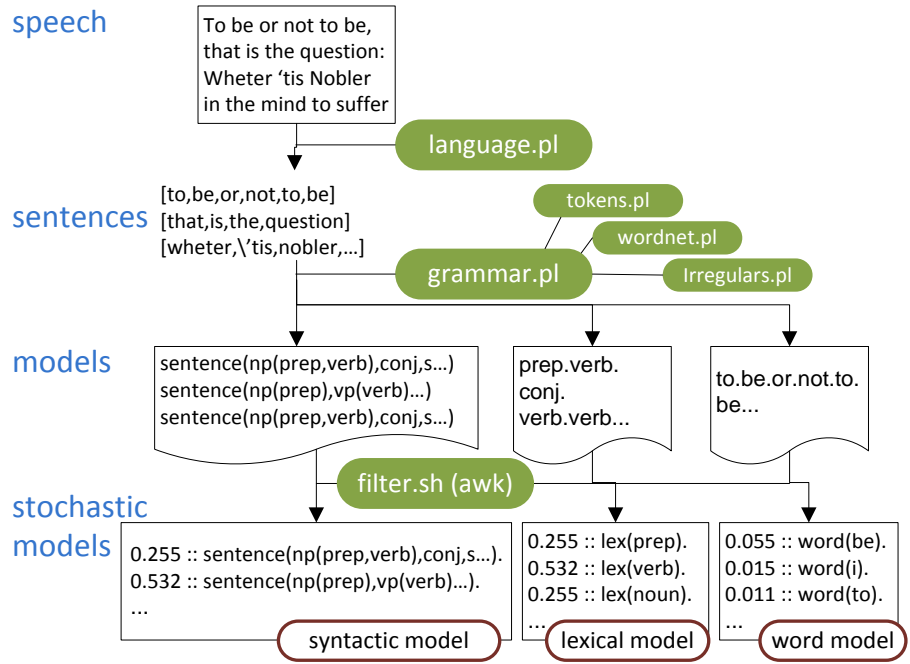

Figure 5.1: Developed parser for language processing

sentences from `grammar.pl` and tries to obtain the syntactic equivalence for this sentence:

```
? s([my,tailor,is,rich],X)

X= attributive_sentence( np(det(my),noun(
    tailor)),
        vp(verb(is),adj(rich)) )
```

The syntactic model of the speech is built using the syntactic equivalence of the recognized sentences.

- Our DCG also uses the ANSI Prolog version of the *WordNet* database [3] in `wordnet.pl`, in order to recognize the nouns, verbs, adjectives and adverbs. The rest of lexical items of the English language are loaded from `tokens.pl` (prepositions, determiners, etc) and from `irregulars.pl` (irregular verbs, which are not present in *WordNet*).

- The sentences from which we are not able to obtain its syntactic rela-

---

[3] WordNet is a lexical database for the English language [34]. It groups English words into sets of synonyms called *synsets*, provides short, general definitions, and records the various semantic relations between these synonym sets. Our purpose with the *WordNet* database is only to have a dictionary of the English language.

tionships are ignored (around two thirds of the total, according to the complexity of the speech).

- Since our intention is to test the MML tool, we are also going to consider two extra different models of the speeches: first, a model of the vocabulary used (named here "word model"). And second, a model of the lexicon of the speech (so a model of how many times a noun, a verb or another lexical item appears in the speech). These models are generated together with the syntactic one with `language.pl` using a different command line modifier.

- The results of `language.pl` are a collection of rules of the three different models, many of them repeated (two sentences with the same syntactic form will generate the same rule, and the same in the lexical model). We need then to apply some filters using tools like `awk`, `uniq` and `sort` to transform the results into a notation for the number of occurrences of each item (and also the item):

```
$ ./language.pl | sort | uniq -c \
  | grep -v "^$" | awk '{ print $1 " :: " $2 }' \
  | sort -g -r

85 :: sentence(np(det,noun)),conj,noun),vp(verb,noun))
53 :: sentence(np(det,noun)),vp(verb))
...
```

  This is not yet a normalized stochastic logic model (the left numbers are not a percentage), but our MML tool is able to normalize this without our intervention, so these values are suitable as input of the MML tool.

The DCG we have developed is able to manage affirmative, negative, interrogative and imperative sentences, with subordinated sentences also. Obviously it is not intended to cover a real speech, especially because the order in a sentence could be altered (increasing the difficulty of recognizing the parts of the sentence) and also some parts could be missed (because they are assumed). Some techniques like gap threading have been applied to solve this, but it makes the DCG more complex.

For example, these two real sentences from George Orwell's novel 1984 are really complex (they are represented in the Prolog syntax we are using):

```
[ [thirty,to,forty,group,!, yapped,a,piercing, female, voice],
  [was,only,a,bundle,of,blankets,that,she,was, carrying] ]
```

The first sentence has an incorrect order, so our DCG is not going to recognize it, and the second one has and assumed subject. To have them recognized, they must be rewritten like this:

```
[ [a,piercing,female,voice,yapped,thirty,to,forty,group,!],
  [it,was,only,a,bundle,of,blankets,that,she,was,carrying]
]
```

Then, the incorrect order of sentences is not going to be recognized by our parser. The use of gap techniques could solve that, since our intention is not really to develop a precise parser for the English language.

Also, it is problematic with our DCG to recognize personal nouns. In order to avoid this, we will transform the speech to lowercase, and consider the personal nouns as simple nouns. Also, shorts forms like "I´ve" are going to be expanded.

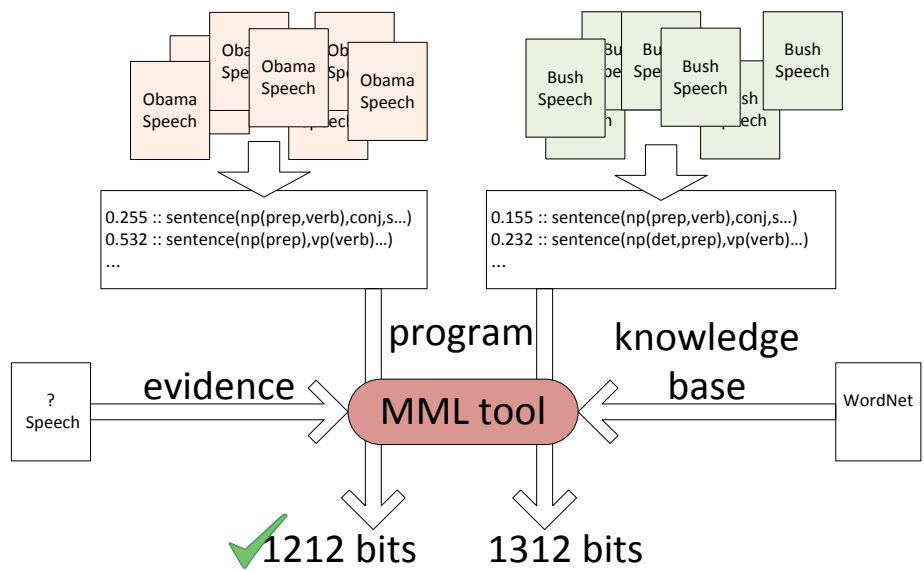

Figure 5.2: Experiment with speeches

## 5.2.2    Experimenting with speeches

With this parser and the MML tool, we are going to analyze some speeches of two American presidents: Barack Obama and George W. Bush (son). The decision of selecting them is because we consider the language used in the speeches rather modern and simple.

We used some speeches of each president, extracted from the Internet:

```
http://www.presidentialrhetoric.com/speeches
http://obamaspeeches.com
```

We will select five of them to create the model and we will use three extra speeches as evidence. Applying the parser to the merge of the five speeches will generate a unique model for each president, one for Obama and another for Bush. A unique syntactic model, but also unique lexical and word models.

Applying the parser to each one of the three example speeches of each president we generate three syntactic models, three lexical and three for the words. If we apply the MML tool to the model of each president together with one of this "evidence" models, the most probable president that has produced this speech introduced as evidence is the one that should generate the lowest cost with the model as program. This is depicted in Figure 5.2 (in the figure, the evidence speech fits better as an Obama's speech, due to the lower cost of 1212 bits).

We could include also `wordnet.pl` and `tokens.pl` as knowledge base in our analysis with the MML tool, but since it is the same to all of them, it is not going to give us any relevant information. It will only increase the cost and also the computational time necessary to compute the results, so we are going to ignore it.

With the first five speeches of each president, we obtain these two syntax models of their language:

```
% Bush syntactic model:
12 :: imperative_sentence(verb_phrase(verb,noun_phrase(noun)))
13 :: affirmative_sentence(noun_phrase(noun),verb_phrase(verb))
 3 :: imperative_sentence(verb_phrase(verb,noun_phrase(noun),
        prop_pred(prep,noun_phrase(det,noun))))
 3 :: imperative_sentence(verb_phrase(verb,noun_phrase(noun),
        prop_pred(prep,noun_phrase(noun))))
 4 :: affirmative_sentence(connector,
        affirmative_sentence(noun_phrase(noun),
        verb_phrase(verb(aux,verb,verb),noun_phrase(noun))))
 4 :: affirmative_sentence(noun_phrase(noun),
        verb_phrase(verb(aux,aux,verb)))
 4 :: imperative_sentence(verb_phrase(verb))
 7 :: affirmative_sentence(noun_phrase(noun),
        verb_phrase(verb,noun_phrase(det,noun)))
...

% Obama:
 2 :: affirmative_sentence(noun_phrase(noun),verb_phrase(
        verb,noun_phrase(det,noun,prop_pred(prep,
        noun_phrase(noun)))))
 2 :: affirmative_sentence(noun_phrase(noun),
        verb_phrase(verb,noun_phrase(noun)))
 2 :: affirmative_sentence(noun_phrase(noun),
        verb_phrase(verb(verb,prep),
        noun_phrase(det,noun)))
 4 :: affirmative_sentence(noun_phrase(noun),
        verb_phrase(verb,noun_phrase(det,noun)))
```

```
 4 :: imperative_sentence(verb_phrase(verb))
 8 :: affirmative_sentence(noun_phrase(noun),
       verb_phrase(verb))
...
```

The same with the models of the lexicon:

```
% Bush lexical model:
 460 :: lex(adj).
 858 :: lex(adv).
 681 :: lex(aux).
 689 :: lex(connector).
2104 :: lex(det).
3423 :: lex(noun).
1906 :: lex(prep).
  36 :: lex(qmark).
4017 :: lex(verb).
189 :: lex(wh).

% Obama lexical model:
 304 :: lex(adj).
 862 :: lex(adv).
 574 :: lex(aux).
 535 :: lex(connector).
1914 :: lex(det).
2657 :: lex(noun).
1627 :: lex(prep).
  24 :: lex(qmark).
3238 :: lex(verb).
 219 :: lex(wh).
```

And finally, a model of frequency of occurrence of each word in the speeches (we discard the words that appear less than ten times):

```
% Bush word model:
 33 :: word(would).
 20 :: word(years).
113 :: word(you).
...

% Obama word model:
24 :: word(would).
17 :: word(years).
69 :: word(you).
```

```
10 :: word(young).
17 :: word(your).
...
```

With all these models, it is now possible to analyze the other speeches we have (three of each president) to ask the MML tool to calculate the lengths of the coding, and discover the attribution to a president. In the results, we provide the cost in bits obtained from the MML tool only for the evidence, because the cost of transferring the model is not relevant for the comparison we want to make here. Remember that in the attribution to a president, the one with the most reduced value is the one that fits best:

|  | Examples | Bush | Obama |
|---|---|---|---|
| Speech Obama #1 | 8 | 67.89835 | **56.89299** |
| Speech Obama #2 | 3 | 30.85755 | **24.31828** |
| Speech Obama #3 | 2 | 22.47885 | **17.89894** |
| Speech Bush #1 | 9 | **54.70721** | 55.10070 |
| Speech Bush #2 | 3 | **24.37596** | 25.64021 |
| Speech Bush #3 | 2 | **14.23174** | 16.20097 |

Table 5.3: Cost in bits of the syntax models

With respect to the syntax models shown in the Table 5.3, the few number of recognized sentences in the speeches (around a quarter of the total) makes that in our comparison, the values are rather similar (in the first model of Bush, the difference is less than a bit). But in all the situations, the MML tool has been able to identify the correct author of the speech.

|  | Examples | Bush | Obama |
|---|---|---|---|
| Speech Obama #1 | 440 | 59.07253 | **57.84381** |
| Speech Obama #2 | 501 | 56.51041 | **52.20659** |
| Speech Obama #3 | 822 | 57.42554 | **57.27029** |
| Speech Bush #1 | 1848 | **69.87457** | 83.95714 |
| Speech Bush #2 | 2031 | **87.94910** | 103.84496 |
| Speech Bush #3 | 1031 | **80.40577** | 91.60670 |

Table 5.4: Cost in bits of the lexicon models

Our second perspective, focused on identifying the authority of the speech attending to the lexicon that the speech has, is shown in Table 5.3. Here the amount of examples is huge because we have used all words, not only the ones from the recognized sentences. The alignment to the right speaker is correct.

And finally, in the analysis of the word models in Table 5.5, we have determined three mismatches that shown that the model based in the frequency of words is not a good classifier of the speeches. This may be because there might be speeches which are thematic or address some specific issues, which makes them align better with speeches from other speakers. In any case, from the number in the three tables, we can say which speeches are more distinctively Obama-like or Bush-like.

|                  | Examples | Bush      | Obama     |
|------------------|----------|-----------|-----------|
| Speech Obama #1  | 87       | 207.21608 | **191.03175** |
| Speech Obama #2  | 138      | **256.59675** | 264.45314 |
| Speech Obama #3  | 243      | 375.55681 | **322.40221** |
| Speech Bush #1   | 842      | 918.87515 | **659.99080** |
| Speech Bush #2   | 827      | 962.05868 | **748.79984** |
| Speech Bush #3   | 317      | 532.42015 | **497.97185** |

Table 5.5: Cost in bits of the word models

Overall, this example shows how a representation of natural language text patterns in the form of stochastic logic programs can be used for model selection.

## 5.3   A probabilistic graph

We are going to use an example extracted from ProbLog [9] to illustrate how the tool works, instead of using it for comparing theories (section 5.1) or evidences (section 5.2). Also this example is going to show the limitations of the developed tool. The example is available for testing ProbLog on the website of the University of Lovaine (http://dtai.cs.kuleuven.be/problog):

Listing 5.10: A probabilistic program

```prolog
%% :- use_module(library(problog)).

% probabilistic facts
0.9::dir_edge(1,2).
0.8::dir_edge(2,3).
0.6::dir_edge(3,4).
0.7::dir_edge(1,6).
0.5::dir_edge(2,6).
0.4::dir_edge(6,5).
0.7::dir_edge(5,3).
0.2::dir_edge(5,4).

% Now comes the background knowledge
% definition of acyclic path using list of visited nodes
path(X,Y) :- path(X,Y,[X],_).

path(X,X,A,A).
path(X,Y,A,R) :-
    X\==Y,
    edge(X,Z),
    absent(Z,A),
    path(Z,Y,[Z|A],R).
```

```
% using directed edges in both directions
edge(X,Y) :- dir_edge(Y,X).
edge(X,Y) :- dir_edge(X,Y).

% checking whether node hasn't been visited before
absent(_,[]).
absent(X,[Y|Z]):-X \= Y, absent(X,Z).
```

In Figure 5.3 it is drawn the network representing the reachability of the nodes of this simple program. So simple that is going to cause a failure in the tool when we look for successful derivations of predicate `edge/2`.

The reason lies in the predicate `absent/2`, whose second rule contains an inequality that blocks the Vanilla meta-interpreter to obtain the successful subgoals. This happens only in our solver (used to estimate the probability of predicate `absent/2`), because we have variables without assignations, and it is not the same to compare two values when Prolog solves the predicate (for instance: $1\bar{1}$), than to compare the meta-interpreter a value and a variable ($X\bar{1}$).

Although the MML tool cans obtain the cost of the rules, if we analyze it together with the evidence, no one of the examples in the evidence will success for our Vanilla meta-interpreter. So to continue with this analysis, we are going to define the predicate `absent/2` in an extensional way (the purpose of the predicate is to check whether node has not been visited before, and then, it is not present in the list of the second argument):

```
absent(1,[2]).
absent(1,[5]).
absent(1,[6,5]).
absent(2,[1]).
absent(2,[1,6,5]).
absent(2,[5]).
absent(2,[6,5]).
absent(3,[1,]).
absent(3,[1,2]).
absent(3,[1,6,5]).
absent(3,[2]).
absent(3,[2,1]).
absent(3,[2,1,6,5]).
absent(3,[5]).
absent(3,[5,6,1,2]).
absent(3,[6,1,2]).
absent(3,[6,5]).
absent(5,[1]).
absent(5,[1,2]).
absent(5,[2]).
absent(5,[2,1]).
absent(5,[3,2,1]).
absent(5,[6,1,2]).
absent(6,[1,2]).
absent(6,[2]).
```

```
absent(6,[5]).
```

This extensional list contains the necessary facts of `absent/2` to work with the probabilistic graph we have. Now we will generate some evidence to test the results:

```
3 # path(1,3).  % this is: three times this
    example
path(1,5).
2 # path(2,3).  % twice this one
path(5,3).
path(1,2).
```

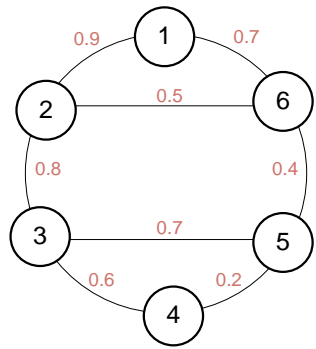

Figure 5.3: Network representing the reachability of the nodes of the probabilistic example

If we use the program without the debugging output, only the cost is returned. But with the debugging option, we will obtain some extra lines (preceded in the original text by double dash, here removed) that we are going to analyze now, dividing them in four blocks:

Listing 5.11: Output of the probabilistic program

```
1  -- Program tesis/examples/prob.pl read as module prob2
2  -- Module with values of probability in some clauses
3  %  Discontiguous definition of examples_prob: (#)/2.
4  %  Discontiguous definition of examples_prob:path/2.
5  -- Example file tesis/examples/examples.prob.pl read as module
       examples_prob
6  -- Normalizing probability of dir_edge/2 dividing by 4.80000
7  -- There are 1 clauses of path/2 that will have a prob of 1.00000
8  -- There are 2 clauses of path/4 that will have a prob of 0.50000
9  -- There are 2 clauses of edge/2 that will have a prob of 0.50000
10 -- There are 2 clauses of absent/2 that will have a prob of 0.50000
11 -- #1 rule cost: Head:7.16993 Body:1.51929 Vars:0.00000 RuleProb
       :25.07072
12 -- 0.18750 :: dir_edge(1,2).
13 -- #2 rule cost: Head:7.16993 Body:1.51929 Vars:0.00000 RuleProb
       :6.92873
14 -- 0.16667 :: dir_edge(2,3).
15 -- #3 rule cost: Head:7.16993 Body:1.51929 Vars:0.00000 RuleProb
       :8.76870
```

```
16 | -- 0.12500 :: dir_edge(3,4).
17 | -- #4 rule cost: Head:7.16993 Body:1.51929 Vars:0.00000 RuleProb
   |    :23.53157
18 | -- 0.14583 :: dir_edge(1,6).
19 | -- #5 rule cost: Head:7.16993 Body:1.51929 Vars:0.00000 RuleProb
   |    :15.28798
20 | -- 0.10417 :: dir_edge(2,6).
21 | -- #6 rule cost: Head:7.16993 Body:1.51929 Vars:0.00000 RuleProb
   |    :9.82745
22 | -- 0.08333 :: dir_edge(6,5).
23 | -- #7 rule cost: Head:7.16993 Body:1.51929 Vars:0.00000 RuleProb
   |    :23.53157
24 | -- 0.14583 :: dir_edge(5,3).
25 | -- #8 rule cost: Head:7.16993 Body:1.51929 Vars:0.00000
26 | -- 0.04167 :: dir_edge(5,4).
27 | -- #9 rule cost: Head:2.00000 Body:13.74811 Vars:9.07682
28 | -- 1.00000 :: path(A,B):-path(A,B,[A],C).
29 | -- #10 rule cost: Head:4.00000 Body:14.07004 Vars:5.95420 RuleProb
   |    :2.51929
30 | -- 0.50000 :: path(A,A,B,B).
31 | -- #11 rule cost: Head:4.00000 Body:27.94753 Vars:34.55796
32 | -- 0.50000 :: path(A,B,C,D):-A \== B,edge(A,E),absent(E,C),path(E,B
   |    ,[E|C],D).
33 | -- #12 rule cost: Head:2.00000 Body:6.84122 Vars:3.80735 RuleProb
   |    :2.51929
34 | -- 0.50000 :: edge(A,B):-dir_edge(B,A).
35 | -- #13 rule cost: Head:2.00000 Body:6.84122 Vars:3.80735
36 | -- 0.50000 :: edge(A,B):-dir_edge(A,B).
37 | -- #14 rule cost: Head:4.58496 Body:10.74811 Vars:0.00000 RuleProb
   |    :2.51929
38 | -- 0.50000 :: absent(A, [] ).
39 | -- #15 rule cost: Head:6.58496 Body:12.41256 Vars:10.81858
40 | -- 0.50000 :: absent(A,[B|C]):-A \= B,absent(A,C).
41 | -- 5 predicates (np): [absent/2,edge/2,path/4,path/2,dir_edge/2]
42 | -- 6 function symbols: [1/0,2/0,3/0,4/0,5/0,6/0]
```

This part gives information about the normalization process of the program, and also returns information about the individual cost of each of the fifteen rules of the program. For example, if we took one:

```
-- #8 rule cost: Header:2.00000 Body
   :1.51854 Vars:0.00000
-- 0.04167 :: dir_edge(5,4).
```

This rule had an original probability of 0.2, but the program has determined that after summing all the values for the predicate `dir_edge/2`, it is normalized to 0.04167.

There are also two lines (3 and 4) preceded by the percentage sign, that are warnings issued by the Prolog interpreter. They do not affect the execution of the tool, and can help us to understand whether the user program introduced is well formed or has mistakes.

### Second block

```
43 | Estimate the probabilities of the predicates
44 |   dir_edge/2 --> prob:1.00000 paths:8
45 |     dir_edge(5,4). (prob:0.04167)
46 |     dir_edge(5,3). (prob:0.14583)
```

```
47      dir_edge(6,5). (prob:0.08333)
48      dir_edge(2,6). (prob:0.10417)
49      dir_edge(1,6). (prob:0.14583)
50      dir_edge(3,4). (prob:0.12500)
51      dir_edge(2,3). (prob:0.16667)
52      dir_edge(1,2). (prob:0.18750)
53   path/2 --> prob:0.56250 paths:1
54     path(_182234,_182235):-path(_182234,_182235,[_182234],_182242). (
            prob:0.56250)
55        path(_182284,_182285,[_182284],_182287):-_182284\==_182285,
              edge(_182284,_182303),absent(_182303,[_182284]),path(
              _182303,_182285,[_182303,_182284],_182287). (prob
              :0.06250)
56           absent(_182361,[_182363]):-_182361\=_182363,absent(
                _182361,[]). (prob:0.50000)
57           edge(_182417,_182418):-dir_edge(_182417,_182418). (prob
                :0.50000)
58              dir_edge(5,4). (prob:0.04167)
59              dir_edge(5,3). (prob:0.14583)
60              dir_edge(6,5). (prob:0.08333)
61              dir_edge(2,6). (prob:0.10417)
62              dir_edge(1,6). (prob:0.14583)
63              dir_edge(3,4). (prob:0.12500)
64              dir_edge(2,3). (prob:0.16667)
65              dir_edge(1,2). (prob:0.18750)
66           edge(_182854,_182855):-dir_edge(_182855,_182854). (prob
                :0.50000)
67              dir_edge(5,4). (prob:0.04167)
68              dir_edge(5,3). (prob:0.14583)
69              dir_edge(6,5). (prob:0.08333)
70              dir_edge(2,6). (prob:0.10417)
71              dir_edge(1,6). (prob:0.14583)
72              dir_edge(3,4). (prob:0.12500)
73              dir_edge(2,3). (prob:0.16667)
74              dir_edge(1,2). (prob:0.18750)
75        path(_183372,_183372,[_183372],[_183372]). (prob:0.50000
            freeVars:2)
76   path/4 --> prob:0.62500 paths:2
77     path(_183504,_183505,_183506,_183507):-_183504\==_183505,edge(
            _183504,_183521),absent(_183521,_183506),path(_183521,_183505
            ,[_183521|_183506],_183507). (prob:0.12500)
78        absent(_183597,[_183599|_183600]):-_183597\=_183599,absent(
              _183597,_183600). (prob:0.50000)
79        absent(_183675,[]). (prob:0.50000 freeVars:1)
80        edge(_183743,_183744):-dir_edge(_183743,_183744). (prob
                :0.50000)
81              dir_edge(5,4). (prob:0.04167)
82              dir_edge(5,3). (prob:0.14583)
83              dir_edge(6,5). (prob:0.08333)
84              dir_edge(2,6). (prob:0.10417)
85              dir_edge(1,6). (prob:0.14583)
86              dir_edge(3,4). (prob:0.12500)
87              dir_edge(2,3). (prob:0.16667)
88              dir_edge(1,2). (prob:0.18750)
89           edge(_184387,_184388):-dir_edge(_184388,_184387). (prob
                :0.50000)
90              dir_edge(5,4). (prob:0.04167)
91              dir_edge(5,3). (prob:0.14583)
92              dir_edge(6,5). (prob:0.08333)
93              dir_edge(2,6). (prob:0.10417)
94              dir_edge(1,6). (prob:0.14583)
95              dir_edge(3,4). (prob:0.12500)
```

```
 96                dir_edge(2,3).  (prob:0.16667)
 97                dir_edge(1,2).  (prob:0.18750)
 98       path(_185112,_185112,_185114,_185114).  (prob:0.50000 freeVars:2)
 99    edge/2 --> prob:1.00000 paths:2
100       edge(_185286,_185287):-dir_edge(_185286,_185287).  (prob:0.50000)
101           dir_edge(5,4).  (prob:0.04167)
102           dir_edge(5,3).  (prob:0.14583)
103           dir_edge(6,5).  (prob:0.08333)
104           dir_edge(2,6).  (prob:0.10417)
105           dir_edge(1,6).  (prob:0.14583)
106           dir_edge(3,4).  (prob:0.12500)
107           dir_edge(2,3).  (prob:0.16667)
108           dir_edge(1,2).  (prob:0.18750)
109       edge(_186110,_186111):-dir_edge(_186111,_186110).  (prob:0.50000)
110           dir_edge(5,4).  (prob:0.04167)
111           dir_edge(5,3).  (prob:0.14583)
112           dir_edge(6,5).  (prob:0.08333)
113           dir_edge(2,6).  (prob:0.10417)
114           dir_edge(1,6).  (prob:0.14583)
115           dir_edge(3,4).  (prob:0.12500)
116           dir_edge(2,3).  (prob:0.16667)
117           dir_edge(1,2).  (prob:0.18750)
118    absent/2 --> prob:1.00000 paths:2
119       absent(_187119,[_187121|_187122]):-_187119\=_187121,absent(
          _187119,_187122).  (prob:0.50000)
120       absent(_187238,[]).  (prob:0.50000 freeVars:1)
```

This second block is dedicated to estimate the probabilities for all the predicates `dir_edge/2`, `path/2`, `path/4`, `edge/2` and `absent/2`. That is, to generate the SSLD-tree to consider the probabilities for all the possible values and later use that information with the evidence done.

The program does not return any graphical output, so the SSLD-tree is tabulated text where each line represents an element in the same level of the tree (if there are the same amount of spaces), or a new child (if there are more spaces). For example, the SSLD-resolution for `path/2` is represented graphically in Figure 5.4, with the probability of each predicate at that level of the tree. In the output of the second block, the analysis of `path/2` starts on line 12, with the resulting probability and number of paths that it has (number of children we are going to have).

The child is `path/4`, which has also two children. The first one that starts on line 14, and the second one on line 34. The lines between these two (14-34) are the unfolding of the first child.

### Third block

```
121 |-- Estimate the probabilities of the examples (Np=5)
122 |-- example path(1,2): times=1 probability=0.00167
123 |--     path(1,2):-path(1,2,[1],_874406).  (prob:0.00417)
124 |--         1 freeVars at right in path(1,2,[1],_874443) [*0.00417]
125 |--     path(1,2,[1],_874479):-1\==2,edge(1,_874495),absent(_874495
        ,[1]),path(_874495,2,[_874495,1],_874479).  (prob:0.00208)
126 |--         absent(5,[1]).  (prob:0.03333)
127 |--         absent(3,[1]).  (prob:0.03333)
128 |--         absent(2,[1]).  (prob:0.03333)
129 |--         edge(1,_874656):-dir_edge(1,_874656).  (prob:0.16667)
130 |--             dir_edge(1,6).  (prob:0.14583)
```

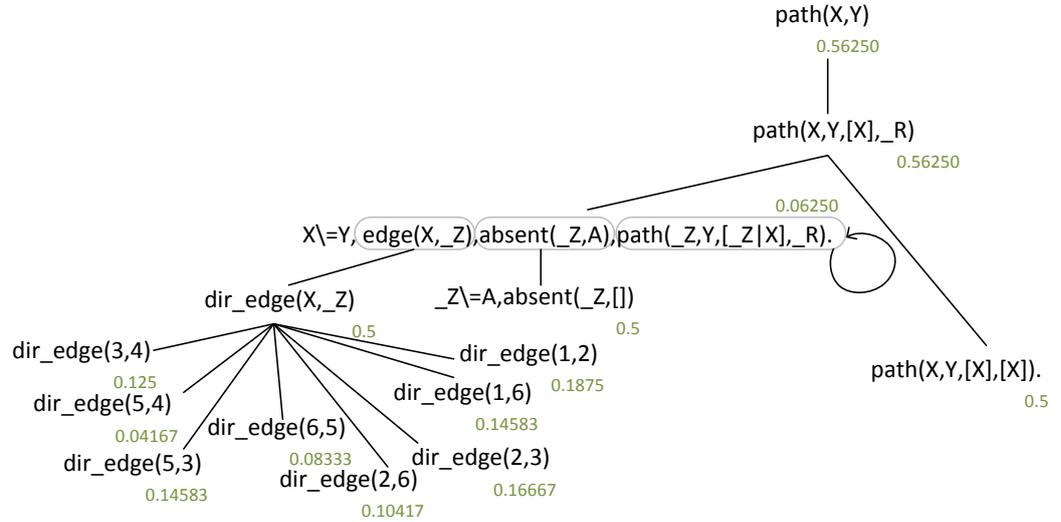

Figure 5.4: SSLD-resolution tree for predicate `path/2`

```
131 --          dir_edge(1,2).  (prob:0.18750)
132 --          edge(1,_874775):-dir_edge(_874775,1).  (prob:0.00000)
133 -- example path(5,3): times=1 probability=0.00181
134 --     path(5,3):-path(5,3,[5],_874865).  (prob:0.00451)
135 --          1 freeVars at right in path(5,3,[5],_874913)  [*0.00451]
136 --     path(5,3,[5],_874960):-5\==3,edge(5,_874976),absent(_874976
         ,[5]),path(_874976,3,[_874976,5],_874960).  (prob:0.00226)
137 --          absent(3,[5]).  (prob:0.03333)
138 --          absent(2,[5]).  (prob:0.03333)
139 --          absent(1,[5]).  (prob:0.03333)
140 --          absent(6,[5]).  (prob:0.03333)
141 --          edge(5,_875231):-dir_edge(5,_875231).  (prob:0.09375)
142 --          dir_edge(5,4).  (prob:0.04167)
143 --          dir_edge(5,3).  (prob:0.14583)
144 --          edge(5,_875386):-dir_edge(_875386,5).  (prob:0.04167)
145 --          dir_edge(6,5).  (prob:0.08333)
146 -- example path(2,3): times=2 probability=0.00306
147 --     path(2,3):-path(2,3,[2],_875557).  (prob:0.00764)
148 --          1 freeVars at right in path(2,3,[2],_875618)  [*0.00764]
149 --     path(2,3,[2],_875678):-2\==3,edge(2,_875694),absent(_875694
         ,[2]),path(_875694,3,[_875694,2],_875678).  (prob:0.00382)
150 --          absent(3,[2]).  (prob:0.03333)
151 --          absent(5,[2]).  (prob:0.03333)
152 --          absent(6,[2]).  (prob:0.03333)
153 --          absent(1,[2]).  (prob:0.03333)
154 --          edge(2,_876014):-dir_edge(2,_876014).  (prob:0.13542)
155 --          dir_edge(2,6).  (prob:0.10417)
156 --          dir_edge(2,3).  (prob:0.16667)
157 --          edge(2,_876208):-dir_edge(_876208,2).  (prob:0.09375)
158 --          dir_edge(1,2).  (prob:0.18750)
159 -- example path(1,5): times=1 probability=0.00167
160 --     path(1,5):-path(1,5,[1],_876418).  (prob:0.00417)
161 --          1 freeVars at right in path(1,5,[1],_876492)  [*0.00417]
162 --     path(1,5,[1],_876565):-1\==5,edge(1,_876581),absent(_876581
         ,[1]),path(_876581,5,[_876581,1],_876565).  (prob:0.00208)
163 --          absent(5,[1]).  (prob:0.03333)
164 --          absent(3,[1]).  (prob:0.03333)
```

```
165  --              absent(2,[1]). (prob:0.03333)
166  --              edge(1,_876890):-dir_edge(1,_876890). (prob:0.16667)
167  --                  dir_edge(1,6). (prob:0.14583)
168  --                  dir_edge(1,2). (prob:0.18750)
169  --              edge(1,_877120):-dir_edge(_877120,1). (prob:0.00000)
170  -- example path(1,3): times=3 probability=0.00167
171  --      path(1,3):-path(1,3,[1],_877284). (prob:0.00417)
172  --          1 freeVars at right in path(1,3,[1],_877369) [*0.00417]
173  --      path(1,3,[1],_877453):--1\==3,edge(1,_877469),absent(_877469
          ,[1]),path(_877469,3,[_877469,1],_877453). (prob:0.00208)
174  --              absent(5,[1]). (prob:0.03333)
175  --              absent(3,[1]). (prob:0.03333)
176  --              absent(2,[1]). (prob:0.03333)
177  --              edge(1,_877822):-dir_edge(1,_877822). (prob:0.16667)
178  --                  dir_edge(1,6). (prob:0.14583)
179  --                  dir_edge(1,2). (prob:0.18750)
180  --              edge(1,_878085):-dir_edge(_878085,1). (prob:0.00000)
181  -- Cost of this 8 examples is 77.73486
```

The third block is dedicated to obtain the probabilities of the given evidence, again by SSLD-resolution using the Vanilla solver. For instance, the evidence `path(5,3)` is solved using two different clauses, the first one using the clause `path(X,X,A,A)` with probability 0.00451, and the second one with the other clause of `path/2` with probability 0.00226. Since there are also free variables (not instantiated during the SSLD-resolution), the multiplier applied (which depends on the size of the signature and the probability of the predicate `path/2`) is applied:

```
example path(5,3): times=1 probability=0.00181
path(5,3):-path(5,3,[5],_874865). (prob:0.00451)
    1 freeVars at right in path(5,3,[5],_874913)
    [*0.00451]
path(5,3,[5],_874960):--5\==3,edge(5,_874976),
absent(_874976,[5]),path(_874976,3,[_874976,5],
_874960).
(prob:0.00226)
    absent(3,[5]). (prob:0.03333)
    absent(2,[5]). (prob:0.03333)
    absent(1,[5]). (prob:0.03333)
    absent(6,[5]). (prob:0.03333)
    edge(5,_875231):-dir_edge(5,_875231). (prob
        :0.09375)
        dir_edge(5,4). (prob:0.04167)
        dir_edge(5,3). (prob:0.14583)
    edge(5,_875386):-dir_edge(_875386,5). (prob
        :0.04167)
        dir_edge(6,5). (prob:0.08333)
```

### Fourth block

```
182  tesis/examples/prob2.pl + examples:tesis/examples/examples.prob.pl
183    Total cost:   1019.48514 bits
184    Cost of program: 441.53204 bits
185      Cost of 15 rules: 8.36748 bits
186      Cost of lexicon: 25.66265 bits
```

```
187        of  5 predicates [absent/2,edge/2,path/4,path/2,dir_edge/2]
188        and 6 function symbols [1/0,2/0,3/0,4/0,5/0,6/0]
189        Cost of heads: 117.35825 bits
190        Cost of bodies: 104.76314 bits
191        Cost of vars: 68.02227 bits
192     Probabilistic cost: 500.21824 bits
193    Cost of 8 examples: 77.73486 bits
```

And finally the last block, where we obtain the cost in bits. If there were also function symbols, they would appear also.

This last block is the only part of the output that was not originally preceded by double dash, because is the normal output returned when the debugging is not used.

# Chapter 6

# Conclusions and future work

## 6.1 Conclusions

The general objective has been achieved: We have developed a MML coding scheme for non-stochastic and stochastic logic programs and a tool implementing it.

With regard to the specific objectives:

- After reviewing the state of the art about the MML principle, we have introduced some changes in the way that the cost is calculated in [13]: the formula used to code the signature is different, and now it is also more clear when the probability of a rule needs to be coded and when not. The problem with the cost of the signature was related to the correction term for the relevance of the order (otherwise it could generate negative results for large signatures).

- The objective of developing a tool has these five achieved subgoals:

  - An interpreter of a logic language (Yap Prolog) has been selected to do the implementation.

  - After reviewing different ways to implement the probabilities, we have selected the most clear to the end user, the one used in ProbLog.

  - The extension of Yap Prolog to manage the probabilities has also been done.

  - The tool is able to manage probabilities.

  - The tool is also able to insert the number of repetitions in the examples, to avoid these repetitions in the input file.

- The tool has been tested using different scenarios, some of them explained in this document. We have shown how it can be used for

model selection, as in the animal example, by calculating $Cost(T) + Cost(E|T)$. We have also seen how to use to just calculate $Cost(E|T)$ in the language example. This problem contains infinite minimal models and many examples, and the evidence has repetitions.

Finally, we have shown how complex programs are dealt with the last example (probabilistic graph). Note that only the first problem can be handled by the $\mathcal{MC}$ and $\mathcal{PC}$ approaches.

All this shows how the tool can be applied to many different scenarios and applications. There is a suite of examples available on the Web page and also in the distributed code.

During the developing of the tool, different particularities and problems have appeared that have affected the process. The major part of those problems was related to the limitations that the use of a particular Prolog compiler introduces: the user programs to be analyzed are also loaded in the Prolog compiler, so their access without any limitation or modification has been a bit difficult many times.

## 6.2 Future work

The implementation done has not covered all the options in [13]: there are other possibly more efficient coding schemes for programs without repeated examples and also for programs with a huge amount of repetitions.

The tool works only in a specific platform of Prolog, the Yap compiler. The reason is the interaction with the user and the I/O, where each compiler uses different syntax, and also because there is not a standard way to manage large numbers with large precision. With some changes, it could be adapted to run with other compiler, like SICStus or SWI-Prolog.

The work conducted is suitable for analyzing stochastic logic programs, as examples in sections 5.2 and 5.3, but there are also other different possibilities:

- Estimate the best probabilities of a stochastic program where the initial values have been randomly (e.g. uniformly) assigned, by brute force.

- Compare ILP platforms, if each different ILP tool (Progol, FOIL...) generates a different model from facts, our MML tool is able to determine the best one.

# Appendix A

# Coding numbers

The cost of coding a number will depend on whether it is an integer or a real or a rational. In computer terminology, we usually refer to real numbers and rational numbers indistinctly, because real numbers are usually represented by floating-point numbers, which are rational numbers in the end. In fact, we are going to approximate real numbers using rational numbers, because the cost of coding the probabilities is calculated from the rational approximation of the real number.

## A.1  Coding positive integers

An integer number $n > 0$ is coded in binary form as a sequence of $k$ binary digits, where $k = (\log_2 n) + 1$. The first digit of this string is 1, but it is also necessary to use a prefix code to distinguish them from other representations. The prefix code used by Wallace [56] is suitable for this. He used a string corresponding to an integer $n$, which is obtained as the function $head(n)$ followed by 0, where $head(n)$ is the following function to transform an integer to a binary representation:

$$
\begin{aligned}
head(1) &= \Lambda && \text{(an empty string)} \\
head(n) &= head(k-1) && \text{followed by the } k = \lceil \log_2(n) \rceil \\
& && \text{digits of } n, \text{ with } n > 1
\end{aligned} \qquad \text{(A.1)}
$$

In this case, the code string length is $CodeLength(n) = length(head(n)) + 1$, and that is optimal and universal, as Wallace explained, because the expected string length required to encode an integer $n$ selected from some finite-entropy distribution is finite using the $CodeLength$ code, which means that the $CodeLength$ code is feasible for this kind of distributions.

The function $CodeLength$, as we have introduced in section 3.1, equals to:

$$
CodeLength(m) = \log^* m + C \qquad \text{(A.2)}
$$

Where $C = \log_2(2.8665064)$ is a normalization constant chosen to satisfy $\sum_{n=1}^{\infty} 2^{-\log^* n} = 1$. Sometimes the authors use $\log^*$ when they are speaking about $CodeLength$. We want to be distinguish those terms to avoid confusion.

The iterated logarithm $\log^*$ is the number of times the logarithm function must be iteratively applied before the result is less than or equal to 1, and it can be obtained using a recursive function:

$$\log^*(m) = \begin{cases} 0 & \text{if } m \leq 1 \\ 1 + \log^*(\log m) & \text{if } m > 1 \end{cases} \qquad (A.3)$$

Since this recursive definition is not optimal for effective calculations, we need to compute a feasible approximation to the value of $\log^*$ for a certain $n$: for that, Rissanen [50] has suggested the following approximation to the $\log^*$ function:

$$\log^* n = \log_2 n + \log_2 \log_2 n + \log_2 \log_2 \log_2 n \qquad (A.4)$$

The approximation is described better in [56, secs. 2.1.12 – 2.1.16]. So, the coding cost of any integer value that we are going to apply in the toolkit will be:

$$CodeLength(n) = \log_2 n + \log_2 \log_2 n + \log_2 \log_2 \log_2 n + \ldots + C \quad (A.5)$$

## A.2    Coding rational numbers

As mentioned elsewhere, we use rational numbers for probabilities. Rational numbers can approximate a real number to any given precision (and this precision is going to be a modifiable value in the toolkit), so there is no effective loss in generality if we assume that probabilities are represented as rational numbers.

In particular, the following definitions will be used to code the probabilities of rules, which will be a rational number between 0 (excluded, because that rule then could be ignored) and 1 (higher values will be normalized, because they represent probabilities).

Despite other coding systems which approximate real numbers, the chosen one is the one given by Dalgleish [7], in which the denominator of the rational number is the relevant part to estimate the cost of coding that number $n/m$, with $n$ and $m$ being natural numbers:

$$Cost(\tfrac{n}{m}) = CodeLength(m) + \log_2 \phi(m), \quad \text{where} \quad \begin{aligned} n &> 0 \\ m &> 0 \\ n &\leq m \end{aligned} \qquad (A.6)$$

From the real number $\frac{n}{m}$, we first encode the denominator $m$, and then there are $\phi(m)$ choices for the numerator, where $\phi$ is the Euler's totient function. The Euler's totient $\phi(m)$ is the cardinality of the set of integers $x$ such that it satisfies $1 \leq x < m$ and the greatest common divisor of $x$ and $m$ is 1.

The first term of the previous formula ($CodeLength$) codes only the denominator $m$ using the technique described for integer numbers. Like the denominator cannot be 0, we do not have to add an extra bit here.

To code the numerator $n$ such that there are no common factors with $m$ we must code only the Euler's totient.

As example, the cost of coding the real number 0.33333 (approximated by the rational 1/3) is the cost of code $CodeLength(3) + \log 2$, since $\phi(3) = |\{1, 2\}| = 2$ .

# Appendix B

# A short introduction to Prolog

The language used to develop the tool is Prolog, because it is "per se" the most suitable logic language to use: the major part of the examples and code interesting to use lately in the tool is available in this language, it is robust and mature.

The name of the Prolog language comes from an abbreviation for *programmation en logique*, the French words for "programming in logic". It is an old language created in 1972 by a group led by Alain Comerauer and Philippe Roussell in Marseille (France). It is a logic programming language and was first associated with artificial intelligence and computational linguistics. It was one of the first logic programming languages created and nowadays remains among the most popular programming languages, at least in academics. The language is now used in many areas, such as: games, expert systems, automated answering systems and control systems.

Prolog is said to be a declarative language, i.e., the logic is expressed by relations and the execution is done by calling queries over these relations (defined by clauses).

The goal is to find a resolution refutation of one negated query. If this negated query is refuted successfully then the query is set to false. Prolog allows the use of impure predicates for checking whether the value of a predicate may have some side effects, such as printing a value to the screen. But this is also one of the most important problems when programming in Prolog: since everything must be solved by resolution, the control the user input/output and the management of streams is a bit harder than with other language paradigms.

## B.1   Syntax of Prolog

Prolog programs describe relations defined in a set of clauses: each clause has a head and a body (then is known as a rule) or only a head (and then is known as a fact). The head must have only a term, but in the body of the rule we could form compound predicates by conjunction of terms (with the special predicate `,/2`) or by disjunction of them (using the special predicate `;/2`).

Due to the relational nature of many built-in predicates, they can typically be used in several directions, so `length/2` can be used to determine the length of a list (`length([a,b,c],X)`), but also to generate a list skeleton of a given length (`length(List,5)`), and also both things if the arguments are not assigned.

The most elemental term is a constant. They are usually words written in Prolog code without any special syntax, although we must surrounded them by single quotes if it starts by capital letter or contain spaces or non-English symbols.

Numbers can be float or integer numbers. In recent implementations of Yap Prolog (versions higher than 6.0), there is no a big limitation to integer numbers precision if we have compiled it with the multi-precision library GMP, that is the common situation in the major part of packages available for Windows/Unix platforms nowadays. To estimate big floating-point computations we will try to use the MPFR library.

Variables are strings starting with a capital letter and followed by more letters, numbers and/or underscore characters. The single underscore is called an anonymous variable and it means "any term". This type of variable does not represent the same value everywhere it occurs within a predicate definition.

## B.2    Programming with Prolog

A Prolog program is a set of procedures (the order does not matter), each of them composed by one or more clauses (and here the order of the clauses is important, due to the evaluation of each of them is performed in a *top-down* fashion. The objective then is to describe relations using the clauses (facts or rules).

A fact is a rule without any body, which could be considered equivalent to a rule with the only body "true". A goal is the opposite concept: a rule without any head.

Once the facts and rules are built it is possible to make queries to that knowledge (using goals), or modify them in the IDB [1].

The reasoning system that makes logical programming different to other paradigms is that it consists in unifying the rules we have introduced with the goals. The idea of unification in Prolog is close to the concept of *matching*. The Prolog compiler matches two terms if they are equal of if they contain variables that can be instantiated in such a way that the resulting terms are equal. So the process of matching for the compiler generates a SLD-resolution[2] with all the possible matches unifying the goal with the rules of the program by backtracking. This operational strategy needs to create a choice-point each time that multiple clause heads could match with a given call, store it, unify with the first clause which has not used yet, and continue with that goal until it success or fails. If it fails in the course of executing the program (or if the user retries for more possibilities) then all variable bindings that were made since the most-recent choice-point with other clauses are undone, and Prolog follows with the next unused clause of that choice-point. This execution strategy is called chronological backtracking.

---

[1]IDB is the Intensional DataBase of a Prolog compiler, where the rules from the program are stored. The other database is the EDB or Extensional Database, where the facts of the logic during execution are analyzed.

The following example shows this:

```
% lectures(X, Y): person X lectures in course Y
lectures(turing, 9020).      lectures(codd,      9311).
lectures(backus, 9021).      lectures(ritchie, 9201).
lectures(minsky, 9414).      lectures(codd,      9314).
% studies(X, Y): person X studies in course Y
studies(fred,       9020).   studies(jack,    9311).
studies(jill,       9314).   studies(jill,    9414).
studies(henry,      9414).   studies(henry, 9314).
% year(X, Y): person X is in year Y
year(fred, 1).               year(jack,    2).
year(jill, 2).               year(henry,  4).
```

In order to illustrate what happens when Prolog tries to find a solution and backtracks, we draw in Figure B.1 a "proof tree" with the goal `lectures(codd,Course),studies(Student,Course)`, where there are four derivations that are true (the bold lines) and many others that are false.

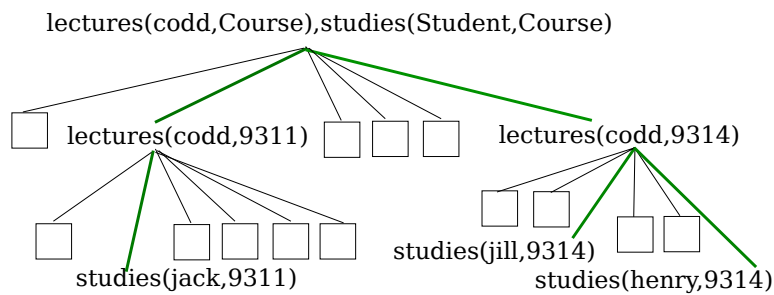

Figure B.1: The backtracking in Prolog

The negation is implemented by the predicate `\+/1`, which drives Prolog to try to prove the goal without the negation. If a proof of that non-negated goal can be found, the original goal fails. If no proof can be found, then the original goal success.

---

[2]Selective Linear Definite clause resolution

# Appendix C

# Prolog interpreters and probabilities

## C.1 Systems implementing probabilities in logic programming

Even though Prolog has been used for implementing the tool, there is no "natural" support for stochastic rules in any widespread compiler of Prolog. This was finally not a problem for our purpose: the idea was to determine the cost of coding probabilistic programs, not to work with them. We only need a basic functionality, as described in section 2.5.

But it is interesting to analyze what other projects and extensions of Prolog have done to integrate probabilities, instead of starting from scratch. Let us see some of these other implementations:

- ProbLog [9]: It has been developed since 2007 by De Raedt's team in the University of Lovaine, by Angelika Kimmig, Bernd Gutmann, Theofrastos Mantadelis, Guy Van den Broeck, Vitor Santos Costa, Gerda Janssens, and the same Luc De Raedt. So this is one of the new attempts to cover stochasticity and so the first we mention.

  It is implemented in Yap Prolog, and we have borrowed from it the way to code the probabilities in the clauses. Despite it is one of the most powerful implementations, with current work and a promising future, ProbLog lacks at the moment the capacity of managing clauses with body, and that is basic for our purpose (programs with body are more difficult to be evaluated as we have seen in section 2.5).

  The way that ProbLog uses probabilistic clauses is by inserting the probability before the rule preceded by `::` (a notation we have copied here):

  ```
  0.8 :: coin(odd).
  0.2 :: coin(X).
  ```

- P-Log [1]:  Proposed in 2007 by Chitta Baral, Michael Gelfond, J. Nelson Rushton and Weijun Zhu.  It extends Answer Set Prolog (a specific compiler of Prolog) with causal probabilities and deliberate actions.  In addition to normal rules, a P-log program may contain a random selection rule that indicates the probability of selecting it unintentionally (because we can also select it by a deliberate action in other rule).  For example, the probability of dying for a certain rat eating arsenic is:

```
random(arsenic).
random(death).
pr(arsenic)=0.4
pr(death | arsenic) =0.8
pr(death | \! arsenic) =0.01
```

  ProbLog way to indicate probabilities is clearer.

- PRISM [27]: Started in 1995, it has been developed in the University of Tokyo by the team led by Taisuke Sato and Yoshitaka Kameya.

  PRISM is an extension of Prolog that allows the expression of complex statistical models as logic programs.  A PRISM program is a usual Prolog program augmented with random variables.  PRISM defines a probability distribution over the possible Herbrand models of a program using predicates `values_x/2` and `msw/2`.

  The execution of a PRISM program is a simulation where values for the random variables are selected stochastically, according to the underlying probability distribution.  PRISM programs can have constraints, usually in the form of equalities between unified logic variables.

  For example, this is the implementation for Bernoulli trials:

```
target(ber,2).
values(coin,[heads,tails]).
:- set_sw(coin, 0.6+0.4).
ber(N,[R,Y]) :-
   N>0,
   msw(coin,R), % Probabilistic choice
   N1 is N + 1,
   ber(N1,Y).    % Recursion
ber(0,[]).
```

  As we can notice, we would need to introduce many changes in the code to allow stochastic models.  It is not so intuitive as ProbLog to attach the probabilities to the code of the program.

- Stochastic Prolog [40]: In the University of Bologna the group of Enrico Oliva developed a framework with promising future, but with no current implementation.  The idea was to use two special predicates for inserting the probabilities: `r/1` for storing the probabilities of a rule and `solve_trace/3` to estimate the probability of a given number of repetitions.

A program in this framework has this aspect:

```
r(49.5):coin(head).
r(49.5):coin(tail).
r(1):coin(manhole).
toss() :- coin(X), continuation(X).
continuation(manhole) :- !.
continuation(_) :- toss().

% ---------------------------
% ?- solve_trace(toss(), 100, T).
```

Again the problem is that the work seems not to be continued since 2008.

- Pepl [29] (by Jianzhong Chen, Stephen Muggleton, James Cussens, and Nicos Angelopoulos) introduces probabilistic choices in the resolution process in 2011. It was used in a framework called MCMCMS (from Nicos Angelopoulosy and James Cussens) to specify models for statistical machine learning in a Bayesian framework. We can consider it an implementation of the failure adjusted maximization algorithm (FAM), but it is far from our purpose to work with these complex programs.

- The CLP(BN) System [5], by Vitor Santos Costa, David Page, and James Cussens (2003). It combines constraint logic programming with Bayesian networks. Again a bit far from our purpose to work with logical programs.

- David Poole [46] worked with the concept of Independent Choice Logic (ILC), also introduced by him in 1997. Essentially it extends logical programming with probability distributions over groups of facts that induce distributions over Herbrand models. This representation can be used as a (runnable) specification for agents that observe the world and have memory. However no more work was conducted.

- CP-logic [55], from Joost Vennekens (2009) merges the basics of CP-logic with the language of Logic Programs with Annotated Disjunctions (LPADs), also used in a similar system known as PITA [48] (2011), of Fabrizio Riguzzi. They share the following syntax for input programs:

```
heads(Coin):1/2 ; tails(Coin):1/2:-
heads(Coin):0.6 ; tails(Coin):0.4:-
fair(Coin):0.9 ; biased(Coin):0.1.

toss(Coin),\+biased(Coin).
toss(Coin),biased(Coin).
```

It computes the probability of queries by transforming the input program into a normal logic program and then calling a modified version of the query on the transformed programs. Again the modification of the original version makes it harder for us to analyze the cost.

## C.2 Which compiler to use

The compiler of Prolog used here has been Yap Prolog, an implementation done by Vitor Santos Costa that has some interesting advantages: It has been written in C++ (so better for efficiency) and it has also been especially improved to support ILP development (many ILP systems have been developed in that compiler, like Aleph and TopLog).

Yap is defined by its developers as a high-performance Prolog compiler developed in the University of Porto [1]. Its Prolog engine is based in the WAM [2], and it follows the Edinburgh tradition [3]. It is largely compatible with the ISO-Prolog standard and with Quintus and SICStus Prolog.

Yap appeared in 1985: its stability is supposed, because its development is still active. Nowadays it is written in C++, but originally it was developed in assembly, C and reflectively in Prolog. At that moment it achieved high performance on *m68k* based machines. Yap is compiled for many 32 and 64 bits platforms and in many different operative systems, such as Sun Solaris, Linux, Microsoft Windows, Mac OS...

Even though SWI-Prolog (other compiler) has more powerful libraries, Yap Prolog is a rather good compiler for this project due to speed. Even with huge infinite SLD-derivations from a clause, it is possible to manage them in a reasonable time. A benchmarking conducted by Bart Demoen [10] demonstrated that among the most known Prolog compilers, Yap obtains a good mark (only outperformed by hProlog):

So if we consider only the most commonly known compilers (SWI-Prolog, SICStus and Yap), the Yap Prolog seems to be a good option for our tool.

---

[1] By the CRACS (`http://cracs.fc.up.pt`) in collaboration with the LIACC (`http://www.liacc.up.pt`)

[2] In 1983, David H. D. Warren designed an abstract machine for the execution of Prolog consisting of a memory architecture and an instruction set. This design became known as the Warren Abstract Machine (WAM) and has become as "de facto" standard target for Prolog compilers, as alternative for the Berkeley Abstract Machine and the Vienna Abstract Machine

[3] Prolog dialect which eventually developed into the standard, as opposed to Marseille Prolog. (The difference is largely syntax.)

| Goal | B-Prolog | ECLIPSe | SWI | SICStus | GNU | Yap | hProlog |
|------|---------:|--------:|----:|--------:|----:|----:|--------:|
| `g(_)` | 130 | 10 | 229 | 90 | 90 | 20 | 10 |
| `g(_,..,_)` | 240 | 120 | 380 | 180 | 130 | 120 | 90 |
| `copy_term/2` | 9030 | 6460 | 20179 | 29530 | 16740 | 5670 | 4980 |
| `findall/3` | - | 1290 | 849 | 2840 | 240 | 450 | 470 |
| Sorting | 373700 | 18850 | 57070 | 175100 | 26610 | 16230 | 25710 |
| Integers | 1040 | 1730 | 18270 | 940 | 1690 | 430 | 480 |
| Floats | 2820 | 3770 | 25139 | 3710 | 3380 | 5280 | 1910 |

Table C.1: Benchmark between Prolog compilers in ms of CPU usage, from [10]

## C.3   Other tools

Nowadays we need an IDE [4] and also a RCS [5] to develop code.

Although Prolog is not so supported by Eclipse natively as other languages, there are some plug ins that allow the integration of the Prolog language into Eclipse [6], the IDE par excellence nowadays:

- Prolog Development Tools (ProDT)
  (`http://prodevtools.sourceforge.net`), the one chosen finally, with the aspect described in Figure C.1.

- The Prolog Development Tool
  (`http://roots.iai.uni-bonn.de/research/pdt`): Rather good, but with some problems when you extend the language of Prolog with new operators (as we are doing), this made us change to the other tool in the middle of the development, with the aspect described in Figure C.2.

- Prolog Plugin, from USP University (Brazil)
  (`http://eclipse.ime.usp.br/projetos/grad/plugin-prolog`).

- SPIDER IDE
  (`http://www.sics.se/sicstus/spider/site/index.html`), a modification of the Eclipse IDE though for SICStus Prolog.

Also for the development of some specific files for the Website we describe later, we needed the Web Development Tools for Eclipse (`http://www.eclipse.org/webtools`), a more common and known plug-in for Eclipse.

---

[4] An integrated development environment (IDE), to have at least a source code editor and a debugger.

[5] Revision control System (RCS): to allow the management of changes to the source code.

[6] Eclipse (`http://www.eclipse.org`) is a multi-language and multi-platform IDE with extensible plug-ins.

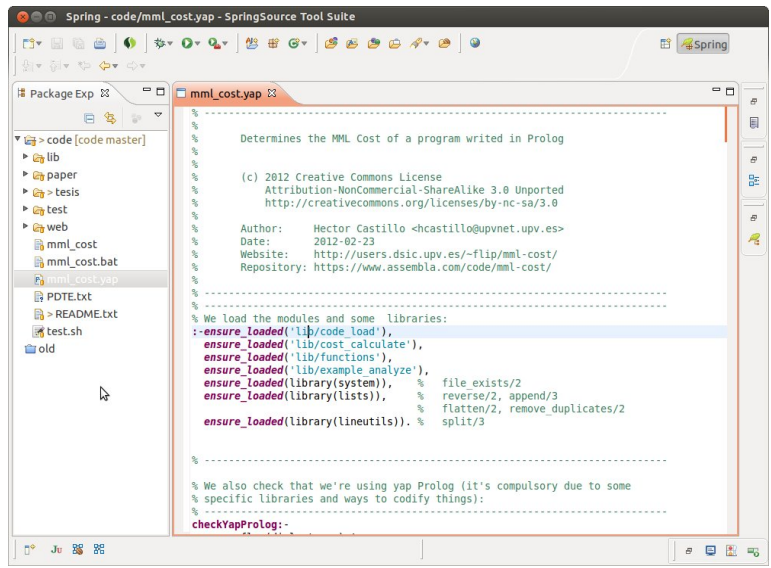

Figure C.1: Aspect of Eclipse with ProDT plug-in

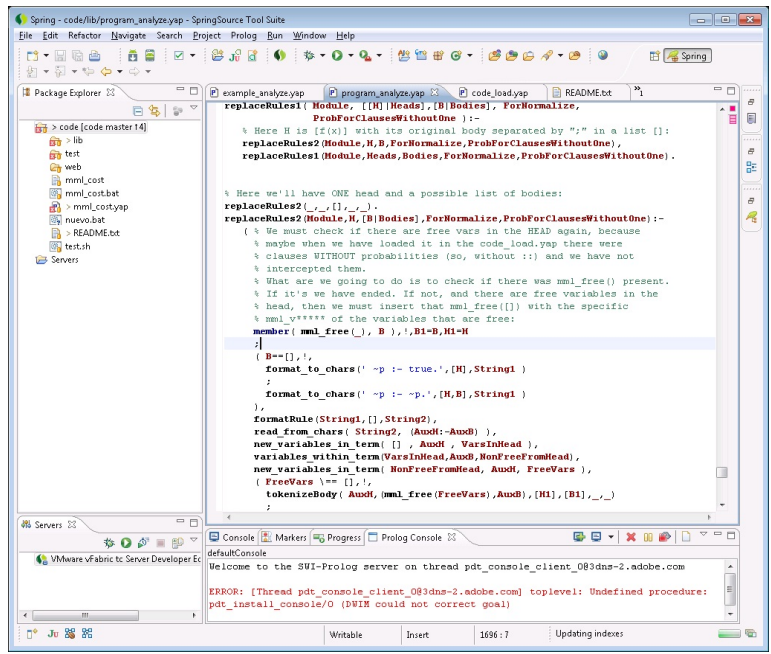

Figure C.2: Aspect of Eclipse with PDT plug-in

The other important tool, as said, it was an RCS. We will use *Git* (`http://git-scm.com`): a distributed revision control and source code management system, created by Linus Torvalds for the Linux kernel development. It is distributed because each copy of the RCS data for control a directory of code is a full repository with complete history and tracking of the code, not dependent on network access or a central server.

Since it is also distributed using the GNU license, it is very common nowadays over the well-known *CVS* and *Subversion*.

The perfect partner for a RCS tool is a Web page that provides a repository to store a back-up or collaborate with others. Even though this second thing was unnecessary (there is no team when you are developing alone), an *Assembla* (`https://www.assembla.com`) repository has been used.

The address of the repository is opened for access to consult the sources of the project (even though it was available at the same time in the Web page of the research group) in `https://www.assembla.com/spaces/mml-cost/new_items`. The change set of the almost three hundred commits done to the source (including Prolog, *PHP* and also the same LaTeX generation of this document) can be viewed there.

## C.4  Restrictions

Because this tool is written in Prolog, its use to analyze Prolog programs needs to set some limitations over the program syntax, due to the limitations of access to the predicates of the code introduced by the user.

For instance, if some user wants to analyze a program, it is loaded as a module in Prolog, the loading introduces some special predicates (`mml_prob/1`, `mml_rep/1`...), and so the use of these words is not allowed in the user code.

Another important limitation: if the user program has modules, the tool will not be able to analyze the code of these other modules. So, the use of modules is forbidden.

As a result, by design, the tool is able to work only with programs introduced by the user that:

- Do not use any predicate starting with the prefix `'mml_'`.

- Do not use modules or different files.

- Do not contain definitions of the same predicate in the knowledge base file and the program file.

# Appendix D

# User manual of the tool

## D.1   How to install the program

The tool is distributed as Prolog source files, so no binary dependence of the platform or compilation is needed. The only requisite to run the tool is to have the Yap Prolog compiler previously installed, and to have it available in the path.

Despite other Prolog compilers, Yap does not include capabilities to calculate big mathematical operations by default, such the GMP library or the MPFR one we use in the tool. So, it is also important to install or compile Yap with the GMP mathematical library. Without the GMP library being active, the tool will also run, but when an operation overloads the machine (easily except for really simple programs), our tool will exit with an error.

The MPFR library is used to calculate large floating-point numbers. It is used only to determine the cost of the evidence. It needs to be available in the system and it is loaded at the start of the program. If this library is not present in the system or the binary used to call it is not executable, then we use an alternative algorithm that fails with more than six hundred examples in the evidence.

Inside the ZIP file there are two executables scripts, one for Windows with the extension `BAT`, and a shell script for Unix/Linux with no extension. These are the entry points to the tool and they must not be removed. It is also a Unix binary file in the subdirectory used to call the MPFR library.

## D.2   Organization

The code is divided in different files: five of them are modules located in a subdirectory. This subdirectory with the modules and the other three files could be moved to any directory in a computer, if there is a Yap compiler available in the local execution path:

This tool could be run on any platform compatible with Yap Prolog (Linux, Windows, Solaris, Mac OS X and HP-UX), but it has been tested only on Windows and Linux.

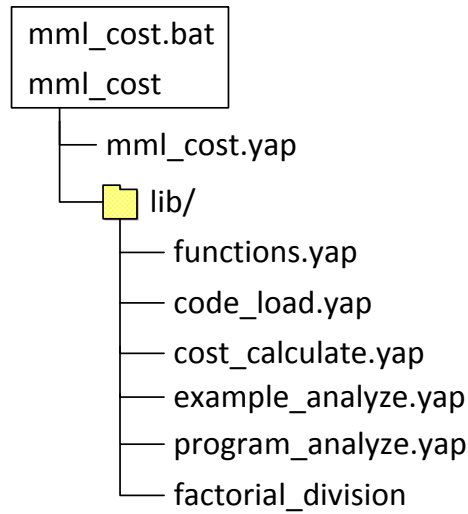

In the main directory, the purpose of the `mml_cost` and `mml_cost.bat` files is only to provide the executables files needed to run the tool in both platforms, and `mml_cost.yap` is the entry point that manages the modules:

- `mml_cost`: Shell script for Linux platforms: it calls `mml_cost.yap`.

- `mml_cost.bat`: Script for Microsoft platforms: it calls `mml_cost.yap`.

- `mml_cost.yap`: Entry point to the tool. It loads the five modules and it is in charge of answering the user with the results of the code, or to notice the errors.

- `lib/functions.yap`: Math involved predicates, global modifiers, print options.

- `lib/code_load.yap`: It manages the loading of the user code. It also intercepts the special function predicates `::/2` and `#/2` to transform them, removing also non-desired predicates involved with debugging.

- `lib/program_analyze.yap`: It obtains the body of all the clauses of the program and processes them (determining the number of variables, number and type of symbols, etc).

- `lib/cost_calculate.yap`: It determines the cost of the user code already loaded as modules and previously tokenized.

- `lib/example_analyze.yap`: It transforms and analyzes the user example files, in a similar way to the `lib/cost_calculate.yap`, but considering only the evidence (it also detects and fails when a body of a clause is found).

- `lib/factorial_division`: A Unix binary that uses the MPFR library to determine the floating-point computations of the fraction described in 4.4.4. It is also possible to compile it to other platforms from the source file `lib/factorial_division.c`.

In any Microsoft/Linux platform, the behavior of the tool is exactly the same: the Yap Prolog compiler is the responsible of transforming the calls to files in order to fit the specific platform format, so there is no necessity to use different code in different platforms.

Rather different is the behavior of the tool depending of which version of Yap Prolog we are going to use: the stable version used in many Linux distributions (Debian, Ubuntu...) is still Yap Prolog 5, instead of Yap Prolog 6.

The old one does not have the predefined predicates `between/3` and `is_list/3`. This problem is detected when we load the code and solved declaring them in `functions.yap`, but there is a more important difference that affects the precision of integer and floating-point calculations: only in Yap Prolog 6 can we have the precision library GMP attached when the code is compiled, and this library is compulsory to make calculations that clearly affect the results (if we have a probability of 0.0019, the result of cost will be hugely different from a probability of 0.002). To avoid problems with this lack of precision, the `functions.yap` will halt when `max_precision` value (a Yap Prolog flag) is not enough to make mathematical calculations.

When we call the program without using any modifier, the tool will fail indicating that you need to call the program with at least one user code file (as a program file) to analyze. It is possible to run the program directly reading from the standard input using the special modifier `--`, in this way:

```
$ echo -e "p(0,1).\np(_,_)." | ./mml_cost --
```

The modifiers will start always with a double dash and the others arguments without them will be considered as files to load as user program code:

```
$ mml_cost --help

It determines the MML cost of a Prolog program
Arguments:
```

```
--examples=e1,e2...   Files with the examples as clauses
--kb=kb1,kb2...       Files with the knowledge base
--debug               Detailed evaluation process
--warnings=off        With 'on', warnings issued by Prolog
                      also abort the execution of the tool
--precision=0.00001   Maximum error allowed in operations
                      You can also use num. of digits (<=15)
--dialect=cprolog     Dialect type: cprolog,iso,sicstus
--predefined          Show the cost of predef predicates
--numbers             Consider the numbers function symbols
--normalize=on        Normalize to 1 the sum of probs with
                      same predicate ('off' to disable)
--rulesprob=zerobitslast  Treatment done to probs of rules:
                      -zerobitslast: Only the relevant (def)
                      -notlast: The last rule prob is ignored
                      -all: All of them
--tabled              Show summarized results
--maxrecursion=20     Max SLD-resolution derivations
[file1.pl..fileN.pl]  At least one file with a prolog
                      program (-- for stdin)
```

```
You can use '::' to indicate the probability of a clause:
    0.25 :: f(X,Y):-X>0,Y is 3.
And also '#' in the examples to indicate the repetitions:
    32 # f(0,1).  ~>  Like writing f(0,1) 32 times
```

So the usual way to call the program will be this (we could omit the examples or the Knowledge Base file):

```
$ mml_cost program_code.pl --examples=example_code.pl \
                           --kb=kb_file.pl
```

This will produce an answer from the tool in which it sums up all the costs, separated by each part:

```
program_code.pl + example_code.pl + kb_file.pl
  Total cost: 106.91945 bits
     Cost of program: 106.91945 bits
        Cost of 5 rules: 5.33713 bits
        Cost of lexicon: 25.66446 bits
              of  3 predicates [sum/3,pos/1,even/1]
              and 2 function symbols [0/0,s/1]
        Cost of heads: 24.92481 bits
        Cost of bodies: 31.10245 bits
        Cost of vars: 19.89060 bits
```

If we need to use more files, we must consider that the cost results will be calculated for each combination of program code and example code. This is applicable to programs and examples, but not to knowledge base (KB). If we indicate more than a KB file, all of the KB files will be merged, so when we call the tool as this:

```
$ mml_cost p1.pl p2.pl --examples=e1.pl,e2.pl \
                       --kb=kb1.pl,kb2.pl
```

That is exactly the same as calling the program in this way:

```
$ mml_cost p1.pl --examples=e1.pl --examples=e2.pl \
                --kb=kb1.pl --kb=kb2.pl p2.pl
```

The same type of results as we have seen in the previous example will be produced, but now it will be repeated four times: `p1-e1`, `p1-e2`, `p2-e1` and `p2-e2`. Taking into account that we have called the program with two program files and two example files, so we must take care that the example files have examples that match with both desired program files.

## D.3   Modifiers

Since there are many modifiers (parameters) available from command line, it is necessary to explain which the purpose of each of them is, and explain which the default values are when we do not indicate one:

- **--examples**: It accepts the file (or files separated by comma) containing the evidence of the program to analyze.

- **--kb**: It accepts the KB file or files separated by comma to load. KB is code which is not present in the same user program code but is required by it. It will be treated separately. Notice that the predefined predicates (part of ISO Prolog and also the specific of Yap Prolog) are ignored and are not the same as the KB. Also notice that any other symbol that appears in the code that is not defined in the KB, in the user program or predefined in Prolog, will be considered a symbol function.

- **--debug**: Detailed evaluation process, where you could see the solving process performed with the evidence, or also the normalization done to the clauses of the programs.

- **--warnings**: Its default value is 'off'. By switching it to 'on', warnings abort the execution of the program. For example, if you introduce a variable which is not assigned, the Prolog compiler will issue a warning that in the normal behavior, is ignored. If you activate this modifier, then the tool will exit abnormally.

- `--precision`: By default the value is 0.00001 (5 digits). This it will be used as a resolution in the output of floating-point values and also (and this is more important) as the maximum error allowed for a rational number. So, with precision 0.0001:

$$0.3333 \rightharpoonup \tfrac{1}{3} \qquad \text{but} \qquad 0.333 \rightharpoonup \tfrac{333}{1000}$$

  You can also use the amount of digits you want (i.e., 6 instead of 0.000001). The maximum resolution is 15 (so $1/10^{15}$).

- `--dialect`: The Yap Prolog developers decided to allow for a change between the specific syntax of CProlog, SICStus and ISO Prolog (ISO/IEC 13211-1)). By default, the "dialect" used is CProlog, but there is the possibility to change to the two others. The most important consideration about the dialects is the incompatibility between SICStus dialect and the use of evidence (`--examples`, due to the impossibility of modifying code).

- `--predefined`: Shows the cost of the predefined predicates in Prolog: without this modifier, they are ignored (sender and receiver know them and we do not need to transfer). These are the predicates like `write/1,nl/0,is/2`...

- `--numbers`: Shows the cost of the numbers, as if they were also function symbols. It is the same situation as before with predefined predicates.

  For instance, if we have this program:

  ```
  p(X):-X is 3+2+r(a)+s(0),write(0),nl.
  ```

  Without these modifiers, the tool reports that the signature is formed by the function symbols `{a/0,s/1}` and the predicate symbols `{p/1,r/1}`. With both modifiers, we obtain new function symbols:

  $$\{\texttt{a/0,s/1,3/0,2/0,0/0}\}$$

  and the predicate symbols will be:

  $$\{\texttt{p/1,r/1,+/2,is/2,write/1,nl/0}\}.$$

- `--normalize`: By default, we normalize the sum of probabilities of rules with same predicate to make them 1. This could be disabled by using the `off` value with this modifier. Let us see first what the normalization implies. If we have this as the whole program:

```
0.3 :: p(0,X):-X > 0.
0.3 :: p(1,X):-X < 0.
```

Since the sum of both probabilities is less than 1, the code will be transformed into:

```
0.5 :: p(0,X):-X > 0.
0.5 :: p(1,X):-X < 0.
```

And the same probability of 0.5 will be assigned in this other one situation (even though is >1, we could deduce that each clause must have a probability of 0.5):

```
1 :: p(0,X):-X > 0.
1 :: p(1,X):-X < 0.
```

When there are clauses without probability (but at least one has probability, otherwise the program will not be considered as a stochastic one), then the tool also deduces the probabilities which are not present by assuming a uniform distribution:

```
0.6 :: p(0,X):-X > 0.
p(1,X):-X < 0.    % --> prob. will be 0.2
p(3,0).           % --> prob. will be 0.2
```

Then if we disable the normalization using this modifier, the tool will fail for any of these examples: whenever there is one clause without probability, the sum of them is not 1, etc.

```
0.8 :: p(a).                    0.8888 :: p(a).
0.1 :: p(b).                    0.1111 :: p(b).
0.4 :: q(a).                    0.4    :: q(a).
q(b).              →            0.3    :: q(b).
q(c).                           0.3    :: q(c).
r(a).                           1      :: r(a).
```

Figure D.1: Example of normalization (left not normalized, right normalized)

- `--rulesprob`: Another important modifier is this involved with the treatment applied to the clauses. Using the Example we are going to explain the differences between the three possible treatments:

  - Value `notlast`: The last rule probability of each different predicate is not code, since it can be inferred from the rest because

probabilities are assumed to be normalized. So, in this situation (Figure , right), we will code 0.8888, 0.4 and 0.3, adding an extra bit for each one of the predicates.

- – Value `all`: All of them are considered (for this example, the six probabilities).

- – Value `zerobitslast`: Only the relevant clauses are considered for evaluating the cost, so in this situation we will only code 0.8888, 0.4 and 0.3, but without the addition of an extra bit per predicate. This is the default behavior.

- `--tabled`: Instead of showing the results as we have seen, they are returned as a tabular list. This option will generate this as output of an execution:

```
;ej4+exampl+kb;Total;Program;CRule,CLexicon;NP;NF;CHeads;
 CBodies,CVars;CProb;CExamples;CKnowledgeBase;
 Predicates;FunctionSymbols;

;prg_8+exampl_8+kb_8;116.34485;116.34485;5.33713;23.41505;1;2;
 78.00000;7.59268;2.00000;0.00000;0.00000;0.00000;
 [sum/3];[0/0,s/1];
;n8_2_3_pr;79.00773;79.00773;4.51854;23.41505;1;2;
 42.00000;6.07414;3.00000;0.00000;0.00000;0.00000;
 [diff/3];[s/1];
```

So all the relevant information of the execution is obtained in a line for each combination of user program and example file (in this case, two) separated by semicolon. The detail of each column is shown in Table D.1.

- `--maxrecursion` Maximum number of SLD-resolution derivations that the tool will generate during the refutation of an example in the evidence. This depth is just use to approximate its probability (currently it is 20 by default if you do not indicate a different value). To understand this, please refer to chapter 2.5.2. In short, this option was introduced to limit the infinite recursion for predicates like `p(X):-p(s(X))` or any other in which a branch could grow the SLD-tree infinitely. If it happens, that branch is discarded when it grows over this limit. So, if the program code does not include many large terms (e.g. long recursive lists), a limit of 20 is enough.

| Column | Name | Description |
|---|---|---|
| 1 | | Name of the modules used in this execution |
| 2 | Total | Total cost (it sums all the other concepts) |
| 3 | Program | Cost of the program |
| 4 | CRule | Cost of the rules |
| 5 | CLexicon | Cost of the lexicon |
| 6 | NP | Number of predicates in the program (also including the ones from the KB files) |
| 7 | NF | Number of function symbols in the program (also including the ones from the example and KB files) |
| 8 | CHeads | Cost of the heads of the clauses in the program |
| 9 | CBodies | Cost of the bodies of the clauses in the program |
| 10 | CVars | Cost of the variables |
| 11 | CProb | Probabilistic cost (zero if there are no probabilities) |
| 12 | CExamples | Cost of the evidence (zero if there is no evidence) |
| 13 | CKnowledgeBase | Cost of the Knowledge Base file (if present) |
| 14 | Predicates | List of the different predicates that appeared in the program and KB files |
| 15 | FunctionSymbols | List of the function symbols that appeared in all the code (KB,examples and program) |

Table D.1: Table of data returned with modifier `--tabled`

## D.4   Code errors

The program shows a detailed message error when it exists abnormally, with a message indicating the source file that it has generated the error or the modifier that is not correctly used. Nonetheless, the program also gives us a return code that could help us to identify the problem, listed in Table D.2.

## D.5   Website organization

As we have mentioned, in section 4.1, there is a Web version of the tool, working as a wrapper for the command-line version.

The website needs a Yap Prolog interpreter (obviously) to work, a web server able to manage PHP files and authorizations to execute the shell. The tool is running currently on a Linux server that uses Debian[1] and HTTPD Server[2] , but it could be also modified to run on another platform with a

---

[1] Debian GNU/Linux, based in the Linux kernel, is a popular and influential Linux distribution urlhttp://www.debian.org

[2] Apache HTTPD is an open-source HTTP server for many operating systems including

PHP interpreter and Yap Prolog installed.

The aspect of the Website is shown in Figure D.2. There are three files:

- **index.html**: The default file which will be accessed by Web users when they will introduce the Web address of the folder where these files are installed.

- **process.php**: It contains the PHP code necessary to call the program from the Website and to format the output for a Web browser.

- **process.sh**: This is the only file we need to set up when we install this tool, to adapt it to other platform (like Windows) or at least to indicate the path to the Yap program or to the tool files. The content of this file is shown in Listing D.1.

Listing D.1: Process to limit in a shell script

```
#!/bin/sh
ulimit -t 15
cd /local_dir_of_the_tool
/path_to_the_compiler/yap -L ./mml_cost.yap
    -- $*
```

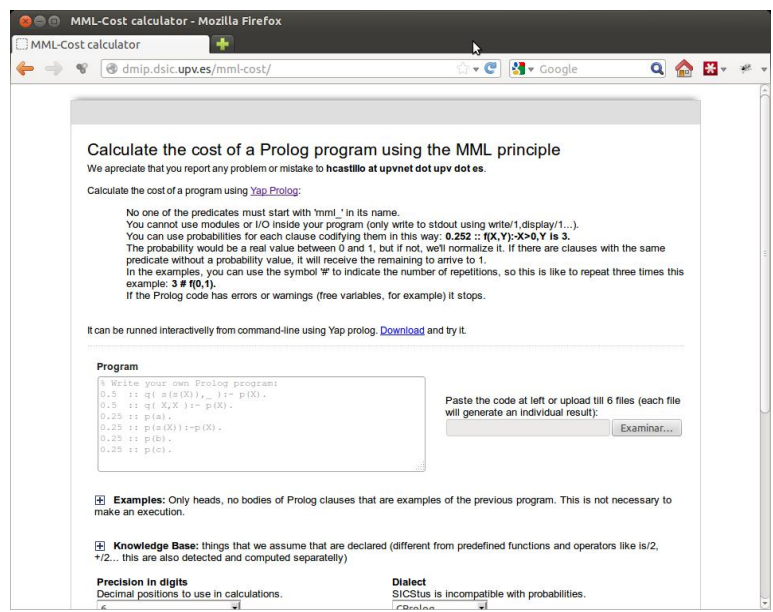

Figure D.2: Aspect of the Website

---

UNIX and Windows: http://httpd.apache.org

The Web version has some limitations but also some extra facilities.

The first is a limitation to avoid the possibility of inserting a huge program (or maybe many of them), trying to run out the resources of the Web server. We then limit the Web server memory and CPU, and also to limit the number of concurrent threads (so, the number of users) that can access to the tool. The second thing is done directly in the configuration of the Web server, whatever we choose.

To obtain this in a Unix-compatible environment is really easy: we introduce the limitation of 15 seconds of maximum execution time in the `process.sh` file (`ulimit -t 15`, as shown in Figure D.1).

The options available in the Website are rather similar to the ones that we have for the tool using it in the command line version of the tool, although we have removed the option to show the results in tabled format.

Another important consideration is that a non-expert user that wants to use this tool needs to be guided with some examples. In this way, the user can understand the cost is calculated, or how the introduction of more evidence or the modification of a probability affects the result.

Another facility is that the Website incorporates a predefined set of examples different from the ones attached to this document, most of them borrowed from the article [13]:

- Case study 2.2, with a simple program and only one example
  `q(s(s(a)),s(s(a)))`:

  ```
  0.5 :: q( s(s(X)),_ ):- p(X).
  0.5 :: q( X,X ):- p(X).
  0.25:: p(a).
  0.25:: p(s(X)):-p(X).
  0.25:: p(b).
  0.25:: p(c).
  ```

- Case study 6, with free variables and also function symbols, but without examples (a non-stochastic example):

  ```
  even(0).
  even(X):-pos(X),even(Y),sum(X,s(s(0)),Y).
  pos(s(_)).
  sum(X,0,X).
  sum(X,s(Y),s(Z)):- sum(X,Y,Z).
  ```

- Case study 7.2, with repeated examples using also a simple program with only `head(coin)` and `tail(coin)`.

- Case study 8.1, where we compare the reachability of nodes with six different programs covering the evidence and comparing the results.

- Case study 8.2, which deals with the sum of natural numbers, using the symbols `0` and `s(_)`.

- Case study 9.2, a known stochastic logic program introduced by Muggleton [38] with animals.

- Case study 9.2, but now using a grammar appeared in J. Cussens [6].

- A final case study to analyze the use of free vars:

```
  1 :: p(X):- r(X,a).
0.5 :: r(a,a).
0.5 :: r(b,b).
0.5 :: r(c,a).

0.5 :: sum(0,Y,Y).
0.5 :: sum(X,s(Y),s(Z)):-sum(X,Y,Z).
```

| Code Error | Description of error code |
|---|---|
| 1 | There have been errors loading user code files: the files are not in the given path, the access to it is forbidden... This error happens when we have not analyzed yet the content of the file. |
| 2 | There is some invalid arguments in the command line. |
| 3 | The Yap compiler does not have the mathematical precision library GMP or the MPFR one. We raise this error when we overload the capacity of calculations without any of these libraries. |
| 4 | The evidence has an example that contains variables, and this can only happen in the program or in the KB. |
| 5 | Example does not match a head of the program or the KB rules. |
| 6 | One example of the evidence has a body (this not possible). |
| 7 | The example has a non-integer value of repetitions before the `#`. |
| 8 | The Prolog code loaded contains errors or warnings: the program loaded, the evidence or maybe the KB (any of them) contains some errors that have been reported previously in lines starting with `%%`. |
| 9 | When we normalize the predicates, we have arrived to an unsolvable situation: there is at least a clause without probability, but the probabilities of the others with same predicate name sum more or equal to 1. |
| 10 | The compiler you are using is not Yap Prolog, or incorrect version or Yap Prolog ($>$5.1, or $>$6.1 for evidence evaluation). |
| 11 | Due to the way we transform the code after loading it, the SICStus dialect is incompatible with the use of examples. |
| 12 | You have used `#` outside of an example file (and this is reserved to indicate the number of repetitions of examples), or maybe you have used `::` in the example file (and this is reserved for code or KB files to indicate the probability). |
| 13 | SICStus dialog used with probabilities (we cannot support this due to the use of `abolish/1` to refurbish the code of the program). Similar to error number 11. |
| 14 | There is a predicate with same functor name declared both in the KB and the program: it could only be declared in one of them, not in both. |
| 15 | One head of the program is repeated in the body without any reduction. |
| 16 | The sum of probabilities of the same predicate is $>$ 1 and you have used the modifier `--dontnormalize` (so you must adjust the probabilities of the program to sum 1 manually). |
| 17 | You have used incompatible options: `--normalize=off` and `--rulesprob=zerobitslast`. It is possible that you have used a program where the probabilities must sum up to 1, but then the normalization has been disabled. |
| 18 | An objective is not valid in the example file. |

Table D.2: Table of program code errors

# Appendix E

# Source code of example speaker recognition

## E.1 Language analysis: source code

Listing E.1: A parser for English language

```prolog
:-use_module(library(readutil)),   % end_of_file/0,
    read_line_to_codes/2
  use_module(library(lineutils)),  % split/3
  use_module(library(system)),     % file_exists/1
  use_module(library(lists)),      % append/3
  use_module(library(terms)).

% [1] Speech load and transformation
% ---------------------------------------------------------
main:- unix( argv(Args) ),
       set_value( mode, syntax ),
       set_value( onlynp, no ),
       set_value( warnings, yes ),
       ( Args==[],
         helpMessage
         ;
         manipulateArgs( Args )
       ).

showMessage( Format, Message ):-
    format( Format, Message ).
showError( Format, Message ):-
    yap_flag(user_error,ErrorStream),
    format( ErrorStream, Format, Message ).

manipulateArgs( ListOfArgs ) :-
    processArgs( ListOfArgs, Files, [] ),
    executeProgram( Files, [], _ ).

processArgs( [ Argument | RestArgs ], Files , Acc ):-
    name( Argument,"--" ),
    processArgs( RestArgs, Files, [user|Acc] ).
```

```prolog
processArgs( [ Argument | RestArgs ], Files , Acc ):-
    (
        name( Argument,"--syntax" ),set_value( mode, syntax )
        ;
        name( Argument,"--lexicon" ),set_value( mode, lexicon )
        ;
        name( Argument,"--word" ),set_value( mode, word )
        ;
        name( Argument,"--np" ),set_value( onlynp, yes )
        ;
        name( Argument,"--nowarnings" ),set_value( warnings, no )
    ),
    processArgs( RestArgs, Files, Acc ).

processArgs( [ Argument | RestArgs ], Files , Acc ):-
    verifyFile(Argument),
    processArgs( RestArgs, Files, [Argument|Acc] ).

processArgs( [], Files, Files ).

% if the file doesn't exists, or is not readable, we exit with error
    :
verifyFile( File ):-
    file_exists( File, 4 ),!
    ;
    exists( File ),!,
    showError('Error:~File~not~readable:~''~p''~n',[File]),
    halt(1)
    ;
    File == '-?',helpMessage
    ;
    File == '-h',helpMessage
    ;
    showError('Error:~File~not~found:~''~p''~n',[File]),
    halt(1).

helpMessage :- nl,write('~Basic~analysis~of~speech'),nl,
    write('~~~--syntax~(by~default)~|~--lexicon~|~--word'),nl,
    write('~~~--np~(also~accept~the~noun_phrase~without~verbs'),nl,
    write('~~~[file1..fileN]~Files~to~analyze~together'),nl,
    write('~'),nl,
    write('~The~content~of~the~file~must~be~English~sentences~
        divided'),nl,
    write('~by~dots.~Non-English~characters~are~forbidden.'),nl,
    write('~Last~line~of~each~file~must~be~an~empty~line~or~its~
        content'),nl,
    write('~will~be~ignored'),nl,nl.

transform1([],[]). % list of sentences
transform1([Elem|List],[Elem1|List1]):-
    transform2(Elem,Elem1),
    transform1(List,List1).

transform2([],[]). % sentence = list of words
transform2([Elem|List],[Elem1|List1]):-
    name(Elem1,Elem),  % word
    transform2(List,List1).

executeProgram( [], Result1, _ ):-transform1(Result1,Result2),
                                   analyzeSentences(Result2),!.
```

```prolog
executeProgram( [File|FileList], Acum, Output ):-
   open(File,'read',Stream),
   % A sentence not ended in a file is considered ended, so a final
        '.' is
   % supposed:
    readLines( Stream, [], Acum, NewAcum ),
   close(Stream),
   executeProgram(FileList, NewAcum, Output ).

readLines( Stream, CurrentSentence, SentencesIn, Result ):-
   % Acum   = [current sentece]
   % Input  = [previous sentences]
   % Output = <result>

   (
      % When this first predicate fails, we have ended:
      read_line_to_codes(Stream,ThisLine),!,
      (
        ThisLine==end_of_file,!,
        ( CurrentSentence==[],!,
          Result=SentencesIn
          ;
          append(  SentencesIn,CurrentSentence, Result )
        )
        ;
        processLine( ThisLine, [],
                     CurrentSentence, NewCurrentSentence,
                     SentencesIn, SentencesOut ),
        readLines( Stream, NewCurrentSentence, SentencesOut, Result )
      )
      ;
      ( CurrentSentence==[],!,
        Result=SentencesIn
        ;
        append(  SentencesIn,CurrentSentence, Result )
      )
   ).

% We have ended the line, so we output the results to readLines/4:
processLine( [], Word, CurrentSentence, NewCurrentSentence,
                        Sentences, Sentences ):-
      Word\==[],!,
      append( CurrentSentence,[Word],NewCurrentSentence )
      ;
      NewCurrentSentence=CurrentSentence.

% New sentence (.), we exclude empty words and empy sentences:
processLine( [46|List],Word,
              CurrentSentence, FinalPendentSentence,
              SentencesIn, SentencesOut ):-
      ( Word\==[],!,
        append( CurrentSentence,[Word],NewSentence)
        ;
        NewSentence = CurrentSentence
      ),
      ( NewSentence \== [],!,
        append( SentencesIn,[NewSentence],NewSentencesIn)
        ;
        NewSentencesIn = SentencesIn
      ),
      processLine( List, [], [], FinalPendentSentence,
                   NewSentencesIn,SentencesOut ).
```

```prolog
% Question mark, so a new sentence also:
processLine( [ 63|List],Word,
              CurrentSentence, FinalPendentSentence,
              SentencesIn, SentencesOut ):-
    ( Word\==[],!,
        append( CurrentSentence,[Word],NewSentence)
        ;
        NewSentence = CurrentSentence
    ),
    ( NewSentence \== [],!,
        append( NewSentence,[[63]],NewSentence1 ),
        append( SentencesIn,[NewSentence1],NewSentencesIn)
        ;
        NewSentencesIn = SentencesIn
    ),
    processLine( List, [], [], FinalPendentSentence,
                 NewSentencesIn,SentencesOut ).

% hyphen also a new word:
processLine( [ 45|List],Word,
              CurrentSentence, FinalPendentSentence,
              SentencesIn, SentencesOut ):-
    ( Word\==[],!,
        append( CurrentSentence,[Word],NewSentence)
        ;
        NewSentence = CurrentSentence
    ),
    ( NewSentence \== [],!,
        append( SentencesIn,[NewSentence],NewSentencesIn)
        ;
        NewSentencesIn = SentencesIn
    ),
    processLine( List, [], [], FinalPendentSentence,
                 NewSentencesIn,SentencesOut ).
% same with ; or :
processLine( [ 58|List],Word,
              CurrentSentence, FinalPendentSentence,
              SentencesIn, SentencesOut ):-
    ( Word\==[],!,
        append( CurrentSentence,[Word],NewSentence)
        ;
        NewSentence = CurrentSentence
    ),
    ( NewSentence \== [],!,
        append( SentencesIn,[NewSentence],NewSentencesIn)
        ;
        NewSentencesIn = SentencesIn
    ),
    processLine( List, [], [], FinalPendentSentence,
                 NewSentencesIn,SentencesOut ).
processLine( [ 59|List],Word,
              CurrentSentence, FinalPendentSentence,
              SentencesIn, SentencesOut ):-
    ( Word\==[],!,
        append( CurrentSentence,[Word],NewSentence)
        ;
        NewSentence = CurrentSentence
    ),
    ( NewSentence \== [],!,
        append( SentencesIn,[NewSentence],NewSentencesIn)
```

```prolog
            ;
            NewSentencesIn = SentencesIn
        ),
        processLine( List, [], [], FinalPendentSentence,
                    NewSentencesIn,SentencesOut ).

% New word, but we exclude empty words ( ) or (,):
processLine( [32|List],[], CurrentSentence, FinalPendentSentence,
            SentencesIn, SentencesOut ):-
        processLine( List, [], CurrentSentence, FinalPendentSentence,
                    SentencesIn, SentencesOut ).
processLine( [32|List],Word, CurrentSentence, FinalPendentSentence,
            SentencesIn, SentencesOut ):-
        append( CurrentSentence,[Word],NewCurrentSentence),
        processLine( List, [], NewCurrentSentence, FinalPendentSentence,
                    SentencesIn, SentencesOut ).
processLine( [44|List],Word, CurrentSentence, FinalPendentSentence,
            SentencesIn, SentencesOut ):-
        append( CurrentSentence,[Word],NewCurrentSentence),
        processLine( List, [], NewCurrentSentence, FinalPendentSentence,
                    SentencesIn, SentencesOut ).

% Any other thing, we add it to the current word:
processLine( [Char|List],Word, CurrentSentence, FinalPendentSentence
        ,
                    SentencesIn, SentencesOut ):-
        ( between(65,90,Char), % we lowercase the char always:
            NewChar is Char + 32
            ;
            NewChar is Char
        ),
        append( Word,[NewChar],NewWord),
        processLine( List, NewWord, CurrentSentence,
            FinalPendentSentence,
                    SentencesIn, SentencesOut ).

% [2] Speech analysis
% ------------------------------------------------------
analyzeSentences(Sentences):-
    length(Sentences,X),
    set_value(counter,0),
    analyzeSentences1(Sentences),
    (
        get_value( mode, syntax ),!,
        ( get_value( warnings, yes ),!,write('% There are '),write(X),
            write(' sentences in the speech'),get_value(counter,Y),
            write('('),write(Y),write(' recognized)'),nl
            ;
            true
        )
        ;
        get_value( mode, lexicon ),!,
        get_value(counter,Y),
        ( get_value( warnings, yes ),!,
            write('% There are '),write(Y),write(' tokens in the speech'),
                nl
            ;
            true
        )
        ;
        get_value(counter,Y),
        ( get_value( warnings, yes ),!,
```

```prolog
        write('%␣There␣are␣'),write(Y),write('␣words␣in␣the␣speech␣'),
            nl
        ;
        true
      )
    ).

analyzeSentences1([]).
analyzeSentences1([Sentence|List]):-
    analyzeSentences2(Sentence),
    ( get_value(mode,syntax),!,
      ( s( Sentence , Result ),
        write(Result),write('.'),increment_counter,nl
        ;
        write('%␣'),write(Sentence),nl
      )
      ;
      true
    ),
    analyzeSentences1(List).

analyzeSentences2([]).
analyzeSentences2([Word|Sentence]):-
    testWord(Word),
    analyzeSentences2(Sentence).

increment_counter :-
    get_value(counter , Value),
    Value1 is Value+1,
    set_value(counter , Value1).

% :-[tokens].
:-[grammar].

% we identify if the token is recognized by our system:

testWord(Word):-
    % we use term_hash/2 to obtain an unique name that represents the
        word,to avoid the
    % problem with word('?'),word('8th') and this kind of problems:
    (get_value(mode,word),write('word('),term_hash(Word,X),write(X),
                          write(').␣%␣'),write(Word),nl,
                          increment_counter;true)
    ,
    det([Word]),!,      (get_value(mode,lexicon),write('lex(det).'),
        increment_counter,nl;true)
    ;
    prep([Word]),!,     (get_value(mode,lexicon),write('lex(prep).'),
        increment_counter,nl;true)
    ;
    wh([Word]),!,       (get_value(mode,lexicon),write('lex(wh).'),
        increment_counter,nl;true)
    ;
    qmark([Word]),!,    (get_value(mode,lexicon),write('lex(qmark).'),
        increment_counter,nl;true)
    ;
    prorel([Word]),!,   (get_value(mode,lexicon),write('lex(prorel).')
        ,increment_counter,nl;true)
    ;
```

```prolog
    aux([Word]),!,       (get_value(mode,lexicon),write('lex(aux).'),
        increment_counter,nl;true)
    ;
    connector([Word]),!,(get_value(mode,lexicon),write('lex(connector)
        .'),increment_counter,nl;true)
    ;
    verb([Word]),!,      (get_value(mode,lexicon),write('lex(verb).'),
        increment_counter,nl;true)
    ;
    adv([Word]),!,       (get_value(mode,lexicon),write('lex(adv).'),
        increment_counter,nl;true)
    ;
    noun([Word]),!,      (get_value(mode,lexicon),write('lex(noun).'),
        increment_counter,nl;true)
    ;
    adj([Word]),!,       (get_value(mode,lexicon),write('lex(adj).'),
        increment_counter,nl;true)
    ;
    ( get_value( warnings , yes ),!,
      write('% not recognized! '),write(Word),nl
      ;
      true
    ).

% [3] Output the model
% ---------------------------------------------------------

% ---------------------------------------------------------
% ---------------------------------------------------------
:-initialization(main).
```

Listing E.2: DCG used in the parser

```
% ------------------------------------------------

s( X , Y ):-
  sentence( sentence(Z), X, [] ),!,transform(Z,Y)
  ;
  get_value( onlynp,yes ),noun_phrase(Z,X,[]),transform(Z,Y).

% ------------------------------------------------

sentence( sentence( X ) ) --> affirmative_sentence( X ) ;
                              subject_interrogative_sentence( X ) ;
                              object_interrogative_sentence( X ) ;
                              imperative_sentence( X ).

imperative_sentence( imperative_sentence( S1,S2 ) ) -->
                         adv_phrase(S1),verb_phrase(S2),!.
imperative_sentence( imperative_sentence( S1 ) ) -->
                         verb_phrase(S1),!.

% they know what they are doing:
affirmative_sentence( affirmative_sentence( S1,S2,S3,S4 ) ) -->
                         noun_phrase(S1),verb1(S2),wh([S3]),
                         affirmative_sentence(S4),!.
affirmative_sentence( affirmative_sentence( S1,S2,S3 ) ) -->
                         adv_phrase(S1),noun_phrase(S2), verb_phrase(S3
                         ),!.

affirmative_sentence( affirmative_sentence( S1,S2,S3 ) ) -->
                         noun_phrase(S1),verb_phrase(S2),adv_phrase(S3)
                         ,!.
affirmative_sentence( affirmative_sentence( S1,S2 ) ) -->
                         noun_phrase(S1), verb_phrase(S2),!.
affirmative_sentence( affirmative_sentence( S1,S2 ) ) -->
                         connector([S1]),affirmative_sentence(S2),!.

object_interrogative_sentence( object_interrogative_sentence( S1,S2,
    S3,S4,S5,S6 ) ) -->
                         wh([S1]),aux([S2]),noun_phrase(S3),verb_phrase
                         (S4),prep([S5]),qmark([S6]),!.
% what progress will we have made?
object_interrogative_sentence( object_interrogative_sentence( S1,S2,
    S3,S4,S5,S6 ) ) -->
                         wh([S1]),noun([S2]),aux([S3]),noun_phrase(S4),
                         verb_phrase(S5),qmark(S6),!.
object_interrogative_sentence( object_interrogative_sentence( S1,S2,
    S3,S4,S5 ) ) -->
                         wh([S1]),aux([S2]),noun_phrase(S3),verb_phrase
                         (S4),qmark([S5]),!.
subject_interrogative_sentence( subject_interrogative_sentence( S1,
    S2,S3 ) ) -->
                         wh([S1]),verb_phrase(S2),qmark([S3]),!.

adv_phrase( adv_phrase( S1,S2 ) ) --> prep([S1]),noun_phrase(S2),!.
adv_phrase( adv_phrase( S1 ) ) --> adv([S1]),!.

% like we remove commas, the adjectives could appear as: "a tiny and
    marvellous" or
```

```
% also "a tiny, wonderful house":
adj_compl( adj_compl( S1,S2,S3 ) ) --> adj([S1]),connector([S2]),adj
    ([S3]),!.
adj_compl( adj_compl( S1,S2 ) ) --> adj([S1]),adj([S2]),!.
adj_compl( adj_compl( S1,S2 ) ) --> adv([S1]),adj([S2]),!.  % manner
    complement
adj_compl( adj_compl( S1 ) ) --> adj([S1]),!.

noun_phrase( noun_phrase(S1,S2,S3,S4,S5) ) --> det([S1]), adj_compl(
    S2),noun([S3]),
                                               connector([S4]),
                                                 noun_phrase(S5)
                                                 ,!.
noun_phrase( noun_phrase(S1,S2,S3,S4) ) --> det([S1]), noun([S2]),
                                           connector([S3]),
                                             noun_phrase(S4),!.
noun_phrase( noun_phrase(S1,S2,S3,S4) ) --> adj_compl(S1),noun([S2])
    ,
                                           connector([S3]),
                                             noun_phrase(S4),!.
noun_phrase( noun_phrase(S1,S2,S3) ) --> det([S1]), noun([S2]),
    prop_pred(S3),!.
noun_phrase( noun_phrase(S1,S2,S3) ) --> adj_compl(S1), noun([S2]),
    prop_pred(S3),!.
noun_phrase( noun_phrase(S1,S2,S3) ) --> det([S1]), adj_compl(S2),
    noun([S3]),!.
noun_phrase( noun_phrase(S1,S2,S3) ) --> det([S1]), noun([S2]),
    rel_phrase(S3),!.
noun_phrase( noun_phrase(S1,S2,S3) ) --> det([S1]), noun([S2]),noun
    ([S3]),!.
noun_phrase( noun_phrase(S1,S2,S3) ) --> noun([S1]), connector([S2])
    ,noun([S3]),!.
noun_phrase( noun_phrase(S1,S2) ) --> det([S1]), noun([S2]),!.
noun_phrase( noun_phrase(S1,S2) ) --> noun([S1]),prop_pred(S2),!.
noun_phrase( noun_phrase(S1,S2) ) --> adj_compl(S1),noun([S2]),!.
noun_phrase( noun_phrase(S1) ) --> noun([S1]),!.
noun_phrase( noun_phrase(S1) ) --> adj_compl(S1),!.

prop_pred( prop_pred(S1,S2) ) --> prep([S1]),noun_phrase(S2),!.
prop_pred( prop_pred(S1,S2) ) --> prep([S1]),verb([S2]),noun_phrase(
    S3),!.  % to reaffirm our situation

rel_phrase( rel_phrase(S1,S2) ) --> prorel([S1]),
    affirmative_sentence(S2).
rel_phrase( rel_phrase(S1,S2) ) --> prorel([S1]),verb_phrase(S2).

% transitive:
verb_phrase( verb_phrase(S1,S2,S3) ) --> verb1(S1),noun_phrase(S2),
    adv([S3]),!.
verb_phrase( verb_phrase(S1,S2,S3) ) --> verb1(S1),noun_phrase(S2),
    prop_pred(S3),!.
verb_phrase( verb_phrase(S1,S2,S3) ) --> verb1(S1),prop_pred(S2),
    noun_phrase(S3),!.  % oi+od
verb_phrase( verb_phrase(S1,S2,S3) ) --> verb1(S1),noun_phrase(S2),
    noun_phrase(S3),!.  % oi+od, but oi without prep
verb_phrase( verb_phrase(S1,S2,S3) ) --> verb1(S1),adj_compl(S2),
    prop_pred(S3),!.  % attributive sentences
verb_phrase( verb_phrase(S1,S2) ) --> verb1(S1),noun_phrase(S2),!.
% intransitive:
verb_phrase( verb_phrase(S1,S2) ) --> verb1(S1),adj_compl(S2),!.  %
    attributive sentences
```

```prolog
verb_phrase( verb_phrase(S1,S2) ) --> verb1(S1),adv([S2]),!.
verb_phrase( verb_phrase(S1,S2) ) --> verb1(S1),prop_pred(S2),!.

verb_phrase( verb_phrase(S1) ) --> verb1(S1),!.

% negation:
verb1( verb( S1,S2,S3 ) ) --> aux([S1]),aux([S2]),verb([S3]),!.
verb1( verb( S1,S2,S3 ) ) --> aux([S1]),verb([S2]),verb([S3]),!.
verb1( verb( S1,S2,S3 ) ) --> aux([S1]),adv([S2]),verb([S3]),!. %
     are now threated
verb1( verb( S1,S2 ) ) --> aux([S1]),verb([S2]),!.
verb1( verb( S1,S2 ) ) --> aux([S1]),prep([S2]),!.
verb1( S1 ) --> verb([S1]),!.

verb([verb(X)]) --> [X], {verb([X])}.
det([det(X)]) --> [X], {det([X])}.
noun([noun(X)]) --> [X], {noun([X])}.
adj([adj(X)]) --> [X], {adj([X])}.
adv([adv(X)]) --> [X], {adv([X])}.
prep([prep(X)]) --> [X], {prep([X])}.
wh([wh(X)]) --> [X], {wh([X])}.
qmark([qmark(X)]) --> [X], {qmark([X])}.
prorel([prorel(X)]) --> [X], {prorel([X])}.
aux([aux(X)]) --> [X], {aux([X])}.
connector([connector(X)]) --> [X], {connector([X])}.

% -----------------------------------------------

:-[tokens].

% -----------------------------------------------

% inside we have only adj(),adj(), then:
transform( A , A ):- atomic(A),!.
transform( A , B ):-
    A =.. [B,C],atomic(C),!.

% [a,b] --> [a',b']
transform( [], [] ):-!.
transform( [Elem|List], [Elem1|List1] ):-!,
    transform(Elem,Elem1),
    transform(List,List1).

transform( A , B ):-
    A =.. [C|List],!,
    transform(List,List1),
    B =.. [C|List1].

% -----------------------------------------------
```

Listing E.3: Lexicon used in the DCG

```prolog
:-use_module(library(lists)).
det([the]).
det([that]).
det([those]).
det([a]).
det([an]).
det([this]).
...

prep([to]).
prep([at]).
prep([on]).
prep([in]).
prep([since]).
...

wh([who]).
wh(['when']).
wh([where]).
wh([how]).
wh([what]).
wh([which]).
wh([whose]).
wh([whom]).

qmark([?]).

prorel([who]).
prorel([that]).
prorel([which]).

aux([does]).
aux([do]).
...
aux([should]).

connector([and]).
connector([or]).
connector([but]).
...

% ----------------------------------------------
% nouns:   s(100004475,1,'organism',n,1,9).
% plural: +s
noun([X]):- wordnet:s(_,_,X,n,_,_),!
                ;
                X \== 'is',remove_es(X,Y),
                wordnet:s(_,_,Y,n,_,_),!.

% verbs: s(200001740,1,'breathe',v,1,22).
% 3rd person: +es|+s
% past: +ed
:-[irregulars].
verb([X]):- wordnet:s(_,_,X,v,_,_),!
                ;
                remove_es(X,Y1),
                wordnet:s(_,_,Y1,v,_,_)
                ;
```

```prolog
                        remove_ed(X,Y2),
                        wordnet:s(_,_,Y2,v,_,_)
                        ;
                        remove_ing(X,Y3),
                        wordnet:s(_,_,Y3,v,_,_)
                        .

% adverbs: s(400048475,2,'now',r,2,182).
adv([X]):- wordnet:s(_,_,X,r,_,_),!.

% adj:     s(300001740,1,'able',a,1,70).
%          s(300003700,1,'dissilient',s,1,0).
adj([X]):- wordnet:s(_,_,X,a,_,_),!
                        ;
                        wordnet:s(_,_,X,s,_,_),!.

remove_es(X,Y):-
    name(X,Z1),append(Z2,[101,115],Z1),name(Y,Z2)
    ;
    name(X,Z1),append(Z2,[115],Z1),!,name(Y,Z2).

remove_ed(X,Y):-
    name(X,Z1),append(Z2,[101,100],Z1),name(Y,Z2)
    ;
    name(X,Z1),append(Z2,[100],Z1),!,name(Y,Z2).

remove_ing(X,Y):-
    name(X,Z1),append(Z2,[105,110,103],Z1),name(Y,Z2).

% we load wordnet, with this structure:
%    s(100002452,1,'thing',n,12,0).
%    s(100002684,1,'object',n,1,51).
%    s(100002684,2,'physical object',n,1,0).
:-[wordnet:wordnet].

% ------------------------------------------------
% Content of irregulars.pl is this:
verb([arisen]).
verb([arose]).
verb([am]).
verb(['are']).
verb([ate]).
verb([beat]).
verb([beaten]).
verb([became]).
verb([become]).
verb([been]).
verb([began]).
...
```

*We can only see a short distance ahead,*
*but we can see plenty there*
*that needs to be done.*

*Alan Turing*